\documentclass{jfm}

\usepackage{graphicx}
\usepackage{lineno}
\usepackage{multirow}
\usepackage{epstopdf, epsfig}
\usepackage{subfigure}
\usepackage{array}
\usepackage[usenames]{color} %
 \usepackage{amssymb} %
 \usepackage{amsmath} %
 \usepackage[table,usenames,dvipsnames]{xcolor}
 \usepackage{bm}
 \usepackage{mathdots}
 \usepackage[utf8]{inputenc} 
 \usepackage{tikz}
 \usetikzlibrary{shapes}
 \usepackage{pgfplots}
 \usepackage{hyperref}
\hypersetup{
	colorlinks=true, %
	linktoc=all,     %
	linkcolor=blue,  %
	citecolor=blue,
	urlcolor = blue
}
\usepackage[noabbrev]{cleveref}

 \usepackage{todonotes}
 \newcommand{\bluelt}{\raisebox{2pt}{\tikz{\draw[blue,solid,line width=0.9pt](0,0) -- (5mm,0);}}}
 \newcommand{\bluedashlt}{\raisebox{2pt}{\tikz{\draw[blue,dashed,line width=0.9pt](0,0) -- (5mm,0);}}}
 \newcommand{\greylt}{\raisebox{2pt}{\tikz{\draw[gray,solid,line width=0.9pt](0,0) -- (5mm,0);}}} 
 
 \newcommand{\greendashlt}{\raisebox{2pt}{\tikz{\draw[black!40!green,dashed,line width=0.9pt](0,0) -- (5mm,0);}}} 
  
 \newcommand{\dottedlt}{\raisebox{2pt}{\tikz{\draw[black,dotted,line width=0.9pt](0,0) -- (5mm,0);}}}
 \newcommand{\dashdotlt}{\raisebox{2pt}{\tikz{\draw[black,dash dot,line width=0.9pt](0,0) -- (5mm,0);}}}
 \newcommand{\dashlt}{\raisebox{2pt}{\tikz{\draw[black,dashed,line width=0.9pt](0,0) -- (5mm,0);}}}

 \newcommand{\greydashedlt}{\raisebox{2pt}{\tikz{\draw[gray,dashed,line width=0.9pt](0,0) -- (5mm,0);}}}
 
 \newcommand{\solidlt}{\raisebox{2pt}{\tikz{\draw[black,solid,line width=0.9pt](0,0) -- (5mm,0);}}}
 \newcommand{\redlt}{\raisebox{2pt}{\tikz{\draw[red,solid,line width=0.9pt](0,0) -- (5mm,0);}}}
 \newcommand{\bluedottedlt}{\raisebox{2pt}{\tikz{\draw[blue,dotted,line width=0.9pt](0,0) -- (5mm,0);}}} 
 \newcommand{\reddashedlt}{\raisebox{2pt}{\tikz{\draw[red,dashed,line width=0.9pt](0,0) -- (5mm,0);}}} 
  \definecolor{Gray}{gray}{0.9}

 \newcommand{\DPP}[2]{\dfrac{\partial{#1}}{\partial{#2}}}
 \newcommand{\DCP}[2]{\dfrac{\bm{\text{d}}{#1}}{\text{d}{#2}}}

 \newcommand{\grad}[1]{\nabla{#1}}

 \newcommand{\Div}[1]{\nabla\cdot{#1}}
 \definecolor{myblue}{rgb}{0.858, 0.188, 0.478}
 \long\def\comment#1{}
 \newcommand*{\Scale}[2][4]{\scalebox{#1}{\ensuremath{#2}}}%

\allowdisplaybreaks[1]
\newcommand{\bI}{\mathsfbi{I}}

\newcommand{\bu}{\bm{u}}

\newcommand{\bB}{\mathsfb{B}}

\newcommand{\bn}{\bm{n}}

\shorttitle{Simulations of turbulence over compliant walls}
\shortauthor{A. Esteghamatian, J. Katz \& T. A. Zaki}

\title{Simulations of turbulence \\ over compliant walls}

\author{Amir Esteghamatian, Joseph Katz 
\and Tamer A. Zaki\corresp{\email{t.zaki@jhu.edu}}}

\affiliation{Department of Mechanical Engineering, Johns Hopkins University,
	Baltimore, MD 21218, USA}

\begin{document}

\maketitle

\begin{abstract}
Direct numerical simulations of turbulent flow in a channel with one rigid and one viscoelastic wall are performed. An Eulerian-Eulerian model is adopted with a level-set approach to identify the fluid-compliant material interface. Focus is placed on the propagation of Rayleigh waves in the compliant material, whose speed depends on the shear modulus of elasticity and whose dominant wavelength depends on the thickness of the viscoelastic layer. These parameters are selected to ensure coupling between the compliant surface and turbulence. When the phase speed of Rayleigh waves is commensurate with the advection velocity of near-wall pressure fluctuations, sheets of vorticity are lifted up and detached near the critical layer and lead to a local pressure minimum.  These events are caused by the inflectional velocity profile near the troughs, and are controlled by the net vorticity flux at the elastic surface. This phenomenon is central to understanding the statistical characteristics of the flow, including the surface deformation-pressure correlation and enhanced stochastic turbulent shear stresses. The pressure work onto the fluid is positive at the surface, which increases the turbulent intensity and wall-normal mixing, without any streamwise momentum gain. Finally, we discuss the influence of three-dimensionality of the surface topography on the generation of streamwise vorticity, secondary motions, and lateral turbulent transport. 
\end{abstract}


\section{Introduction}

When a compliant wall bounds a turbulent flow, the hydrodynamic stresses can lead to deformation of the surface which, in turn, can modify the near-surface flow. 
This two-way coupling has been the subject of active study due to the potential impact of material compliance on laminar-to-turbulence transition \citep{metcalfe1988compliant}, skin friction \citep{bushnell1977effect}, and noise generation \citep{nisewanger1964flow}. 
In the present study, direct numerical simulations (DNS) are performed to examine the two-way interactions between turbulence in channel flow and a viscous hyper-elastic wall, with a particular focus on the role of wave propagation in the compliant material. 

Early investigations of compliant surfaces were inspired by the potential drag-reducing effects of the skin of dolphins \citep{kramer1960boundary,kramer1962boundary}. 
The original idea was that the compliance damps instabilities and hence can delay breakdown of laminar boundary layers to turbulence. 
Theoretical studies confirmed the reduction in the growth rate of classical Tollmien-Schlichting waves, and that material damping inhibits flow-induced surface instabilities \citep{carpenter1985hydrodynamic,carpenter1986hydrodynamic}. These findings were confirmed in experiments \citep{lee1993investigation} and direct numerical simulations \citep{wang2006two}. 
In a turbulent boundary layer, however, there is no consensus regarding the effectiveness of wall compliance to reduce drag. Various experimental studies were performed, some confirming the reduction in drag \citep{fisher1966turbulent,choi1997turbulent}, and others reporting little change compared to the rigid wall \citep{mcmichael1980experimental,llssaman1969turbulent} or even a drag increase \citep{boggs1962performance}\textemdash see \citet{bushnell1977effect} and \citet{gad2003drag} for a comprehensive review. 

An important property of compliant materials is their capacity to sustain the propagation of waves whose speed depends on the shear modulus of elasticity of the compliant layer and whose dominant wavelength depends on the layer thickness. 
\citet{gad1984interaction} investigated the spanwise-oriented structures that travel at wave speeds $U_c$ much smaller than the free-stream velocity $U_0$, typically $U_c\approx 0.05U_0$. 
These static-divergence waves appear when the free-stream velocity exceeds a certain threshold, and are reportedly non-existent in the laminar regime.
\citet{duncan1985dynamics} theoretically confirmed the slow propagation of static-divergence waves when $U_0>2.86U_s$ where $U_s\equiv \sqrt{G/\rho_s}$ is the elastic shear-wave speed, $G$ is shear modulus of elasticity and $\rho_s$ is the density of the compliant material. 
Traveling wave flutter is another type of instability which travels at an advection speed of approximately $0.7U_0$ \citep{gad1986response,duncan1985dynamics}. 

\citet{kulik1991experimental} investigated the frequency band of resonant interactions between turbulent flow and a viscoelastic coating. 
It was concluded that for a hydrodynamically smooth interaction, the surface deformation must be smaller than the thickness of the viscous sublayer, while for an effective drag reduction the band of the interaction frequencies must be in the region of energy-carrying frequencies. 
These conclusions hint at simultaneous effects of the wall compliance on stabilizing/destabilizing the flow near the surface.
Understanding the nature of these interactions can shed light on the impact of wall properties on turbulence and drag.

More recently, \citet{zhang2015integrating} performed tomographic particle image velocimetry (TPIV) of the time-resolved three-dimensional flow in a turbulent boundary layer, and simultaneous Mach Zehnder Interferometry of the two-dimensional deformation at the surface of the compliant wall.
In the one-way coupling regime, where surface deformations are smaller than one wall unit,  \citet{zhang2017deformation} reported two classes of surface motions: (i) non-advected low-frequency component; (ii) ``slow" and ``fast" travelling waves with advection speeds approximately $0.72U_0$ and $U_0$. In addition, the deformation–pressure correlation reached its peak in the log layer at the same location as the  Reynolds shear stress maximum, with the surface deformation lagging the pressure.  This streamwise lag was attributed partly to variations of pressure phase with elevations, and partly to the material damping.
Complementary experiments were performed by \citet{wang2020interaction} to investigate the two-way coupling regime where the surface deformation exceeds several wall units. The authors reported streamwise traveling waves at the fluid-material interface with advection speeds approximately $0.66U_0$, and spanwise waves with an advection speed equal to the material shear speed. The surface deformation, therefore, exhibited a repeated pattern of waves with a preferential spanwise orientation. The most important effects of these waves on the flow were the increase in the near-wall turbulence intensity and a sharp decrease in the streamwise momentum. 

Direct numerical simulations (DNS) were also performed to study drag modification due to compliance. Early investigations modeled the compliant wall as a mass, damper and spring system \citep{endo2002direct,xu2003turbulence,kim2014space,xia2017direct}. In this model the wall pressure fluctuations determine the hydrodynamic forcing on the wall; the surface displacement and wall-normal velocity are obtained by solving the spring-and-damper equations and are used as time-evolving boundary conditions to the flow equations. Using this model, \citet{endo2002direct} reported that the in-phase wall velocity and pressure result in a modest drag reduction of approximately $2.7\%$.  \citet{xu2003turbulence} performed similar simulations, and observed insignificant changes in the near-wall turbulence and drag compared to the rigid-wall simulations. They concluded that it is not possible to obtain an in-phase wall velocity and pressure with a uniform compliant wall, and that the drag reduction reported by \citet{endo2002direct} is possibly a transient effect due to the short simulation time. \citet{kim2014space} studied softer materials with larger surface deformations. They confirmed the out-of-phase correlation between the wall velocity and pressure, and the drag increase due to the additional form drag on the wall. The authors also reported large-amplitude quasi two-dimensional waves propagating in the downstream direction with an advection speed of approximately $0.4U_0$. 
While the spring-and-damper model has been insightful in understanding some space-time characteristics of the compliant wall, it does not account for tangential wall motions which are an important part of wave motion in an elastic layer attached to a rigid wall \citep{rayleigh1885waves}. In addition, since the vorticity flux at the boundary depends on the tangential acceleration of the surface \citep{morton1984generation}, neglecting the tangential wall motion is potentially an unjustified  simplification. 

\citet{rosti2017numerical} performed DNS of turbulent flow over a hyper-elastic compliant wall in a Eulerian-Eulerian framework, where they employed a volume-of-fluid approach to distinguish the fluid and solid phases. Their approach, thus, accounts for the full wall motion and implicitly satisfies the no-slip boundary condition at the interface. The authors reported a drag increase which is inversely proportional to the rigidity of the compliant material. They discussed a correlation between the wall-normal velocity fluctuations and a downward shift in the log-layer profile which also becomes steeper. They related their observations to flows over porous media \citep{breugem2006influence} and discussed them as an extension of flow over rough walls. Within the compliant material, the authors reported that the two-point velocity correlations exhibit oscillating behaviour, and attributed this effect to the typical near-wall flow structures above porous media and rough walls. There was no discussion of wave propagation in the compliant wall. 

Although the roughness effect of surface undulations is important in understanding the impact of wall compliance, a more precise approach must account for wave propagation and material acceleration \citep{jozsa2019active,fukagata2008evolutionary}.  The approach adopted herein is inspired by previous analyses of turbulence-wave interaction.  This topic of research has a long and rich tradition, starting with the seminal work by \citet{miles1957generation,miles1959generation} who examined the critical-layer mechanism for the generation of water-waves, and followed by a large body of work on the interactions of winds and currents with surface gravity waves (see \citet{sullivan2010dynamics} for a review).
For instance, the influence of wave kinematics on mean velocity profile, vertical flux of streamwise momentum, Reynolds stresses, and surface pressure has been subject of various studies in air-sea interactions \citep{sullivan2000simulation,yang2010direct,aakervik2019role,yousefi2020momentum}, and their analysis was aided by introducing appropriate surface-fitted coordinates \citep{hara2015wave,yousefi2020boundary}.  Similar techniques will be adopted herein to examine the implications of wave propagation in the solid material on the adjacent turbulent flow, and conversely, the impact of the flow on the surface motion.

In this work, we perform DNS of turbulent flow interacting with a neo-Hookean material that satisfies the incompressible Mooney–Rivlin law \citep{rivlin1997large} and compare the results to flow over a rigid wall. The material properties are designed to trigger two-way coupling, and the effects of the Reynolds number, compliant-layer thickness and elastic modulus are examined. The flow configuration, governing equations and computational set-up are described in \S\ref{sec:method}. The main body of results, including the mean-flow and turbulence modifications, surface spectra, and phase-averaged statistics are reported in \S\ref{sec:results}. Section \S\ref{sec:conclusion} contains the discussion and concluding remarks.

\section{Methodology \label{sec:method}}
\begin{figure}		
	\centering
	\includegraphics[width =0.7\textwidth,scale=1]{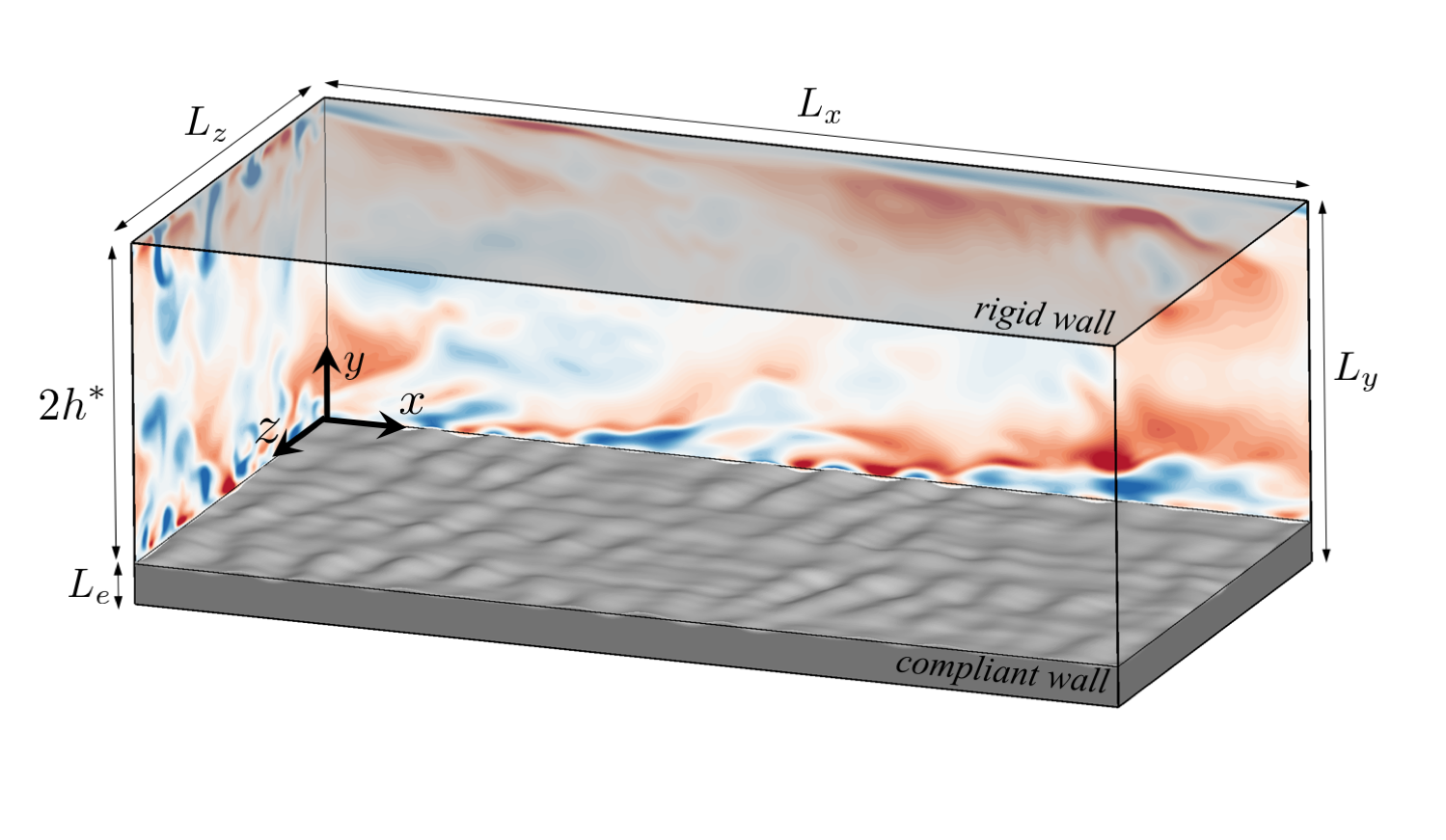}	
	\caption{Turbulent flow  in a channel with a viscoelastic bottom wall. No-slip boundary conditions $\bm{u} = 0$ are imposed at $y = \{-L_e, 2\}$, and periodicity is enforced in the $x$ and $z$ directions. \label{fig:config}	} 
\end{figure}
\subsection{Problem set-up and governing equations} 

The flow configuration is a plane channel with one rigid and one viscoelastic wall (figure \ref{fig:config}), operated at constant mass flux. 
The nominal half-height of the flow region $h^\star$ is selected as the reference length, and the bulk flow speed $U_b^\star$ is adopted as the reference velocity; here and throughout, the `star' symbol indicates dimensional quantities.  
The streamwise, wall-normal and spanwise coordinates are $\{x, y, z\}$. 
Undisturbed, the bottom viscoelastic layer occupies $y\in[-L_e, 0]$, and is attached to a rigid backing at $y=-L_e$.
The bulk Reynolds number is $\Rey \equiv \rho_f^\star U_b^\star h^\star /\mu^\star_f$, where $\rho_f^\star$ and $\mu_f^\star$ are the fluid density and dynamic viscosity.  The friction Reynolds number is therefore $\Rey_\tau \equiv u_\tau^\star \Rey/U_b^\star $, where $ u_\tau^\star$ is the friction velocity $u_\tau^\star \equiv \sqrt{\tau_\text{w}^\star/\rho_f^\star}$ and $\tau_{\text{w}}^\star$ is the mean shear stress at $y=0$.  When $u_\tau^\star$ is adopted for viscous scaling, variables will be designated by superscript `$+$'. When beneficial to scale variables by the friction velocity from a reference rigid-walls simulations, they will distinguished by superscript `$\ast$'.

We adopt an Eulerian-Eulerian model to simulate the motion and deformation of the incompressible, viscous, hyper-elastic layer interacting with an incompressible flow \citep{sugiyama2011full}. In order to identify the fluid-solid interface, we employ a conservative level-set approach  \citep{jung2015effect,you2019conditional}. The non-dimensional mass conservation and momentum equations in terms of the velocity field $\bu$, pressure $p$ and the stress tensor $\bm{\sigma}$ are unified over the entire domain $\varOmega = \varOmega_f \cup \varOmega_s$,
\begin{align}
	& \Div{\bu} = 0, \label{eq:mass}\\
	& \rho \left( \DPP{\bu}{t} + \bu \cdot \grad{\bu}\right) = -\grad{p} + \Div{\bm{\sigma}}. \label{eq:momentum}
\end{align}
The velocity, density and stress fields in the fluid and compliant solid material are denoted by subscripts ``$f$" and ``$s$", and are related to the unified quantities by,
\begin{equation}
	\bu = (1-\Gamma)\bu_s + \Gamma\bu_f, 
	\qquad	 
	\rho = (1-\Gamma)\rho_r + 1, 
	\qquad
	\bm{\sigma} = (1-\Gamma)\bm{\sigma}_s + \Gamma\bm{\sigma}_f,
\end{equation}
where $\Gamma$ is a phase indicator function that is zero in the solid and unity in the fluid phase, and $\rho_r\equiv \rho_s^{\star}/\rho_f^{\star}$ is the solid-to-fluid density ratio. The deviatoric components of the stress in the Newtonian fluid and the compliant solid material are,
\begin{equation}
	 \bm{\sigma}_f = \dfrac{2}{\Rey} \bm{\mathsf{D}}  , \qquad	
	 \bm{\sigma}_s =  \dfrac{2\mu_r}{\Rey}\bm{\mathsf{D}} + \bm{\tau}_e \label{eq:div_stress}
\end{equation}	
where $\mu_r\equiv \mu_s^{\star}/\mu^{\star}_f$ is the ratio between the solid and fluid dynamic viscosities, and $\bm{\mathsf{D}} $ is the strain-rate tensor. In this work, a matching density ($\rho_r=1$) and dynamic viscosity ($\mu_r=1$) in the fluid and solid phases are assumed. 
The neo-Hookean material is modelled as a particular case of the linear Mooney-Rivlin constitutive equation, and the elastic stress $ \bm{\tau}_e$ is \citep{rivlin1997large}
\begin{equation}
    \bm{\tau}_e = G (\bB-\bI)
\end{equation}
where $\bB$ is the left Cauchy–Green deformation tensor, and $G$ and $\bI$ are the modulus of transverse elasticity and the unit tensor. Since the upper convective time derivative of $\bB$ is identically zero \citep{bonet1997nonlinear}, a transport equation can be solved to obtain $\bB$ in a Eulerian manner \citep{sugiyama2011full},
\begin{equation}
    \DPP{\bB}{t} +\bu \cdot \grad{\bB} = \bB \cdot \grad{\bu} + (\bB \cdot \grad{\bu})^
    \top. \label{eq:deformation}     
\end{equation}

A hyperbolic level-set function $\psi$, which varies sharply from zero to unity across the interface between the compliant wall and the fluid \citep{desjardins2008accurate}, is used to track the interface. The phase indicator is thus $\Gamma = 1$ when $\psi \ge 0.5$ in the fluid phase, and $\Gamma = 0$ when $\psi < 0.5$. 
The transport equation for $\psi$ is, 
\begin{align}
    \DPP{\psi}{t} + \Div{\bu \psi} = 0. \label{eq:levelset}
\end{align}
The hyperbolic level-set function is related to a conventional distance function $\varphi$ by, 
\begin{align}
    \psi = \dfrac{1}{2}\left(\tanh \left(\dfrac{\varphi}{2\epsilon}\right) + 1\right)
\end{align}
where $\epsilon \equiv 0.5 \min(\Delta x, \Delta y, \Delta z)$ determines the thickness of the interface marked by $\psi = 0.5$, and $\Delta x, \Delta y, \Delta z$ are the grid sizes in the three physical directions. A re-initialization step is adopted to avoid spurious oscillations at the interface, 
\begin{equation}
    \DPP{\psi}{t'} + \Div{(\psi(1-\psi)\bm{n})} = \Div{(\epsilon (\grad{\psi}\cdot \bm{n})\bm{n})} \label{eq:reinit},
\end{equation}
where $t'$ and $\bm{n}$ are a pseudo-time and the interface normal vector, respectively. 

No-slip boundary conditions $\bm{u} = 0$ are imposed at $y = \{-L_e, 2\}$, and periodicity is enforced in the horizontal $x$ and $z$ directions. Due to the continuity of the velocity and traction at the interface, the no-slip boundary condition is implicitly imposed, and the interfacial tensions are assumed to be zero at $\psi = 0.5$:
\begin{equation}
\bm{u}_f = \bm{u}_s, \hspace{10pt} \bm{\sigma_f \cdot \bn} = \bm{\sigma_s \cdot \bn}.
\end{equation}

\begin{figure}		
	\centering
	\makebox[\linewidth][c]{%
		\subfigure[]{\label{fig:rigidwall_spectra_k_om}
			\includegraphics[height=120pt,scale=1]{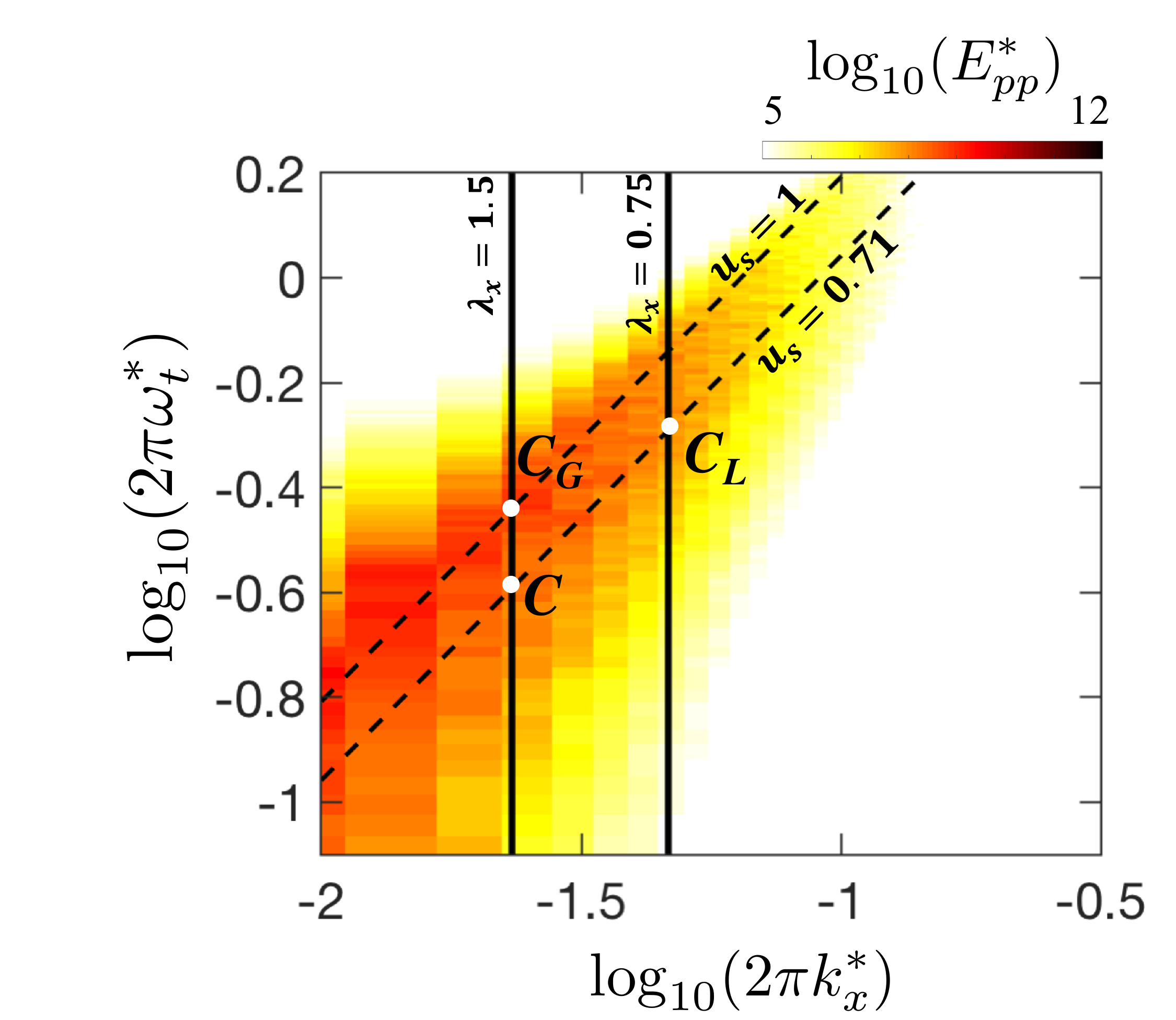} }
		\hspace{0pt}
		\subfigure[]{\label{fig:rigidwall_spectra_uw}
			\includegraphics[height=120pt,scale=1]{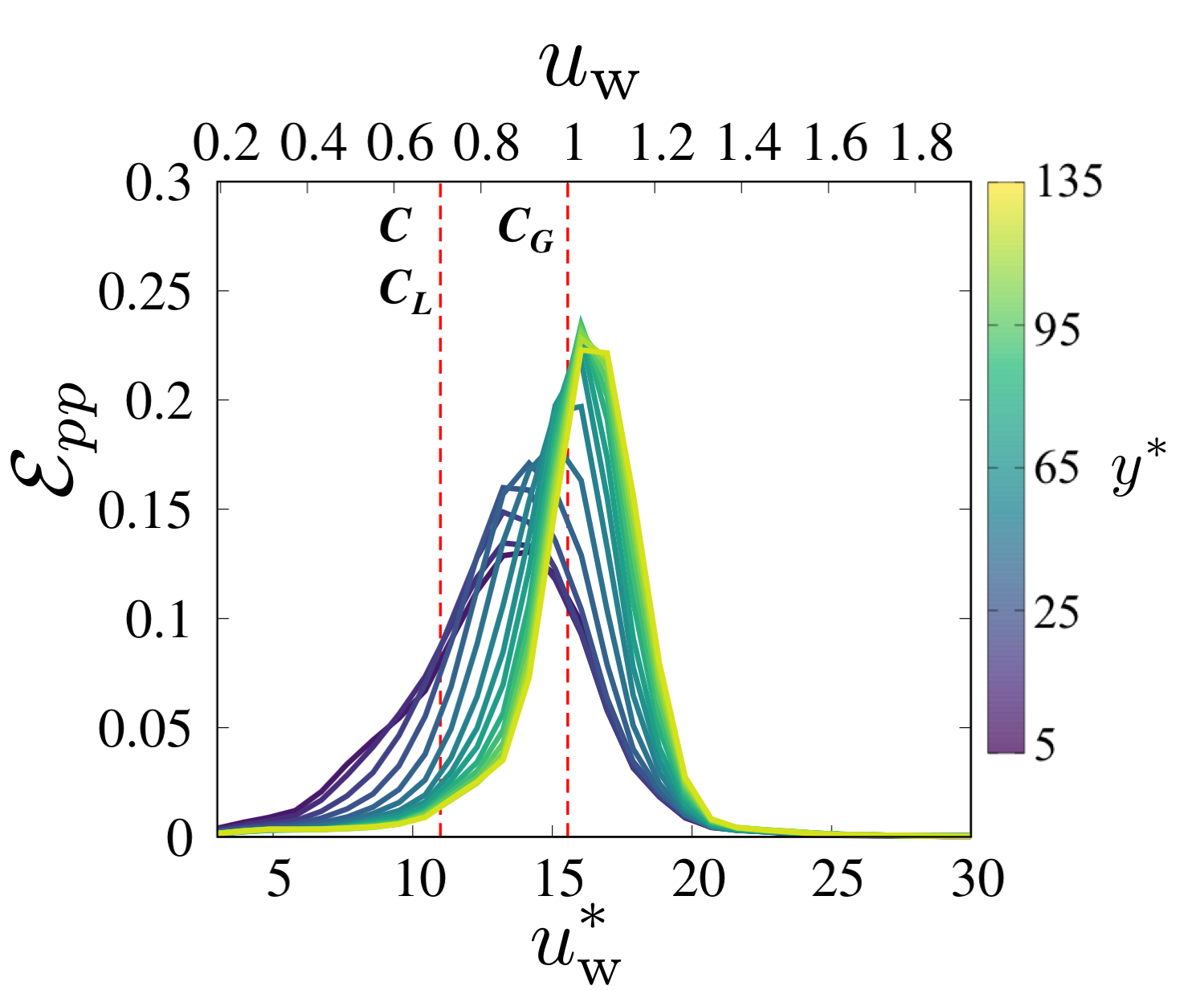} } } 
	\caption{ Pressure spectra from rigid-wall simulation at $\Rey=2800$, case $R_{180}$.  
	(a) Wavenumber-frequency power spectra $E^+_{pp}(k_x,\omega_t)$, at $y^+\approx 5$.  Marked on the contours are estimates from linear models \citep{chase1991generation,benschop2019deformation} for the shear wave speeds in the designed compliant material $u_s$ ({\protect\dashlt}, black) and the peak wavenumber for a given compliant layer thickness $L_e$ ({\protect\solidlt}, black). (b) Profiles of $\mathcal{E}_{pp}$ as a function of wavespeed in viscous (lower axis) and outer (upper axis) units. Shear wave speeds in the designed compliant material are shown by vertical lines ({\protect\reddashedlt}, red).} 
\end{figure}
\subsection{Material properties and flow conditions}
The majority o bf our discussion will focus on a compliant-wall simulation (case $C$) that was designed to ensure two-way coupling with the turbulence at $\Rey=2{,}800$.  Results will be compared to a reference rigid-wall simulation designated $R_{180}$, where the subscript reflects the associated friction Reynolds number.  
We will also examine the impact of the material parameters and the Reynolds number using three additional compliant-wall cases $\{C_G, C_L, C_H\}$ and a rigid-wall simulation $R_{590}$ at a higher bulk Reynolds number $\Rey=10{,}935$. 
In this section, we discuss the design of the simulations and in particular the motivation for our choices of material properties.  The case designations and associated physical parameters are summarized in table \ref{tab:physical_params}.

The design of the main case $C$ attempts to promote interaction between the surface modes and the turbulent fluctuations. According to linear compliant-material models \citep{chase1991generation,benschop2019deformation}, uniform pressure fluctuations lead to a peak surface response at wavelength $\lambda_x^\star = 3L_e^\star$ traveling at the free shear-wave speed $u_s^\star = \sqrt{G^\star/\rho_s^\star}$. Based on these estimates, values of $G$ and $L_e$ can be adjusted such that the peak surface mode is excited at a desired pair of streamwise wavenumber and frequency  ($k_x,\omega_t$). For case $C$, we attempt to have this peak ($k_x,\omega_t$) coincide with the energetic range of the pressure spectra for rigid-wall turbulence, 
\begin{align}
    & E_{pp}(k_x,\omega_t,y) = \langle \hat{p}(k_x,\omega_t,y,z) \times \hat{p}^\dagger(k_x,\omega_t,y,z) \rangle_z, \label{eq:power_spectra_p}\\
    & \hat{p}(k_x,\omega_t,y,z) = \int_{-\infty}^{\infty} \int_{-\infty}^{\infty} p(x,t,y,z) e^{-2\pi i(k_x x + \omega_t t)} \text{d} x \text{d} t, \label{eq:fourier_transform_p}
\end{align}
where $\dagger$ denotes complex conjugate and $\langle \cdot \rangle_z$ indicates averaging in the $z$ direction.
Contours of $E_{pp}$ for the rigid-wall simulation $R_{180}$ at $y^{\ast}  \approx 5$ are plotted in figure \ref{fig:rigidwall_spectra_k_om}.  
Also shown on the figure is the estimated ($k_x,\omega_t$) for peak compliant surface response when $G=0.5$ and $L_e=0.5$, which are the parameters for case $C$; the associated wave speed is $u_s=\sqrt{G/\rho_r}=0.71$ and the wavelength is $\lambda_x = 3L_e =1.5$. 

Two additional configurations are also marked on the figure, namely cases $C_L$ and $C_G$, where the dominant wavelength and shear-wave speed are varied independently. In the former, case $C_L$, the compliant layer thickness $L_e$ is halved compared to case $C$, and in turn the wavenumber of the peak surface mode is doubled (vertical solid lines in figure \ref{fig:rigidwall_spectra_k_om}). For the latter, case $C_G$, the shear modulus of elasticity $G$ is doubled relative to the main case $C$, and therefore the free shear-wave speed is increased to $u_s = 1$ (dashed lines in figure \ref{fig:rigidwall_spectra_k_om}).

The contours of the spectra capture the preferential phase speed $u_w = k_x/\omega_t$ of pressure fluctuations in the rigid channel, which is important in the context of coupling to propagating waves in the material. In order to highlight this connection, we integrate the pressure power spectra for each $u_\text{w}$ and normalize by the total value,
\begin{equation}
    \mathcal{E}_{pp}(u_\text{w},y) = \dfrac{\int_{-\infty}^{\infty} \int_{-\infty}^{\infty} E_{pp}(k'_x,\omega'_t,y)\delta(\omega'_t / k'_x - u_\text{w})\text{d}k'_x\text{d}\omega'_t}{\int_{-\infty}^{\infty} \int_{-\infty}^{\infty}  E_{pp}(k'_x,\omega'_t,y) \text{d}k'_x\text{d}\omega'_t}.   \label{eq:e_pp_function_u}
\end{equation} 
where $\delta$ denotes the Dirac delta function. The resulting $\mathcal{E}_{pp}(u_w)$ is plotted in figure \ref{fig:rigidwall_spectra_uw}, evaluated at different heights in the rigid channel. 
Naturally, the phase speed of the pressure fluctuations increases with height from the wall.  What is important to note, however, are the marked shear-wave speeds for cases \{$C$, $C_L$, $C_G$\}.  All three configurations should be able to couple to the traveling pressure fluctuations in the channel, although the extent of coupling will depend on the amount of energy within specific $(k_x,\omega_t)$ pairs in the coupled simulation; figure \ref{fig:rigidwall_spectra_k_om} only provides a rudimentary but informative guide.

For the influence of Reynolds number, we also considered a compliant case $C_H$ and a corresponding rigid-wall simulation $R_{590}$ at a higher bulk Reynolds number, $\Rey=10{,}935$.  The wall properties for $C_H$ were selected to match those from the main case $C$, in viscous units.  Since the friction velocity is not know \emph{a priori}, the material design was performed using the friction velocities of the corresponding rigid-wall simulations $R_{180}$ and $R_{590}$ (`$\ast$' variables).  The appropriateness of such scaling will be discussed in \S\ref{sec:spectra}, where the  compliant wall responses in cases $C$ and $C_H$ are compared. 

\begin{table}
	\begin{center}
		\def~{\hphantom{0}}
		\setlength\extrarowheight{2	pt}
		\begin{tabular}{l|c|c|c|c|c}	
			Case & $\Rey$ & $G$ & $G^{\ast}$ & $L_e$ & $L_e^{\ast}$  \\ [3pt] \hline  
\rowcolor{blue!20}
            $C$ & 2800 &0.5 & 121 & 0.5 & 90 \\
			$C_G$ & 2800 & 1.0 & 242 & 0.5 & 90 \\
			$C_L$ & 2800 & 0.5 & 121 & 0.25 & 45 \\
            $C_H$ & 10935 & 0.352 & 121 & 0.152 & 90 \\  \hline  
			$R_{180}$ & 2800 & $\infty$ & $\infty$ & $0$ & $0$ \\ 
			$R_{590}$ & 10935 & $\infty$ & $\infty$ & $0$ & $0$ \\ 
		\end{tabular}
	\end{center}
	\caption{Case designations and physical parameters of compliant- and rigid-wall simulations.  \label{tab:physical_params} }
\end{table}

\subsection{Computational details}

The flow equations \eqref{eq:mass} and \eqref{eq:momentum} were solved using a fractional step algorithm on a staggered grid with a local volume-flux formulation  \citep{Rosenfeld1991,wang2019discrete}.
The deformation and level-set transport equations were both advanced using a third-order accurate Runge-Kutta method. 
The viscous terms in \eqref{eq:div_stress} were treated implicitly using the Crank-Nicolson scheme, while an explicit Adams-Bashforth scheme was adopted for advection and stretching terms in \eqref{eq:momentum} and \eqref{eq:deformation}. 
The advection term in \eqref{eq:levelset} was discretized in space using a fifth-order upstream central scheme, while a second-order central differencing was adopted for the compression and diffusion terms in \eqref{eq:reinit}. The level-set equations were solved in a narrow band around the interface only \citep{peng1999pde} to accelerate the computations. Furthermore, equation \eqref{eq:reinit} was invoked every twenty time steps and solved to a steady-state in pseudo-time.
Following \citet{yap2006global}, a global mass correction was employed for the level-set function to preserve the initial compliant material mass. 

Our numerical method has been extensively validated for studies of transition and turbulence in Newtonian and viscoelastic flows \citep{Lee2017,esteghamatian2019dilute,esteghamatian2020viscoelasticity,esteghamatian2021dynamics}; the latter feature the upper convective derivative seen in the evolution equation \eqref{eq:deformation} for $\bB$.  Validation of the interface tracking algorithm was reported by \citet{jung2015effect} who computed the evolution of the Zalesak disc \citep{zalesak1979fully} and the evolution of linear and nonlinear instability waves in two-fluid flows \citep{cheung2010linear,cheung2011nonlinear}. In Appendix \ref{app:validation}, we present an additional validation case to show the accuracy of our two-phase solver in predicting the deformation of a neo-Hookean elastic particle in shear. 

A Cartesian grid was adopted with uniform spacing in the streamwise and spanwise directions, and with cosine stretching in the wall-normal coordinate outside the range $-\delta_m \le y \le \delta_m$. The value of $\delta_m$ was selected such that the deformed material surface remains within this range, which is resolving using a fine uniform grid with $\Delta y^{\ast} = \Delta y^{\ast}_\text{min}$. The domain sizes, grid resolutions and other simulation parameters are summarized in table \ref{tab:numerical_params}. 
The maximum surface displacement $d^{\ast}_\text{max}$ and the height of the uniform-grid region $\delta_m^{\ast}$ are also reported in wall units, using the friction velocity of the rigid-wall simulations (`$\ast$' variables). 

\begin{table}
	\begin{center}
		\def~{\hphantom{0}}
		\setlength\extrarowheight{2	pt}
		\begin{tabular}{l|c|c|c|c|c|c|c|c|c|c|c}	
			Case & $\Rey$ & $L_x$  & $L_y$ & $L_z$ & $N_x \times N_y \times N_z$ & $\Delta x^{\ast}$ & $\Delta y^{\ast}_\text{min}$ & $\Delta y^{\ast}_\text{max}$ & $\Delta z^{\ast}$ &  $d^{\ast}_\text{max}$ & $\delta_m^{\ast}$ \\ [3pt] \hline  
\rowcolor{blue!20}
            $C$   & 2800  & $2\pi$  &$2.5$ & $2\pi$&   $324 \times 364 \times 224$ & 3.5 & 0.8 & 2.6 & 5 & 28 & 60 \\
			$C_G$ & 2800  & $2\pi$ &$2.5$ & $\pi$ & $324 \times 258 \times 112$ & 3.5 & 0.6 & 3 & 5 & 2 & 10 \\
			$C_L$ & 2800  & $2\pi$ &$2.25$ & $\pi$ & $324 \times 296 \times 112$ & 3.5 & 0.6 & 3 & 5 & 12 & 27 \\
			$C_H$ & 10935 & $2\pi$ &$2.152$ & $\pi$  & $384 \times 526 \times 384$ & 9.6 & 0.7 & 7.2 & 4.8 & 28 & 90 \\ \hline
			$R_{180}$ & 2800 & $2\pi$ &$2.0$ & $\pi$ & $ 192 \times 192 \times 112 $ & 5.9 & 0.35 & 3.5 & 4.7 & 0 & 0 \\
			$R_{590}$ & 10935 & $2\pi$ &$2.0$& $\pi$  & $384 \times 384 \times 384$ & 9.6 & 0.55 & 5.6 & 4.8 & 0 & 0 
		\end{tabular}
	\end{center}
	\caption{Domain size, grid resolution, maximum surface displacement $d^{\ast}_\text{max}$ and the height of the uniform grid region $\delta_m^{\ast}$. \label{tab:numerical_params} }
\end{table}

The velocity field in the rigid-wall cases was initialized using a superposition of laminar Poiseuille flow and small-amplitude random fluctuations which trigger breakdown to turbulence. Results were only collected after the flow reaches a statistically stationary state.  The compliant-wall simulations were initialized with a flat material-fluid interface. The initial velocity field was interpolated from a snapshot of the statistically stationary turbulence over a rigid wall. Here too an initial transient elapsed before statistics were collected for sufficiently long duration, e.g.~$T = 550$ convective time units for case $C$, in order to ensure convergence which was verified by comparing results from half and the total number of samples.

\begin{figure}		
    \centering
    \includegraphics[width =0.85\textwidth,scale=1]{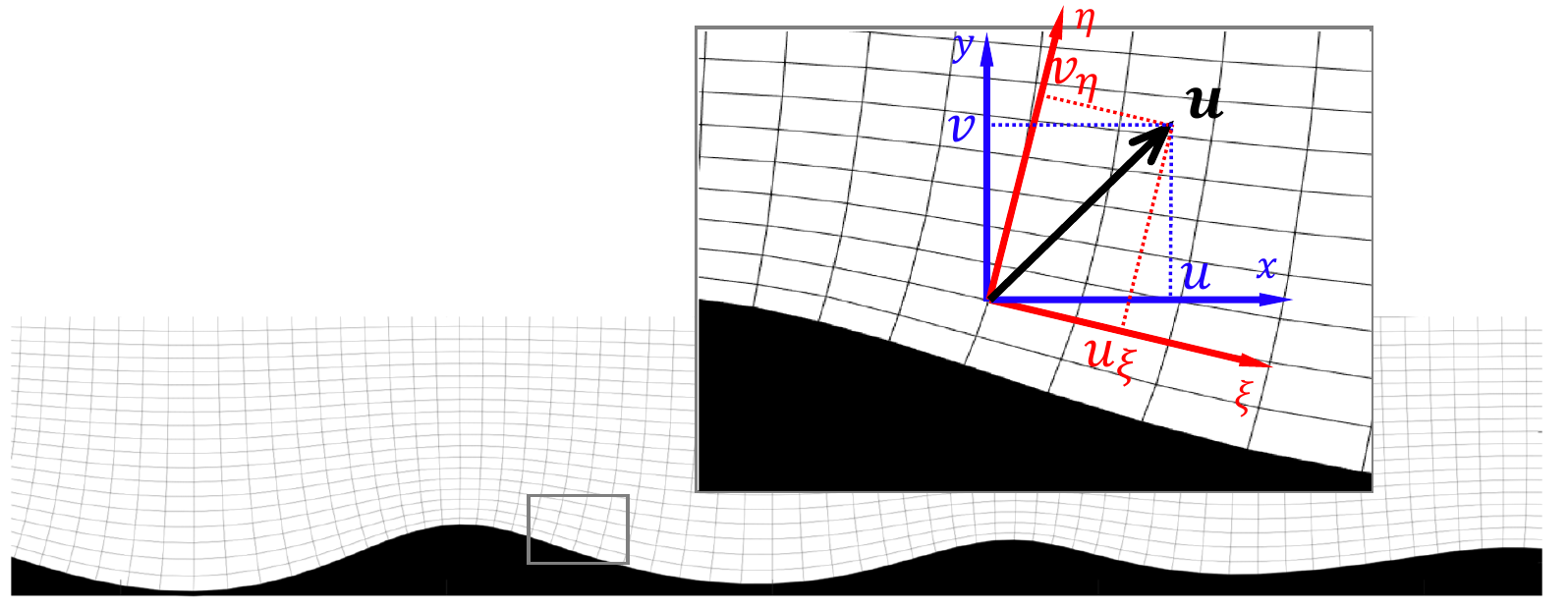}		\caption{Schematic of the velocity vector $\bm{u}$ in Cartesian coordinates ($(u,v)$, blue) and in surface-fitted coordinates ($(u_\xi,v_\eta)$, red). \label{fig:schematic_coord}	} 
\end{figure}

The capacity of a compliant surface to sustain propagating waves is important.  In order to examine how quasi two-dimensional waves interact with the adjacent turbulent flow, we introduce a surface-fitted coordinate system in the $x$-$y$ plane (a detailed description is provided in appendix \ref{app:wallcoordinate}). Figure \ref{fig:schematic_coord} shows the representation of a velocity vector $\bm{u}$ in both the original Cartesian and the adopted surface-fitted coordinates. The contravariant components tangent and normal to the surface in this orthogonal curvilinear coordinate system are $u_\xi$ and $v_\eta$, respectively. 
Phase-averaging was adopted, 
\begin{equation}
    \Phi = \overline{\Phi} + \Phi'' = \langle \Phi \rangle + \underbrace{\tilde{\Phi} + \Phi''}_{\equiv \Phi'} \label{eq:3pl_decomp} 
\end{equation}
where $\overline{\Phi}$ is the phase average and $\Phi''$ denotes the pure stochastic term. The second equality is the triple decomposition \citep{hussain1970mechanics}, where $\overline{\Phi}$ is further decomposed into the average across all phases $\langle \Phi \rangle$ and the wave-correlated part $\tilde{\Phi}$. 
For phase-averaging, crests of streamwise propagating waves $(x_c,d,z_c)$ were identified by satisfying two conditions: (i) the surface displacement $d$ being larger than its root-mean-square, $d>d_\text{rms}$, and (ii) $\partial{d}/\partial{x}$ changing sign. 

\section{Results \label{sec:results}}
\subsection{Global flow modifications  \label{sec:global_changes}}

\begin{figure}		
    \makebox[\linewidth][c]{%
    \subfigure[]{\label{fig:stress_barchart}
        \includegraphics[height=100pt,scale=1]{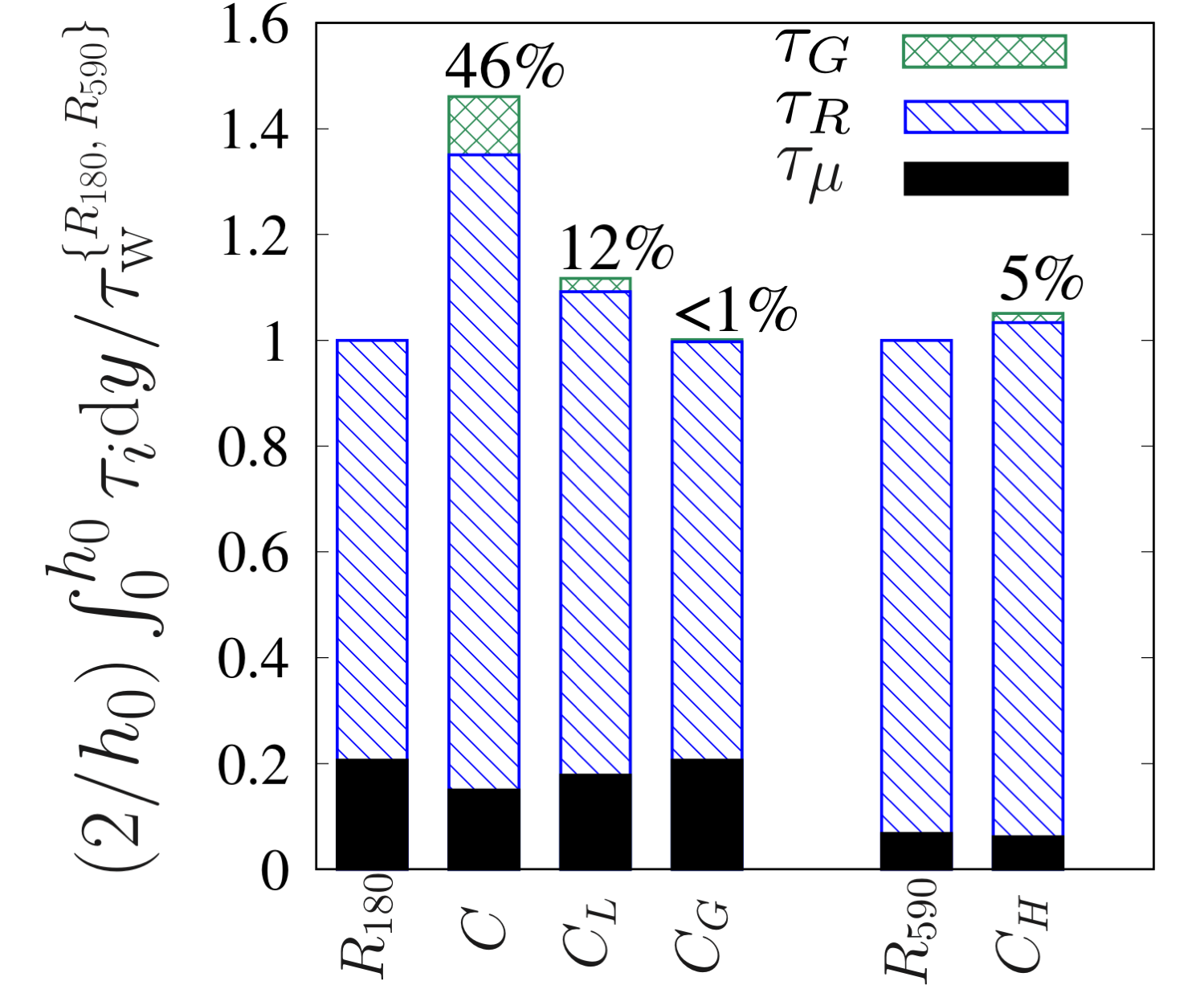} }
    \hspace{-10pt}
    \subfigure[]{\label{fig:stress_profile}
        \includegraphics[height=100pt,scale=1]{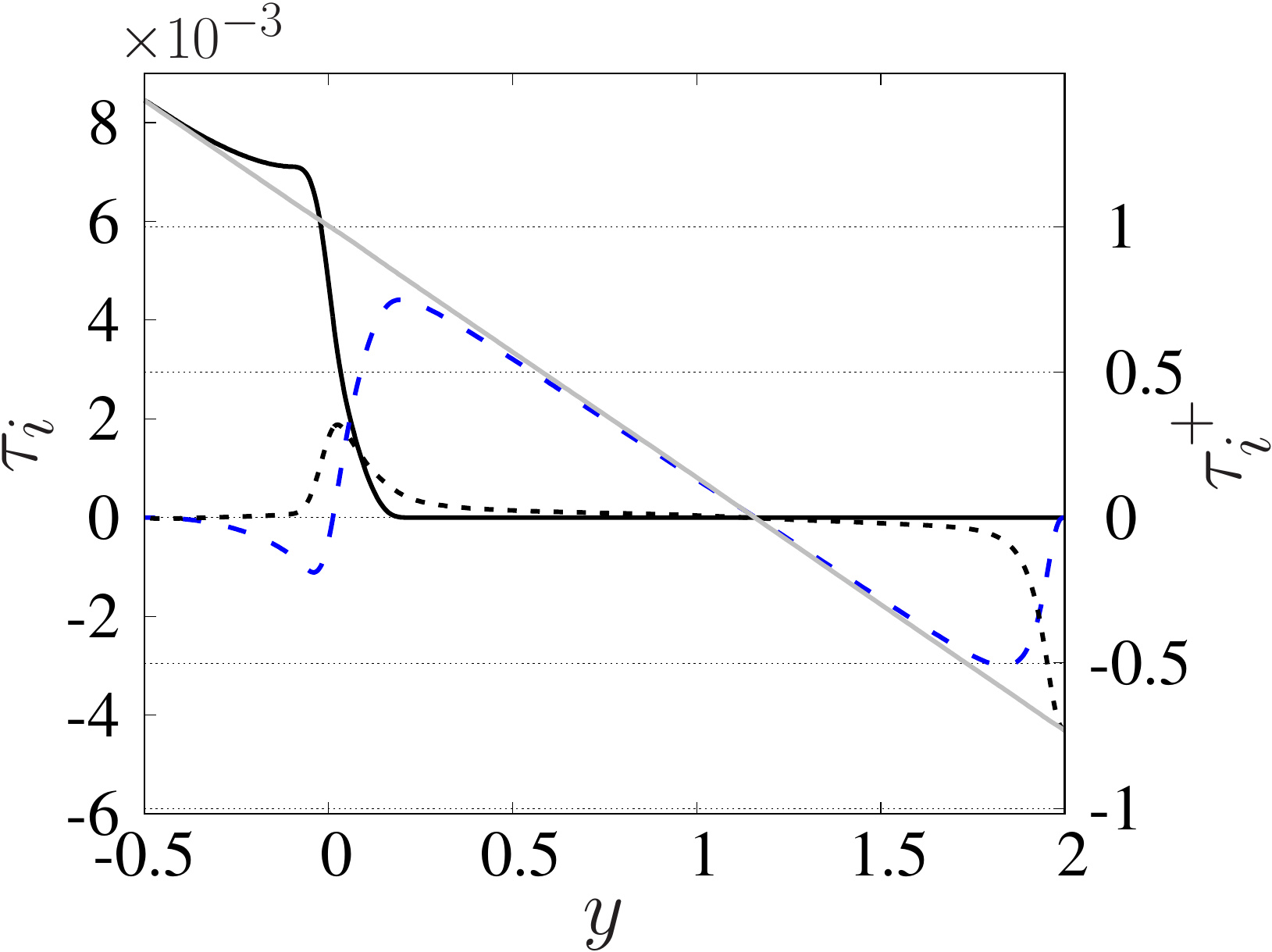} }
    \hspace{-8pt}
    \subfigure[]{\label{fig:mean_profile_Cart}
        \includegraphics[height=100pt,scale=1]{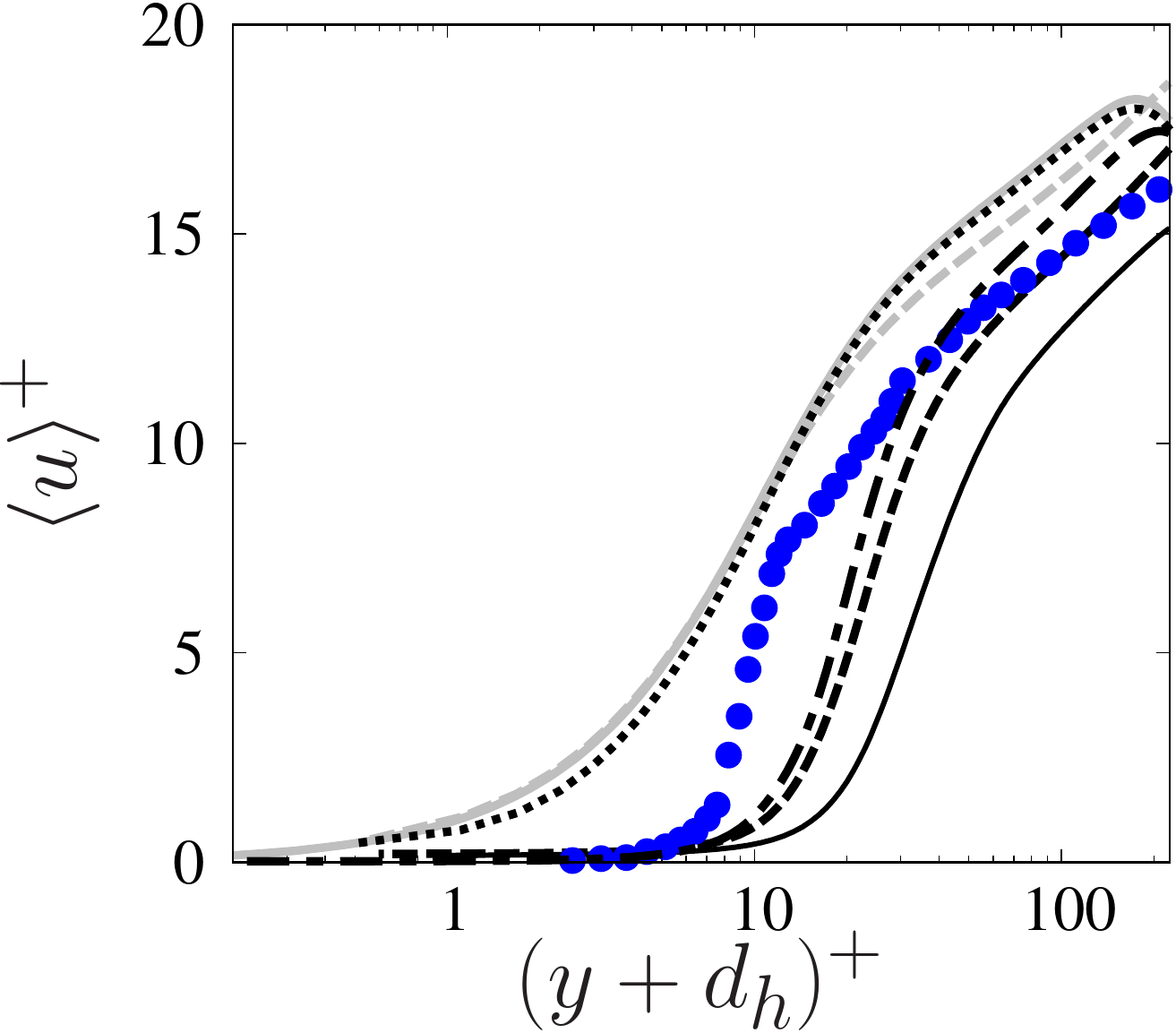} }   } 
    \caption{ (a) Contribution of different stress components to the wall drag (\ref{eq:stressbudget}, \ref{eq:stressbudget_integrated}). 
        (b) Profiles of the stresses for case $C$: 
        viscous stress $\tau_\mu$ ({\protect\dottedlt}, black);
        turbulent Reynolds stress $\tau_{R}$ ({\protect\bluedashlt}, blue);
        elastic stress $\tau_\beta$ ({\protect\solidlt}, black);
        sum of the three components ({\protect\greylt}, grey).
        The stresses are plotted in (left axes) outer and (right axes) wall units.
        (c) Mean streamwise velocity profiles for cases $C$ ({\protect\solidlt}, black), $C_L$ ({\protect\dashdotlt}, black), $C_G$ ({\protect\dottedlt}, black), $C_H$ ({\protect\dashlt}, black), $R_{180}$ ({\protect\greylt}, grey), and $R_{590}$ ({\protect\greydashedlt}, grey), compared to experimental data by \citet{wang2020interaction} at $Re_\tau = 5179$ and $G^\star/(\rho_f^\star {U_0^\star}^2)=0.797$ (blue circle). The $y^+$ coordinate is shifted vertically in order to account for the effect of roughness \citep{jackson1981displacement}.   \label{fig:stress_umean}	}   
\end{figure}

Starting from the Eulerian-Eulerian formulation of the momentum equation \eqref{eq:momentum}, the mean stress in the streamwise direction can be expressed in terms of the unified field variables, 
\begin{align}
    \overbrace{\dfrac{1}{Re} \DCP{\langle u \rangle}{y}}^{\Scale[1.2]{\tau_{\mu}}} ~
    \overbrace{-\langle u' v' \rangle}^{\Scale[1.2]{\tau_{R}}} +  
    \overbrace{G\langle \mathsf{B}_{xy}\rangle }^{\Scale[1.2]{\tau_{e}}} = \left(1 - \dfrac{y}{2}\right) \tau_{\text{w}} +  \dfrac{y}{2} \tau_{\text{w,t}}.
\label{eq:stressbudget}
\end{align}
From left to right, the total stress is comprised of the viscous contribution $\tau_\mu$, the turbulent Reynolds stress $\tau_R$ and the elastic term $\tau_e$. On the right-hand side, $\tau_{\text{w}}$ is the mean shear stress at the nominal height of the compliant surface $y=0$, and $\tau_{\text{w,t}}$ is the mean stress at the top wall $y=2$. We multiply both sides of \eqref{eq:stressbudget} by $2/h_0$, where $h_0$ is the height at which the total stress changes sign, $h_0=1 + (\tau_{\text{w}}+\tau_{\text{w,t}})/(\tau_{\text{w}}-\tau_{\text{w,t}})$, and integrate over $0<y<h_0$,
\begin{align}
    \dfrac{2}{h_0} \int_0^{h_0} \left( \tau_\mu + \tau_R + \tau_e \right)\text{d}y = \tau_{\text{w}}. \label{eq:stressbudget_integrated}
\end{align}
The right-hand side of \eqref{eq:stressbudget_integrated} expresses the wall shear stress at the mean location of the compliant surface, and the left-hand side shows the contribution of different stress constituents. By normalizing \eqref{eq:stressbudget_integrated} with mean wall shear stress in rigid simulation, $\tau_{\text{w}}^{\{R_{180},R_{590}\}}$, we can directly compare the contributors to the stress in flow over a compliant surface with that of a rigid wall (figure \ref{fig:stress_barchart}).
Similar to the recent experimental \citep{wang2020interaction} and numerical \citep{rosti2017numerical} studies, wall compliance increases the drag. The drag increase is associated with an increase in the Reynolds shear stress $\tau_R$, and is largest in the main case $C$. The stress budget is similar in cases $R_{180}$ and $C_G$, which indicates that the stiffer compliant material tends to a rigid wall, and its impact on the turbulence is minimal. 
The average stresses in \eqref{eq:stressbudget} are evaluated in Cartesian coordinates, and therefore differ from averages performed at locations that are equidistant to the surface.
In order to resolve this issue, in \S\ref{sec:formdrag} we adopt wave-fitted coordinates which also allow us to compute the stress due to the pressure acting on the deformed interface.

Note that the drag increase in case $C_H$ relative to the rigid wall $R_{590}$ is only $5\%$, compared to the $46\%$ drag increase in case $C$ relative to $R_{180}$.  This difference is despite channels $C_H$ and $C$ being designed to have the same amplitude and wavenumber of surface displacement in viscous units. From a roughness perspective, both cases $C_H$ and $C$ belong to a `transitionally rough' regime with $d^+<28$.  Therefore, the Reynolds number is expected to influence the normalized drag \citep{nikuradse1950laws}. A similar trend was observed in the experiments by \citet{wang2020interaction}. The authors reported that the drag increase due to wall compliance relative to a rigid wall reduced from $10.7\%$ to $5.0\%$ when the Reynolds number was increased by $84\%$. 

Figure \ref{fig:stress_profile} shows the stress profiles in case $C$ where the drag increase is most substantial. Except in case $C_G$ which is almost in the one-way coupling regime, the trends are similar in other compliant cases and therefore are omitted for brevity. The total stress profile, plotted in outer (left axis) and wall (right axis) units, shows the sum of left-hand side terms in \eqref{eq:stressbudget}. The total stress varies linearly with $y$, and its magnitude is larger by approximately $33 \%$ at $y=0$ than at $y=2$. An important observation is that the turbulent shear stress changes sign near the surface. This effect is consistently observed in all compliant cases, and will be discussed in detail in \S\ref{sec:formdrag}.  

Figure \ref{fig:mean_profile_Cart} shows the mean velocity profiles in a semi-logarithmic coordinate in wall units. 
For the sake of consistency with the previous literature, the data are first presented in a standard Cartesian coordinate and without any phase-specific conditional sampling. Since the effective location where the mean drag is exerted on the surface may not coincide with the nominal surface height, we use a vertical displacement $d_h$ proposed by \citet{jackson1981displacement} to shift the coordinate. This model is widely used in the study of turbulent flows over rough walls \citep{leonardi2010channel,ismail2018effect}, permeable walls \citep{breugem2006influence} and, more recently, compliant surfaces \citep{rosti2017numerical}. The value of $d_h$ is chosen in a way to attain a constant slope in the inertial range, i.e. $(y+d_h)^+ \left({d\langle u^+ \rangle}/{dy^+}\right)$ remains approximately constant over the log-layer.  The enhanced drag is accompanied by a downward shift in the logarithmic region, even if $d_h=0$. The reduced momentum over compliant material is evident in the shown experimental data by \citet{wang2020interaction} at $\Rey_\tau = 5179$ and $G^\star/(\rho_f^\star {U_0^\star}^2)=0.797$, where $U_0^\star$ is the free-stream velocity. 
These data were sampled in the flow only and, therefore, are not contaminated by samples within the material; they are also plotted with $d_h=0$.  

Unlike the experiments, the computational results are not conditioned on the fluid phase, and hence include samples from the compliant material.  In addition, both the experimental and numerical results are plotted in Cartesian coordinates with reference to the nominal interface height, as opposed to the instantaneous interface position. Therefore, averages at a fixed $y$-location include samples from a range of distances to the surface, which most significantly affects the statistics near the interface. In order to better capture the mean flow in the viscous sub-layer, we adopt a surface-fitted coordinate which follows the interface near the compliant surface and smoothly transitions to a Cartesian coordinate with distance from the interface (see appendix \ref{app:wallcoordinate} for details of the surface-fitted coordinates).

\begin{figure}		
    \subfigure[]{\label{fig:umean_180}
        \begin{minipage}[b]{0.4\textwidth}
        \begin{center}
        \includegraphics[width =\textwidth,scale=1]{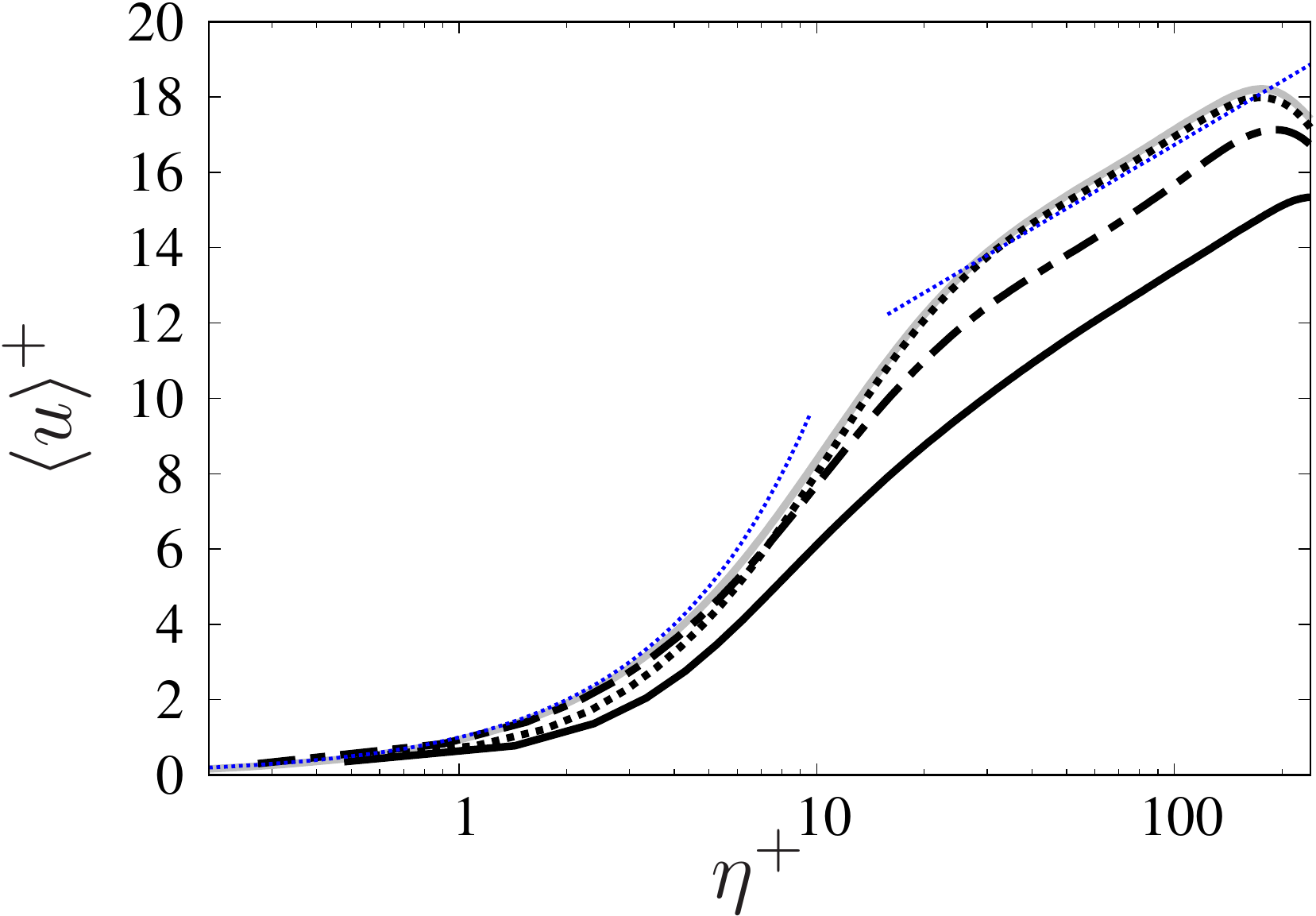}
        \end{center}
        \end{minipage} }
    \subfigure[]{\label{fig:umean_590}
        \begin{minipage}[b]{0.4\textwidth}
        \begin{center}
        \includegraphics[width =\textwidth,scale=1]{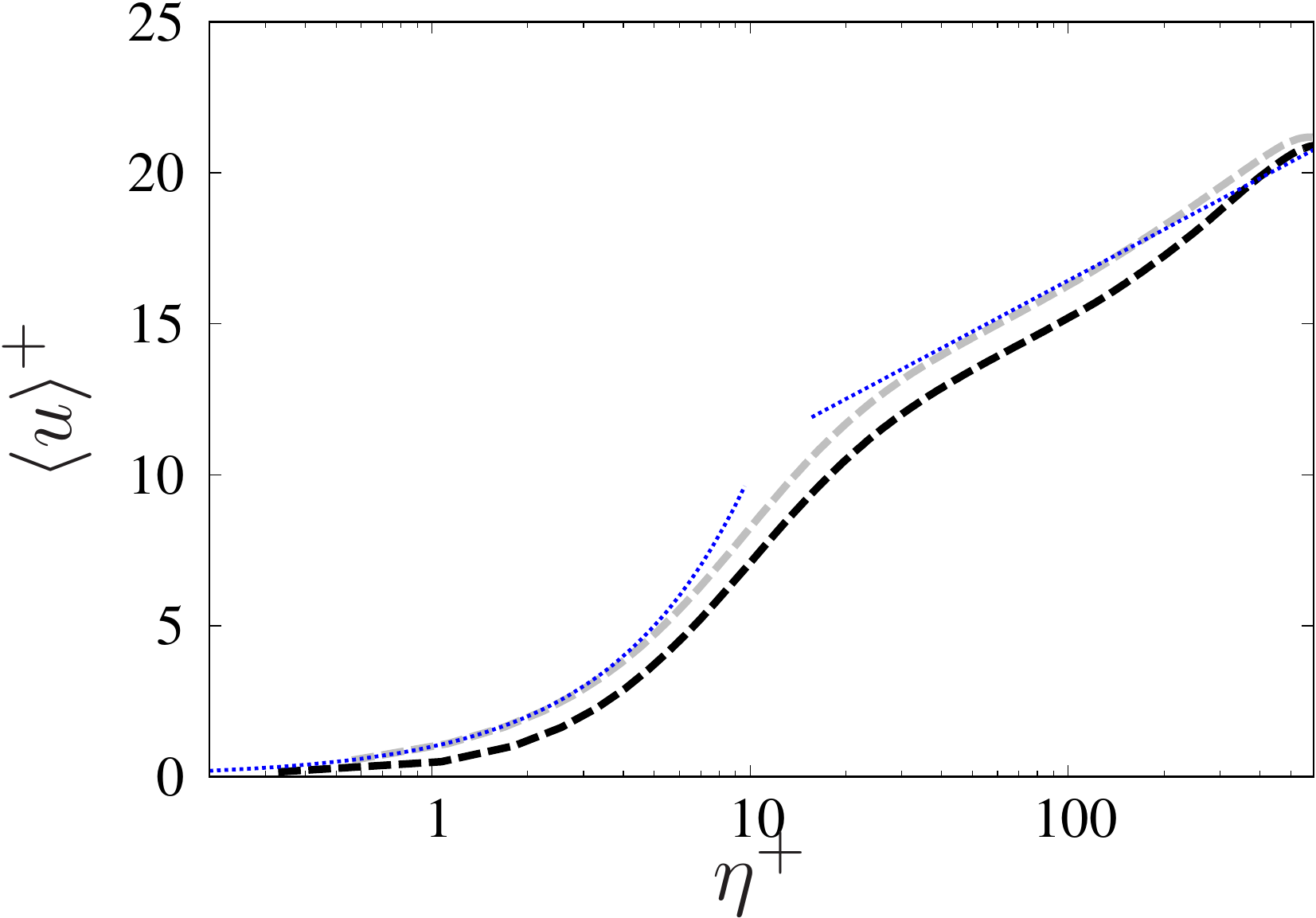}
        \end{center}
        \end{minipage} }
    \caption{Mean streamwise velocity profiles in a surface-fitted coordinate: (a) $C$ ({\protect\solidlt}, black), $C_L$ ({\protect\dashdotlt}, black), $C_G$ ({\protect\dottedlt}, black), $R_{180}$ ({\protect\greylt}, grey), (b) $C_H$ ({\protect\dashlt}, black) and $R_{590}$ ({\protect\greydashedlt}, grey). The log-law for a smooth wall, $u^+=(1/0.41)\log \eta^+ + 5.5$, and the viscous sub-layer velocity profile, $u^+ = \eta^+$, are also plotted for reference ({\protect\bluedottedlt}, blue).    \label{fig:umean}	}   
\end{figure}

Figure \ref{fig:umean} shows the mean-velocity profiles compared to the smooth-wall simulations. The momentum deficit in the log-layer is still observable, particularly in cases $C$, $C_L$ and $C_H$. The slope of the log-layer, however, does not change significantly, similar to the experimental observations by \citet{wang2020interaction}. The viscous sublayer is still retained for the most part, and a decrease in momentum in the buffer layer is observed only in cases $C$ and $C_H$ which, as will be discussed, experience large surface displacements $d^+ \approx 20$. In case $C_G$ the mean profile tends to that from the rigid-wall simulations, similar to trends reported by \citet{wang2020interaction} and \citet{rosti2017numerical} for stiffer material.  

\begin{figure}		
    \subfigure[]{\label{fig:vis_C0}
        \begin{minipage}[b]{0.45\textwidth}
        \begin{center}
        \includegraphics[width =\textwidth,scale=1]{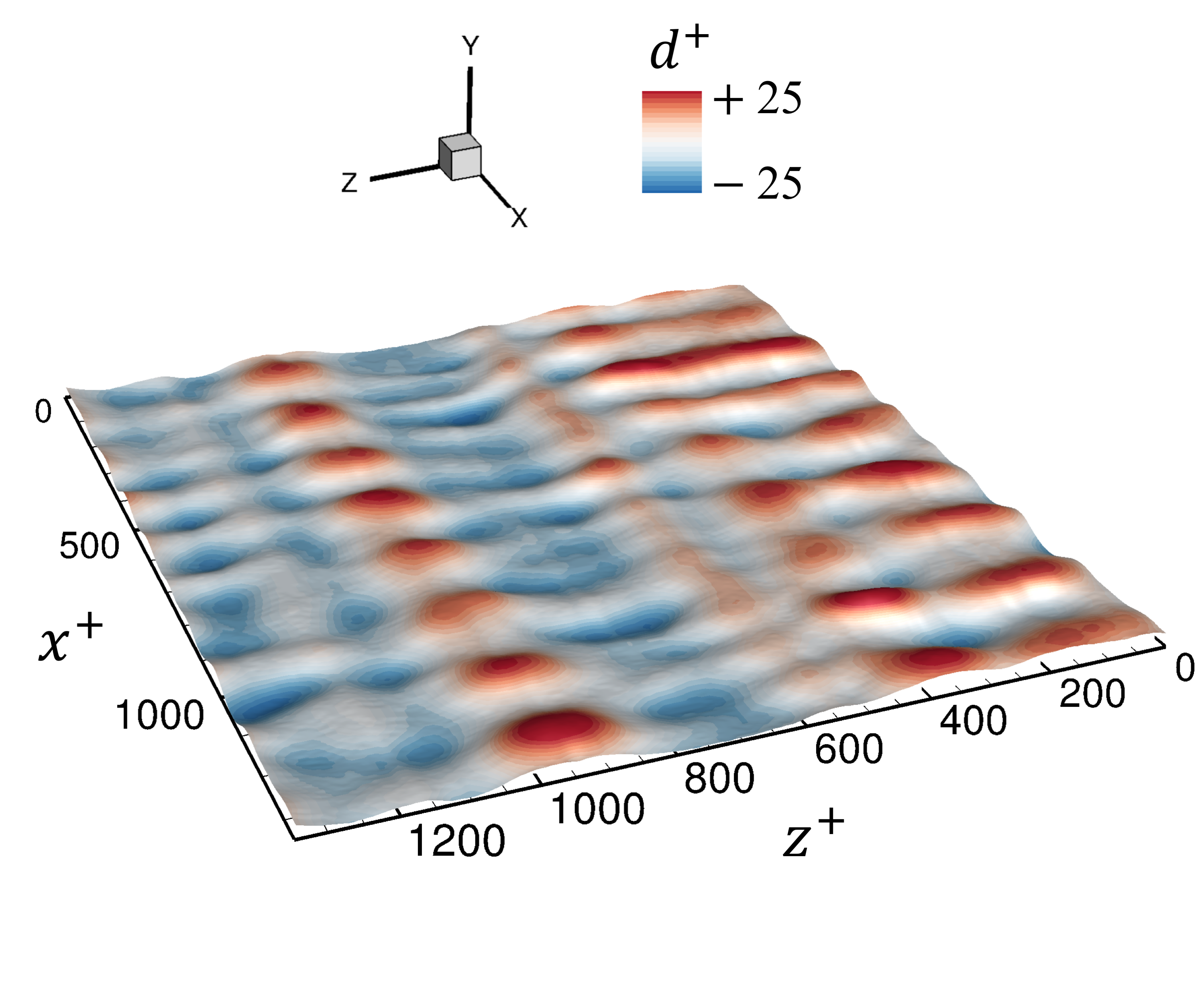}
        \end{center}
        \end{minipage} }
    \subfigure[]{\label{fig:vis_CL}
        \begin{minipage}[b]{0.4\textwidth}
        \begin{center}
        \includegraphics[width =\textwidth,scale=1]{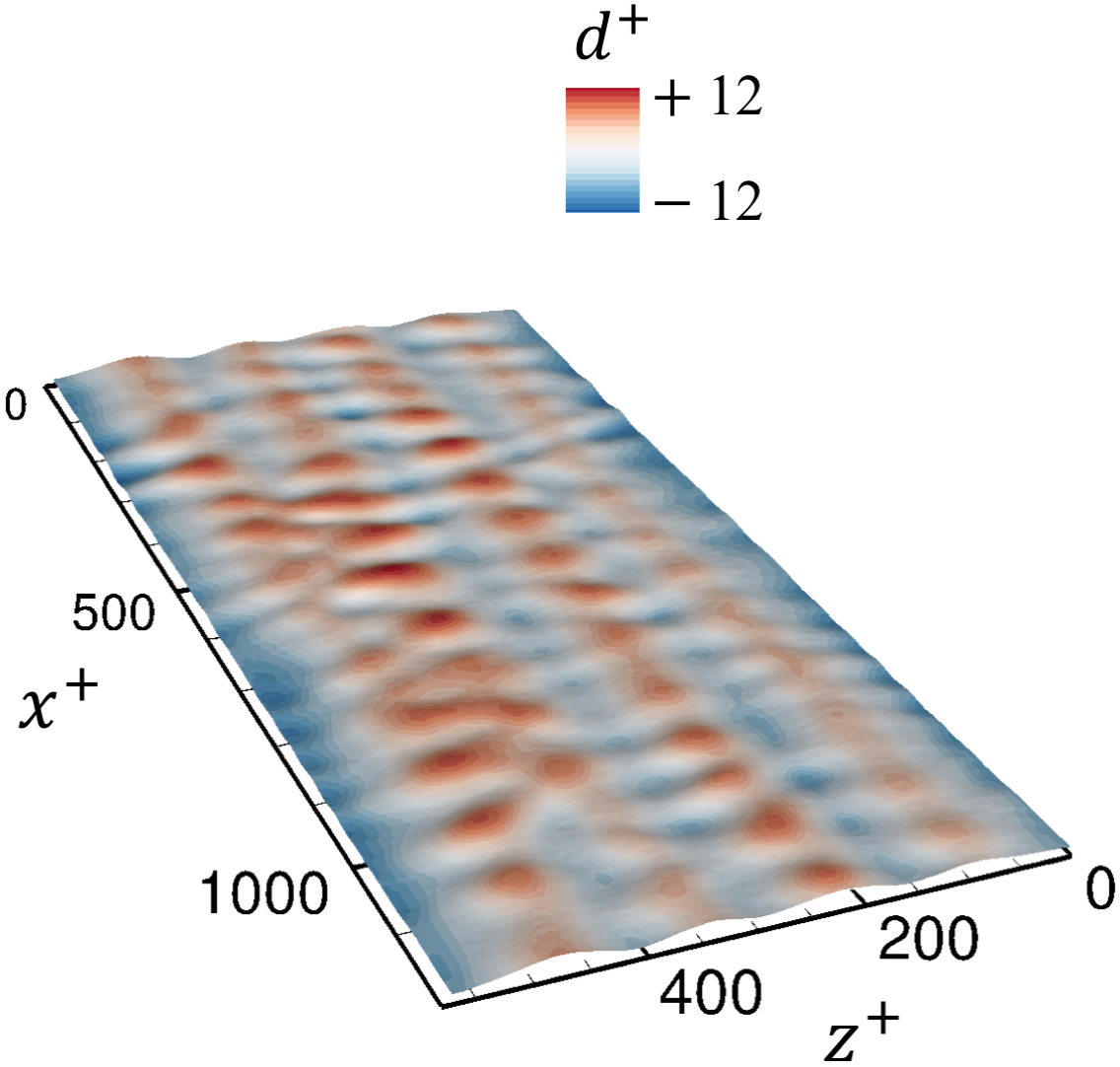}
        \end{center}
        \end{minipage} } \\
    \subfigure[]{\label{fig:vis_CG}
        \begin{minipage}[b]{0.4\textwidth}
        \begin{center}
        \includegraphics[width =\textwidth,scale=1]{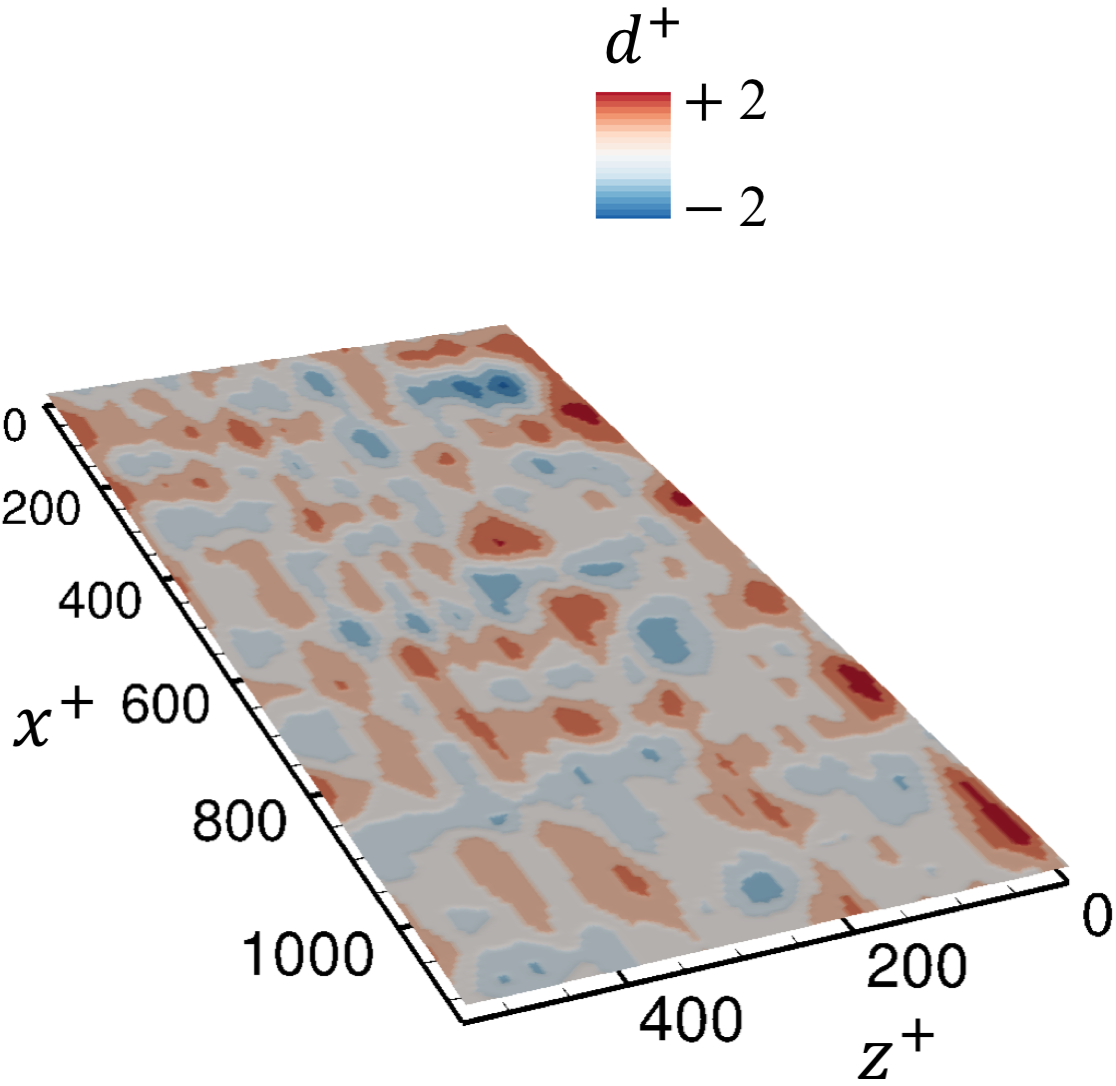}
        \end{center}
        \end{minipage} }
        \hspace{30pt}
    \subfigure[]{\label{fig:vis_CR}
        \begin{minipage}[b]{0.4\textwidth}
        \begin{center}
        \includegraphics[width =\textwidth,scale=1]{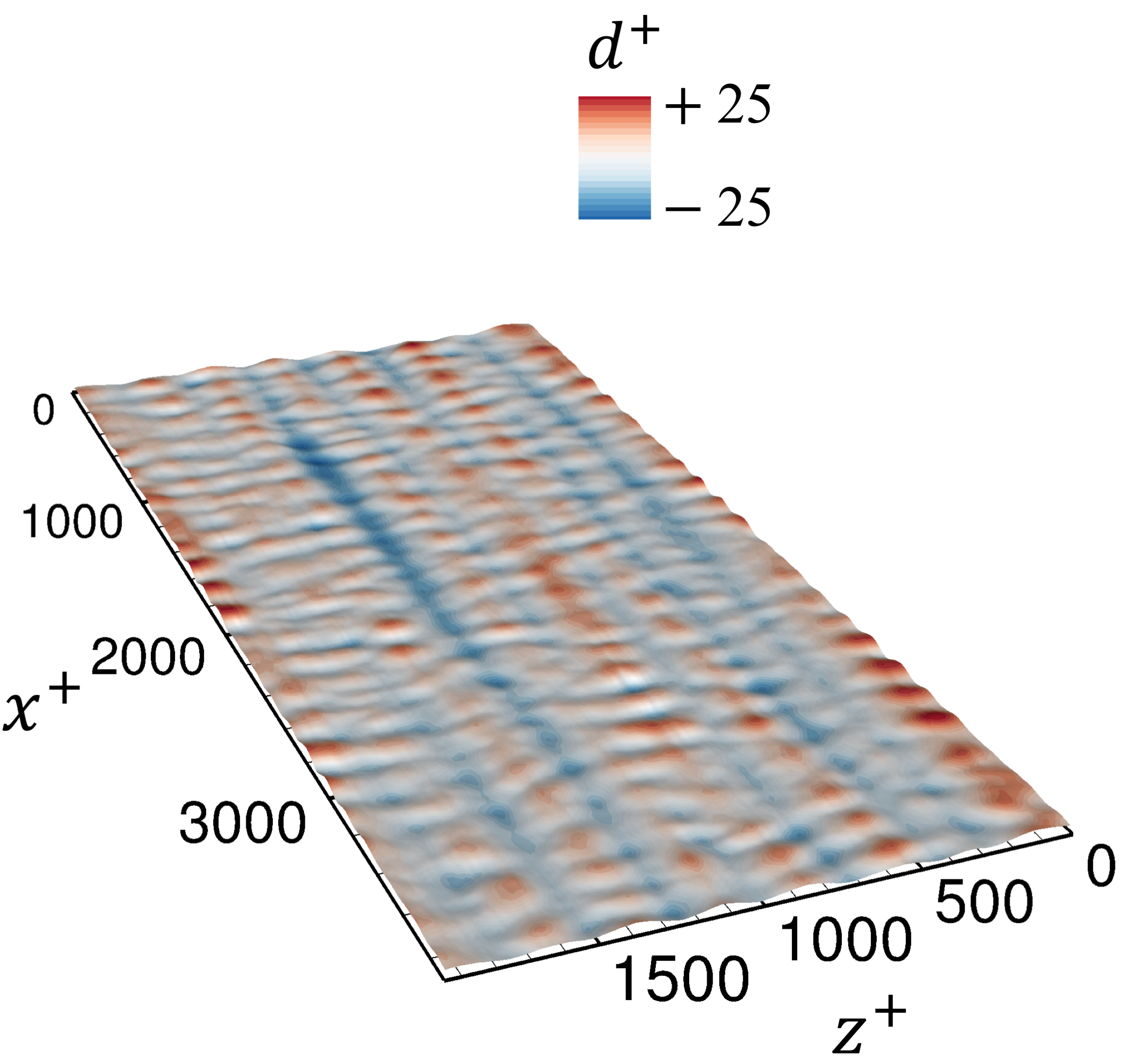}
        \end{center}
        \end{minipage} }
    \caption{ Instantaneous visualization of compliant wall surface, coloured by displacement in wall units, for cases (a) $C$, (b) $C_L$, (c) $C_G$ and (d) $C_H$.  \label{fig:vis}	} 
\end{figure}

\begin{figure}
    \centering
    \includegraphics[width =0.75\textwidth]{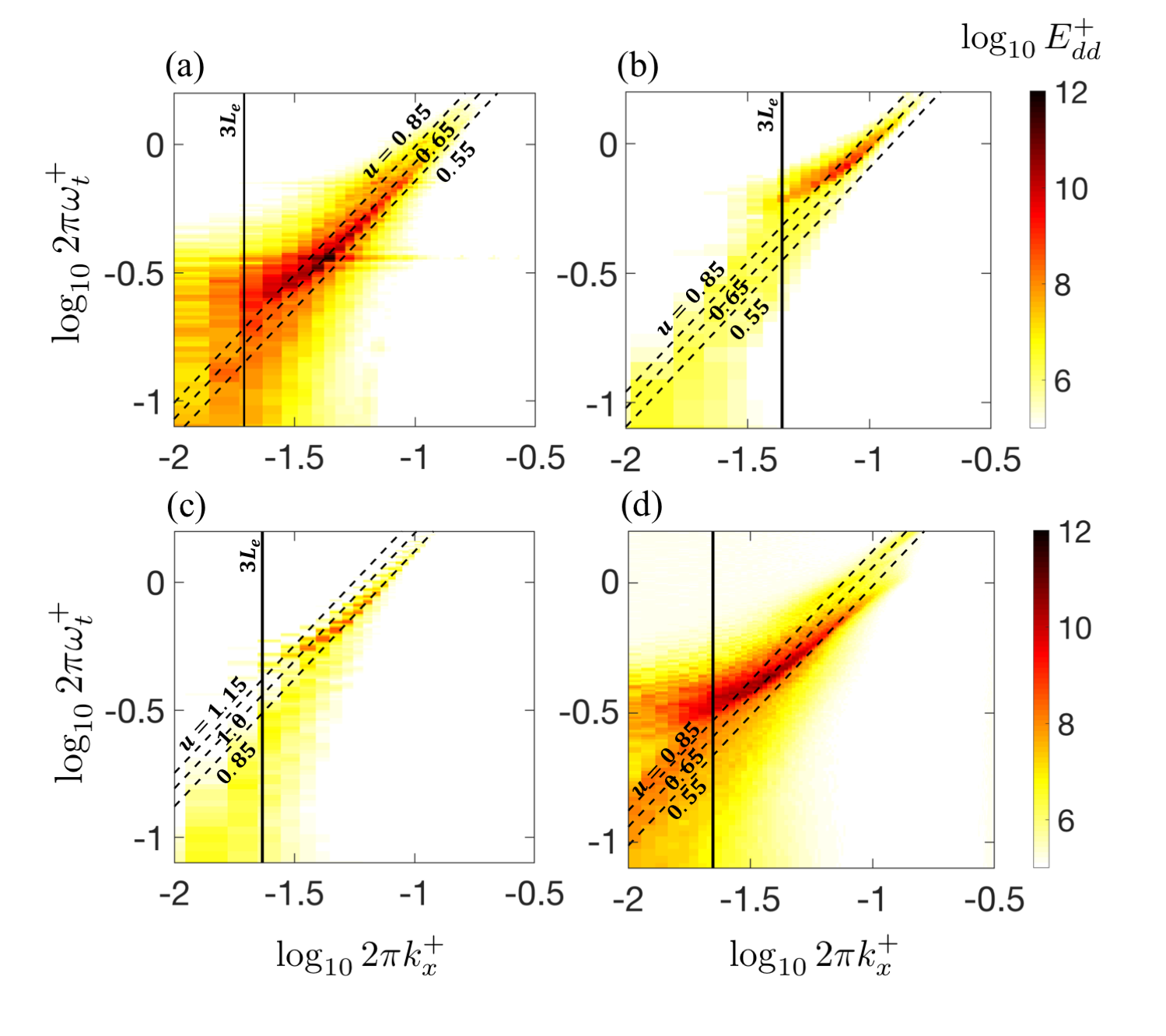}
    \caption{Streamwise wavenumber-frequency spectra of surface deformation for cases (a) $C$, (b) $C_L$, (c) $C_G$ and (d) $C_H$. Vertical lines indicate the wavenumber corresponding to $3L_e$ and inclined dashed lines indicate different phase speeds.   }
    \label{fig:spectra_surf_x}
\end{figure}

\subsection{Deformation and pressure spectra \label{sec:spectra}}

Visualizations of the instantaneous surface deformation from the compliant wall simulations are shown figure \ref{fig:vis}. 
The amplitude of the displacements are relatively large in cases $C$ and $C_H$, while the material properties were selected to strongly couple with the turbulence in the channel. Case $C_G$, on the other hand, was designed to have a high material shear-wave speed and, as a result, decouple from the turbulence; this case has the smallest displacements which are on the order of one wall unit. The most salient feature in all cases is the formation of spanwise-oriented surface-displacement patterns that propagate in the streamwise direction. The visualizations also capture a streamwise-oriented pattern with relatively low spanwise wavenumber. 
The co-presence of the spanwise and streamwise undulations gives rise to a complex topography which is reminiscent of ripples on the water surface. Spanwise-oriented deformation patterns were observed in the experiments by \citet{wang2020interaction}, although in their case the surface displacements were more chaotic. Similar to the experiments, the length and width of surface displacements do not vary appreciably with $G$, which is in agreement with the presumption that the peak-wavenumber material response is controlled by the thickness of the layer rather than its modulus of elasticity \citep{chase1991generation,benschop2019deformation}. 

\begin{figure}
    \centering
    \includegraphics[width =0.75\textwidth]{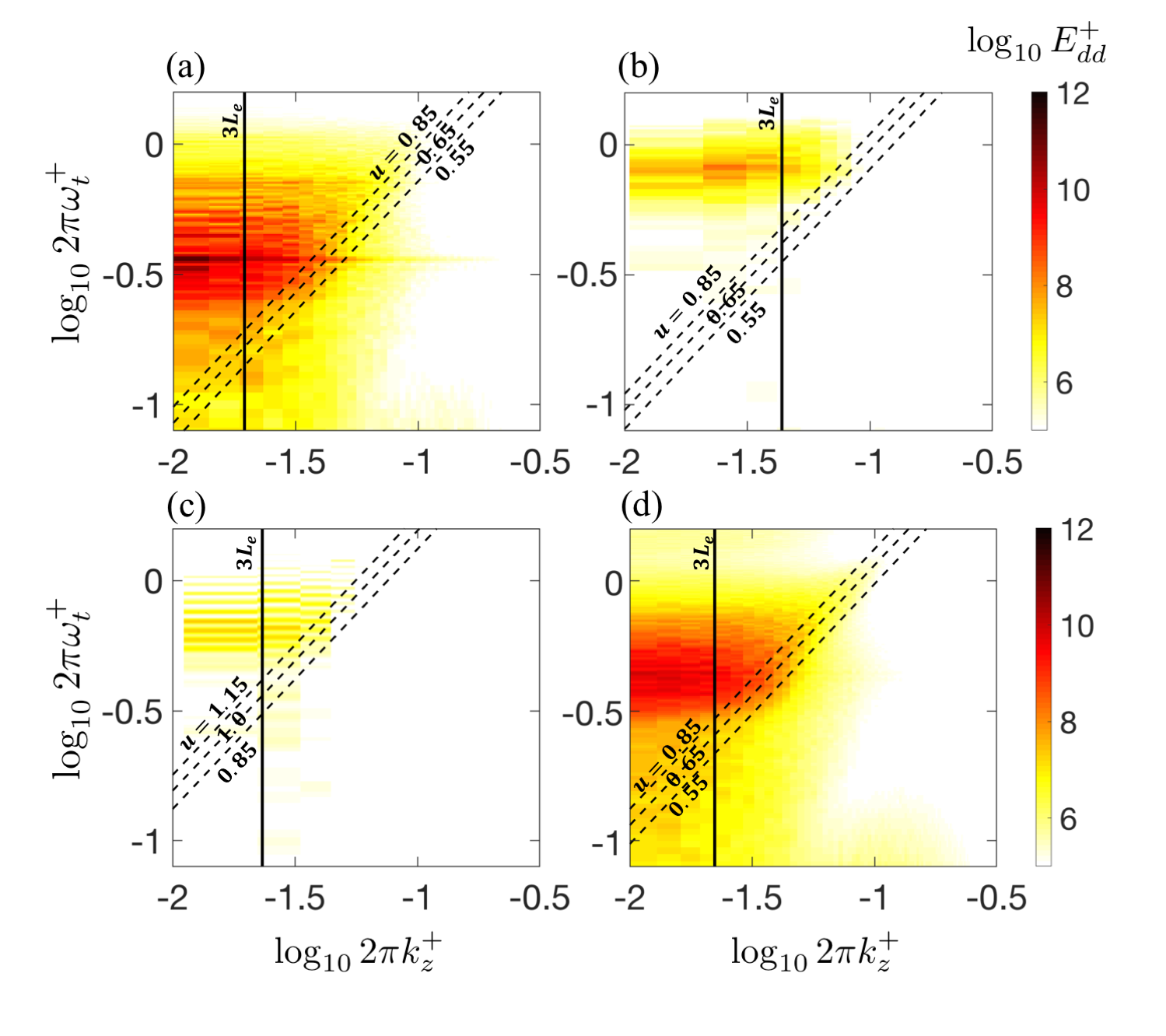}
    \caption{Spanwise wavenumber-frequency spectra of surface deformation for cases (a) $C$, (b) $C_L$, (c) $C_G$ and (d) $C_H$. Vertical lines indicate the wavenumber corresponding to $3L_e$ and inclined dashed lines indicate different phase speeds. }
    \label{fig:spectra_surf_z}
\end{figure}

In order to demonstrate the wave propagation at the material-fluid interface, we examine both the streamwise and spanwise wavenumber-frequency spectra of the surface displacement (figures \ref{fig:spectra_surf_x} and \ref{fig:spectra_surf_z}). 
In the first figure, streamwise travelling modes are observed in all cases with speeds marginally slower than those of the shear waves in the compliant materials, $\sqrt{G^\star/\rho_s^\star} = \{0.7, 0.7, 1.0, 0.59\}$ in cases $\{C, C_L, C_G, C_H\}$. 
As will be discussed in \S\ref{sec:instability}, the wave motion is very similar to the Rayleigh wave propagating in a viscoelastic material \citep{rayleigh1885waves}, whose advection speed is similarly slightly smaller than the shear wave, i.e.~$0.954\sqrt{G^\star/\rho_s^\star}$. 
In case $C_G$ with stiff material, the phase-speed is relatively high and the resonance is weak, since pressure fluctuations near the wall have lower phase speeds and hence weakly couple to the material deformation.  
In all cases, the range of dominant wavenumbers are clearly controlled by of the material thickness $L_e$, and the peak response shifts to higher wavenumbers in case $C_L$ with the thinner compliant layer (compare figures \ref{fig:spectra_surf_x}(a) and \ref{fig:spectra_surf_x}(b)).  
This trend is in qualitative agreement with the predictions by linear models \citep{chase1991generation,benschop2019deformation} and the experiments by \citet{wang2020interaction}. 
Relative to the main case $C$, the peak frequencies are higher in cases $C_L$ and $C_G$, the former due to higher range of triggered wavenumbers and the latter due to the larger $u_\text{w}$. 

In the high Reynolds number case $C_H$, the shear-wave speed and layer thickness match with those of case $C$ in wall units. 
The amplitude and wavenumber-frequency range of excited modes are similar in the two cases, which confirms that selecting the material properties ($G,L_e$) to be matched in inner scaling was appropriate.

The spanwise wavenumber-frequency spectra (figure \ref{fig:spectra_surf_z}) do not show clear travelling modes with constant speed, due to the lack of a preferential direction of disturbance propagation in the span for the channel configuration. Pressure fluctuations can still trigger spanwise-traveling waves with equal probability of positive and negative velocities, which can give rise to nearly stationary waves. This description is consistent with the time-evolution of the surface deformation from the simulations. 
The observed peak energy at  high frequencies in figure \ref{fig:spectra_surf_z} is associated with the streamwise waves, as can be verified by comparing them to the streamwise wavenumber-frequency spectra. 

The general picture of the spanwise wavenumber-frequency spectra is quite similar to the experimental observations by \citep{wang2020interaction}. However, there is a slight difference when compared to their low fluid velocity case (equivalent to large non-dimensional $G$) in which the excited modes are distributed across both low and high frequencies. They attributed the energy in the low-frequency range of the spectra to the spanwise-travelling modes, which are only dominant in cases with higher values of $G$. In our simulations, some weak traces of travelling waves at the shear-wave speed can still be found particularly in case $C_H$ (figure \ref{fig:spectra_surf_z}(d)). Similar to the experiments, the energy is spread across a range of wavenumbers, and the peak wavelength is smaller than $3L_e$. The peak modes are also at much lower wavenumbers compared with the $k_x$-$\omega_t$ spectra, implying that the surface structures are primarily spanwise oriented. 

\begin{figure}		
    \subfigure[]{\label{fig:spectra_pp_C}
    \begin{minipage}[b]{0.27\textwidth}
        \includegraphics[width =\textwidth,scale=1,trim={0cm 0cm 0cm 0}]{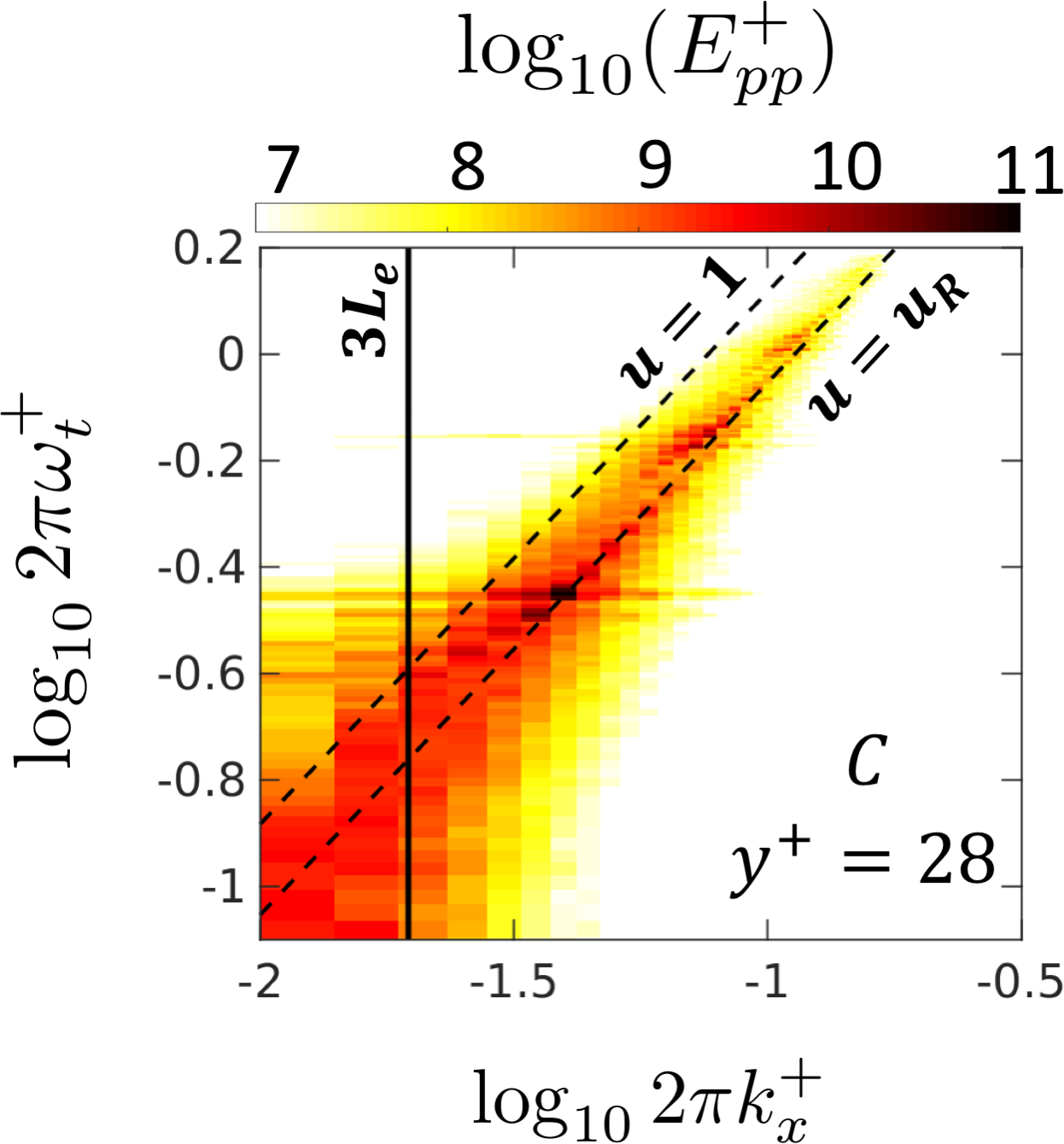}
     \end{minipage}          }
         \hspace{-10pt}
    \subfigure[]{\label{fig:spectra_pp_CL}
    \begin{minipage}[b]{0.23\textwidth}
        \includegraphics[width =\textwidth,scale=1,trim={0cm 0cm 0cm 0}]{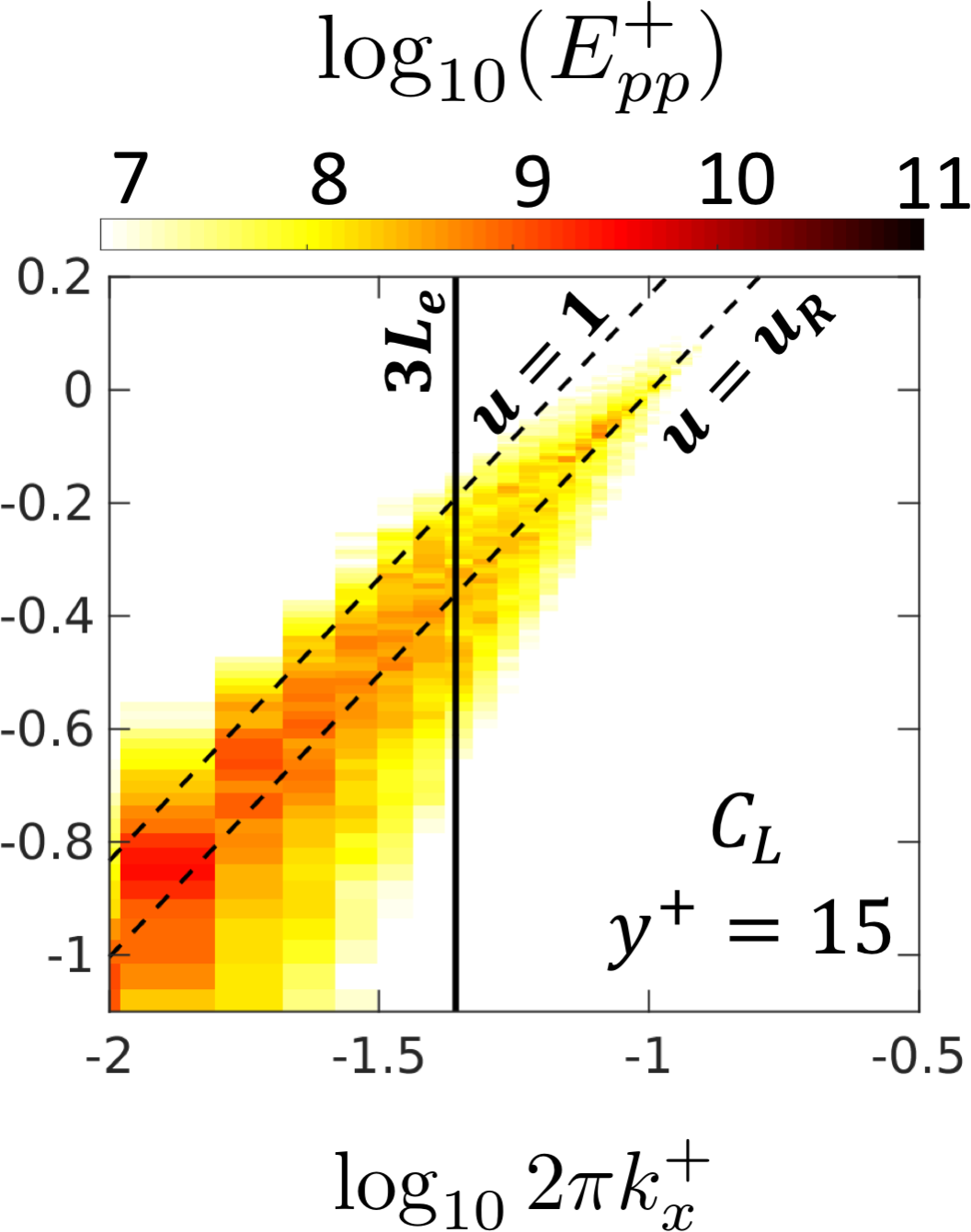} 
        \end{minipage}  }
        \hspace{-10pt}
    \subfigure[]{\label{fig:spectra_pp_CG}
    \begin{minipage}[b]{0.23\textwidth}
        \includegraphics[width =\textwidth,scale=1,trim={0cm 0cm 0cm 0}]{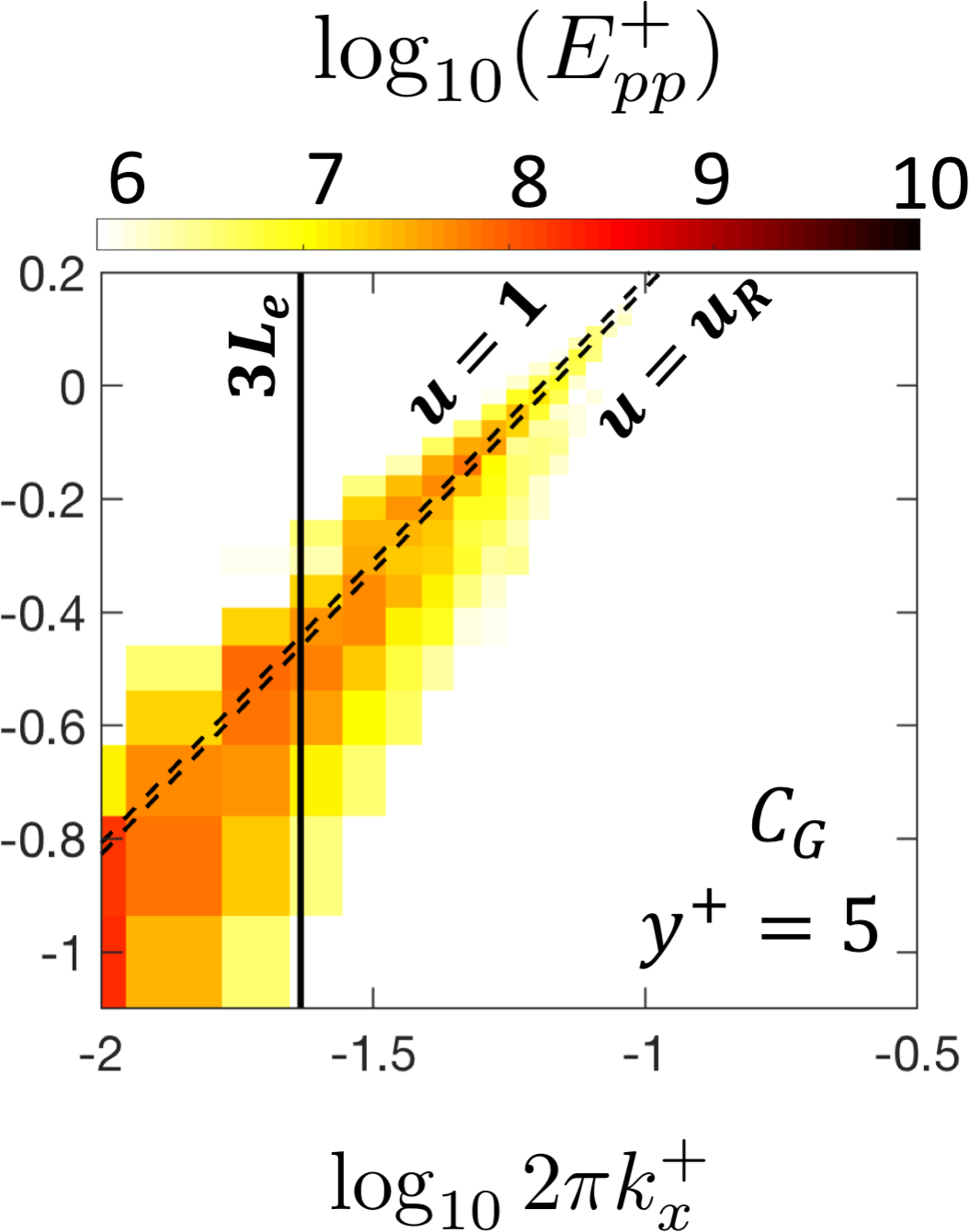}
        \end{minipage}  }
        \hspace{-10pt}
    \subfigure[]{\label{fig:spectra_pp_CH}
    \begin{minipage}[b]{0.23\textwidth}
        \includegraphics[width =\textwidth,scale=1,trim={0cm 0cm 0cm 0}]{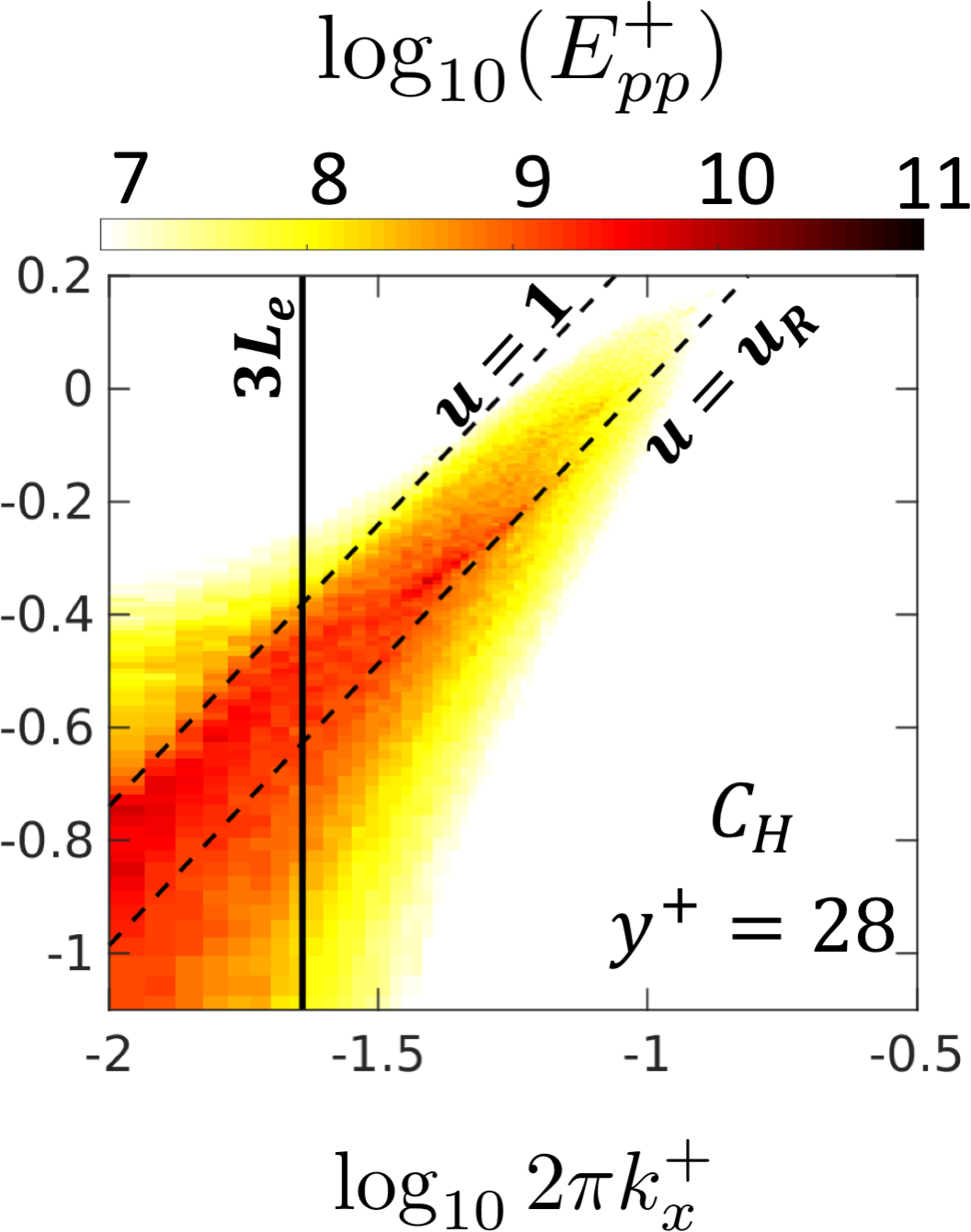}
        \end{minipage}  } \\ 
            \subfigure[]{\label{fig:spectra_pp_R180_25}
    \begin{minipage}[b]{0.27\textwidth}
        \includegraphics[width =\textwidth,scale=1,trim={0cm 0cm 0cm 0}]{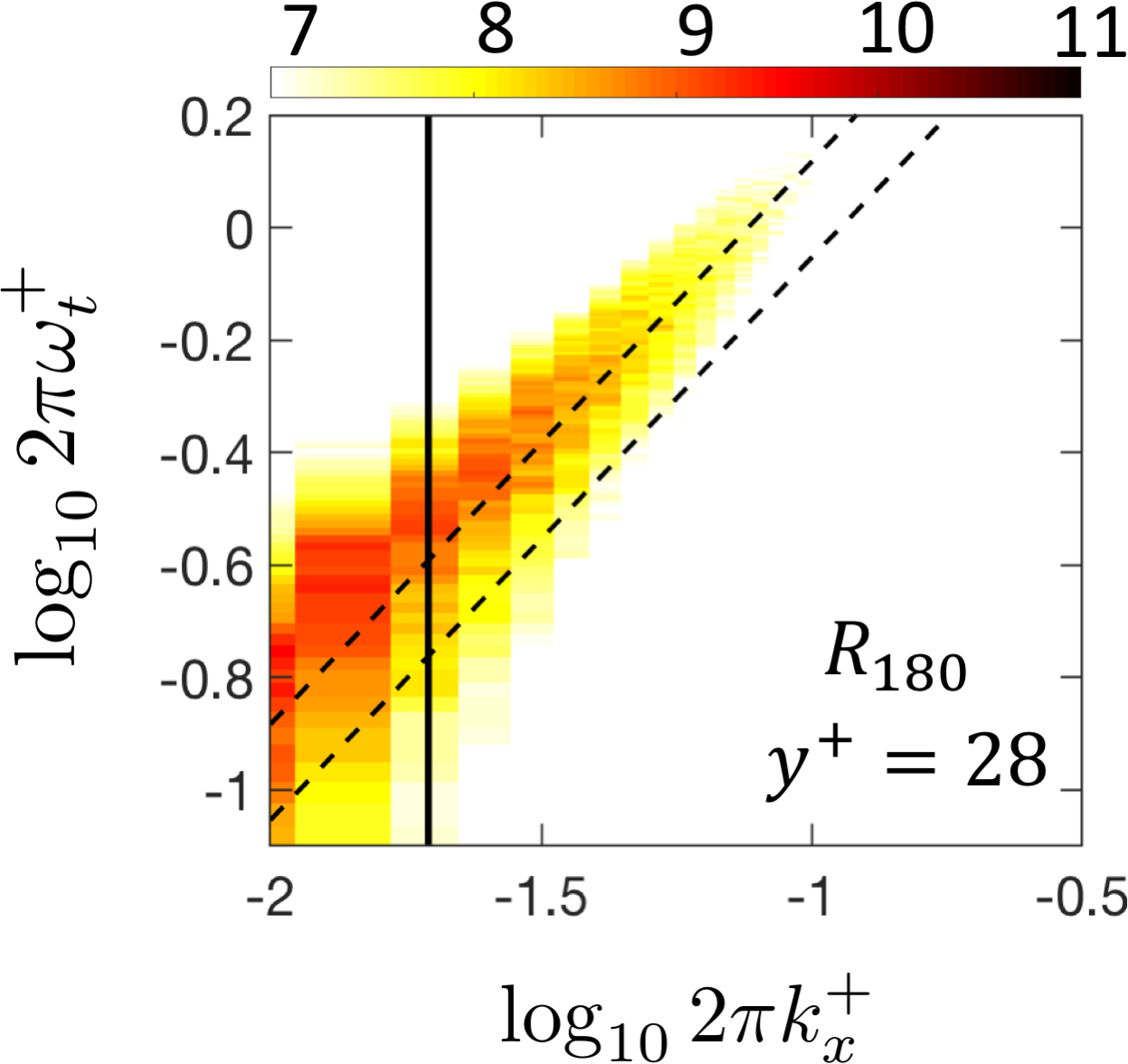}
     \end{minipage}          }
         \hspace{-10pt}
    \subfigure[]{\label{fig:spectra_pp_R180_15}
    \begin{minipage}[b]{0.23\textwidth}
        \includegraphics[width =\textwidth,scale=1,trim={0cm 0cm 0cm 0}]{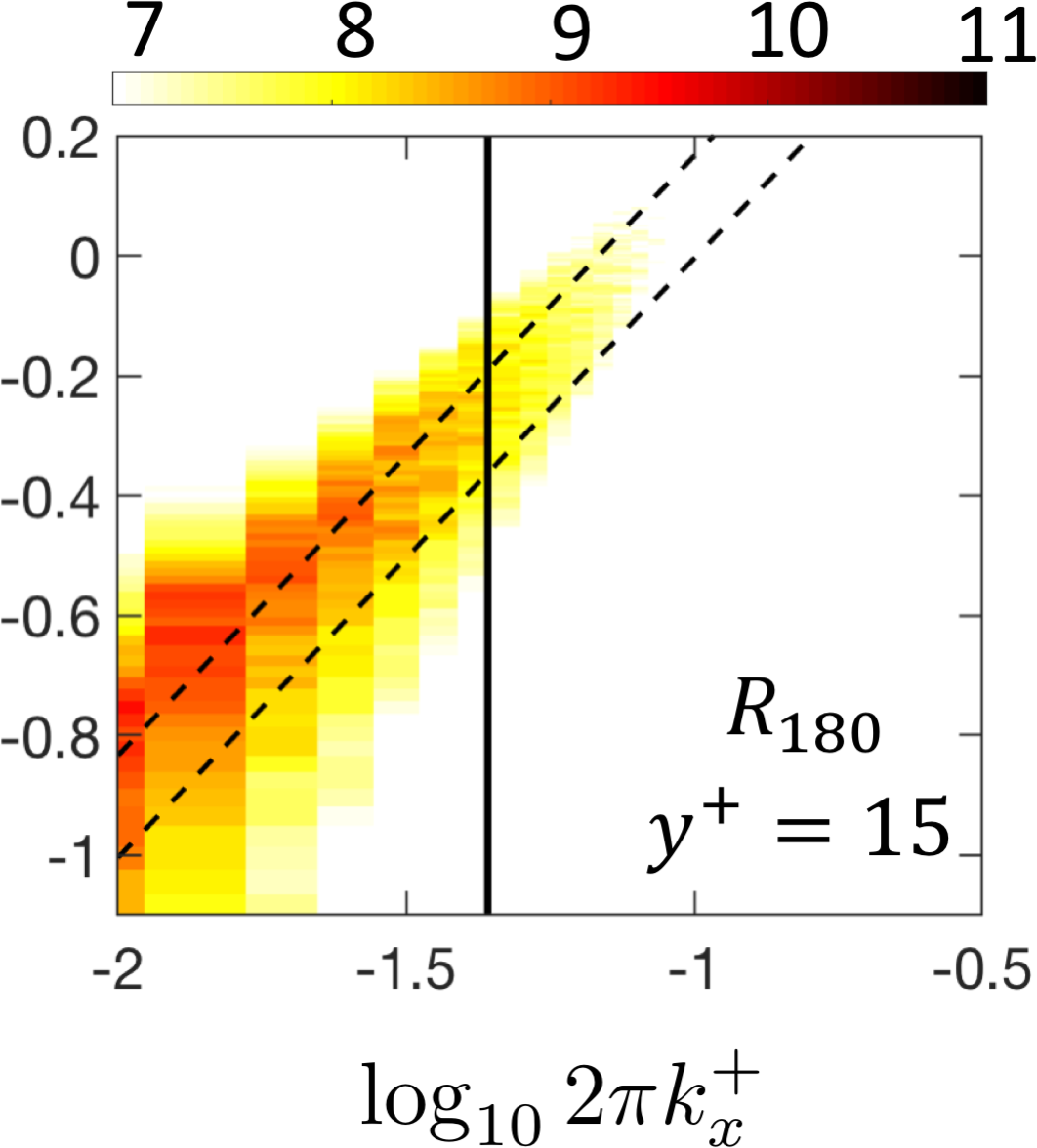} 
        \end{minipage}  }
        \hspace{-10pt}
    \subfigure[]{\label{fig:spectra_pp_R180_5}
    \begin{minipage}[b]{0.23\textwidth}
        \includegraphics[width =\textwidth,scale=1,trim={0cm 0cm 0cm 0}]{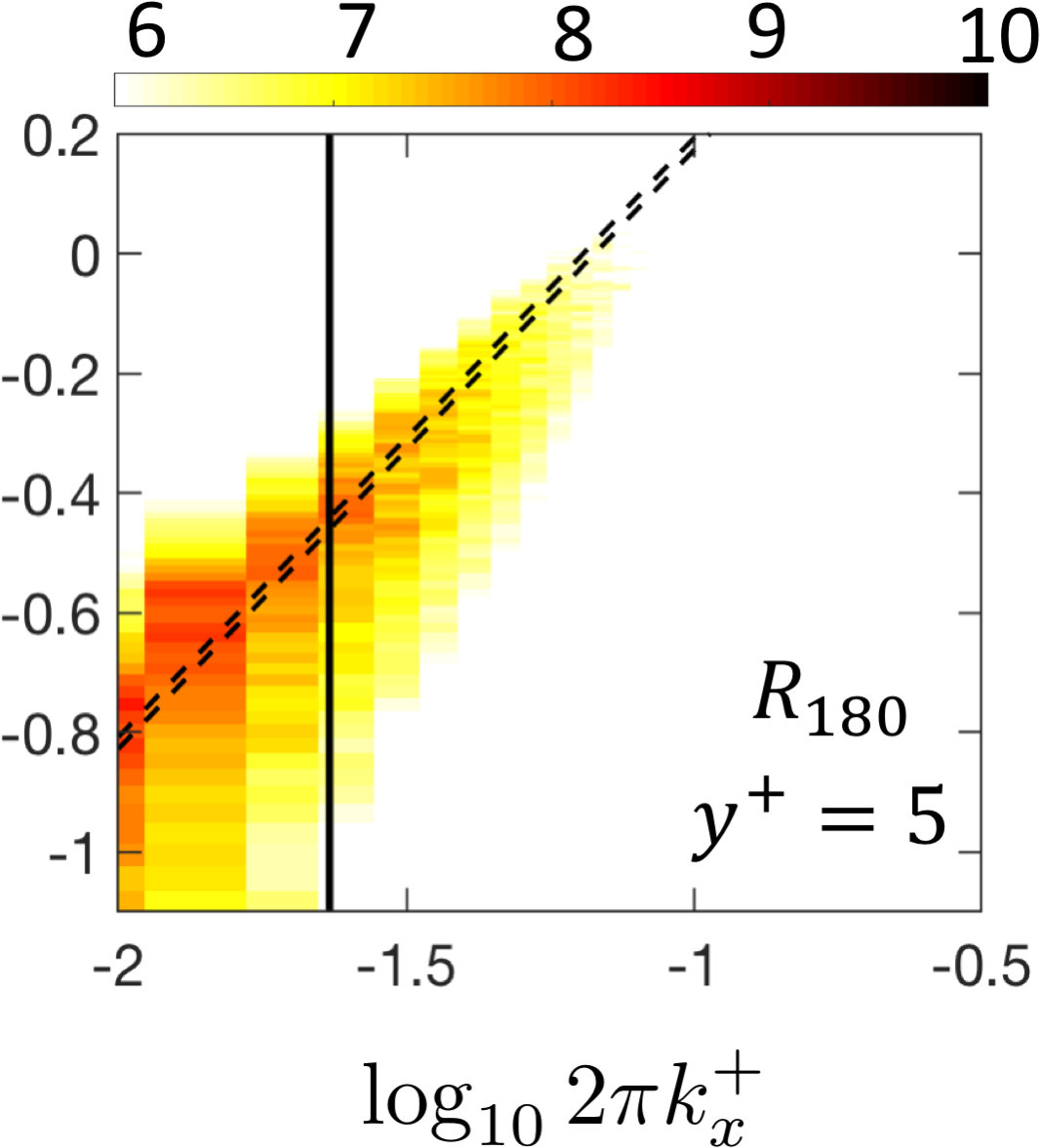}
        \end{minipage}  }
        \hspace{-10pt}
    \subfigure[]{\label{fig:spectra_pp_R590_25}
    \begin{minipage}[b]{0.23\textwidth}
        \includegraphics[width =\textwidth,scale=1,trim={0cm 0cm 0cm 0}]{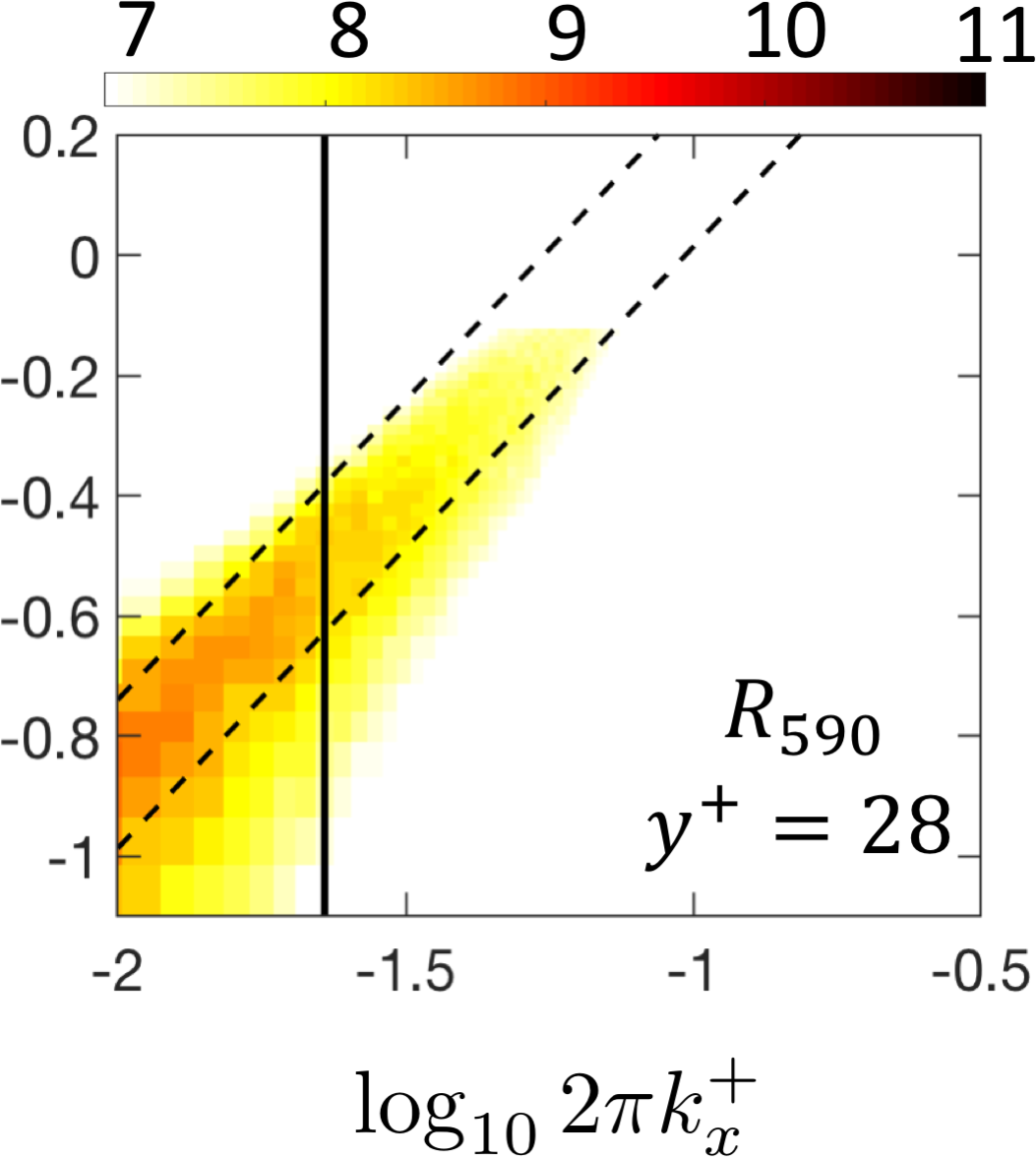}
        \end{minipage}  }

    \caption{Wavenumber–frequency power spectra of pressure for cases (a) $C$, (b) $C_L$, (c) $C_G$, (d) $C_H$, (e-g) $R_{180}$, (h) $R_{590}$. The $y^+$ location is above the the crest of surface waves in compliant cases. Vertical lines indicate the wavenumber corresponding to $3L_e$ and inclined dashed lines indicate the bulk velocity $u=1$ and the phase speed of the Rayleigh wave in elastic material, $u_R=0.954\sqrt{G^\star/\rho_s^\star}$.  \label{fig:spectra_pp}	}   
\end{figure}

\begin{figure}		
    \subfigure[]{\label{fig:spectra_pd_C0_NW}
    \begin{minipage}[b]{0.255\textwidth}
        \includegraphics[width =\textwidth,scale=1,trim={0cm 0cm 0cm 0}]{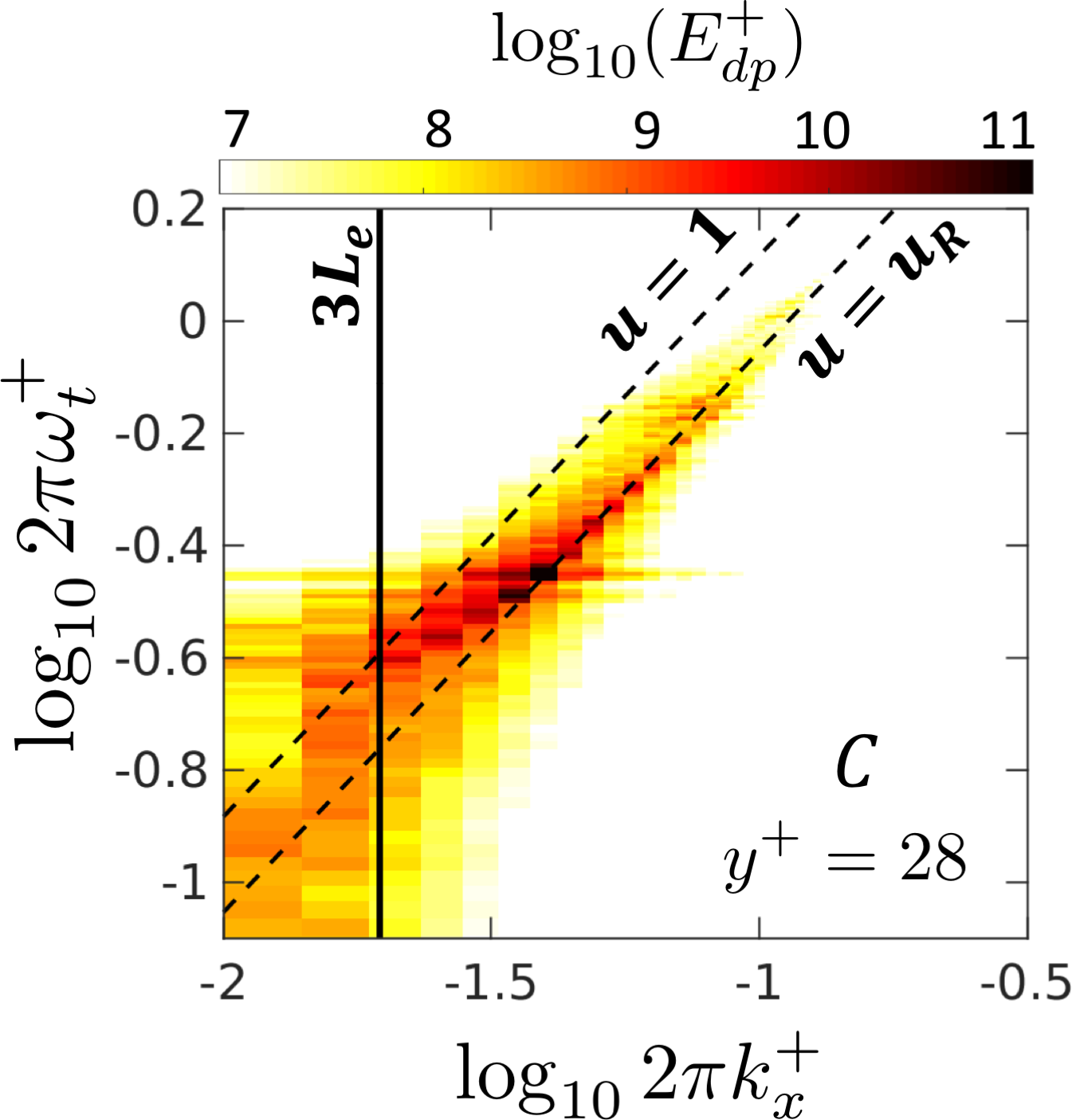}
     \end{minipage}          }
         \hspace{-10pt}
    \subfigure[]{\label{fig:spectra_pd_CL_NW}
    \begin{minipage}[b]{0.23\textwidth}
        \includegraphics[width =\textwidth,scale=1,trim={0cm 0cm 0cm 0}]{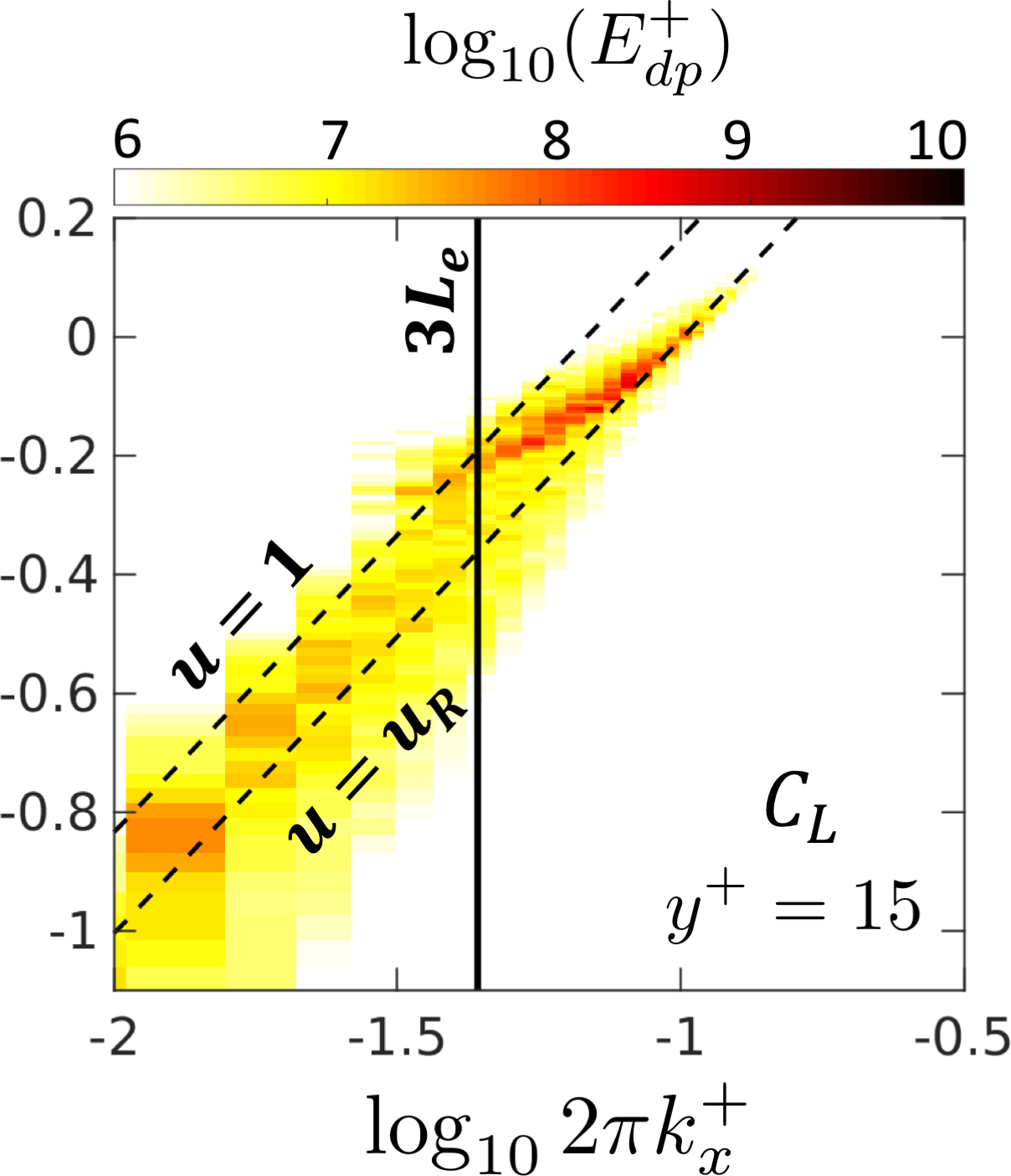} 
        \end{minipage}  }
        \hspace{-10pt}
    \subfigure[]{\label{fig:spectra_pd_CG_NW}
    \begin{minipage}[b]{0.23\textwidth}
        \includegraphics[width =\textwidth,scale=1,trim={0cm 0cm 0cm 0}]{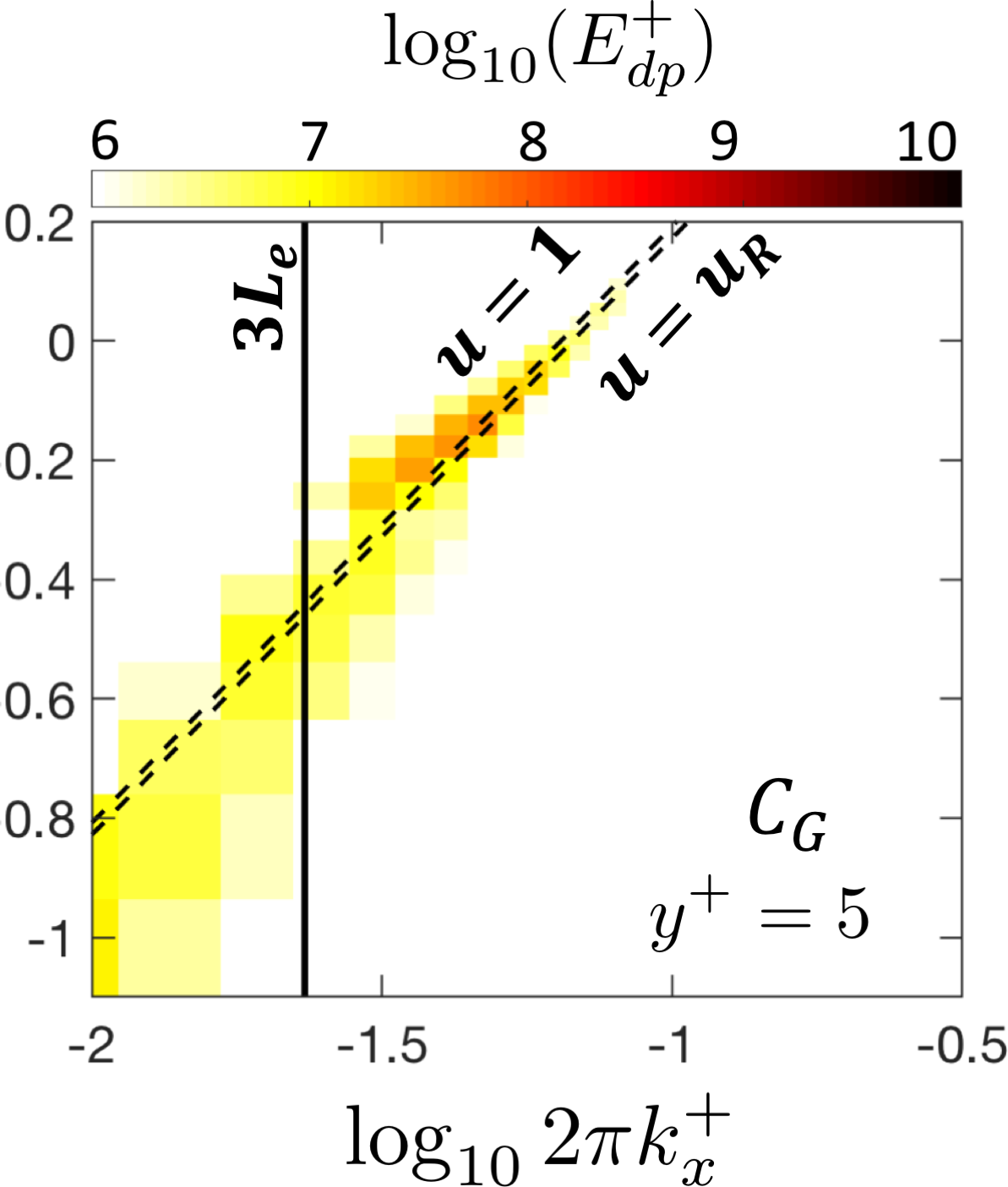}
        \end{minipage}  }
        \hspace{-10pt}
    \subfigure[]{\label{fig:spectra_pd_CH_NW}
    \begin{minipage}[b]{0.23\textwidth}
        \includegraphics[width =\textwidth,scale=1,trim={0cm 0cm 0cm 0}]{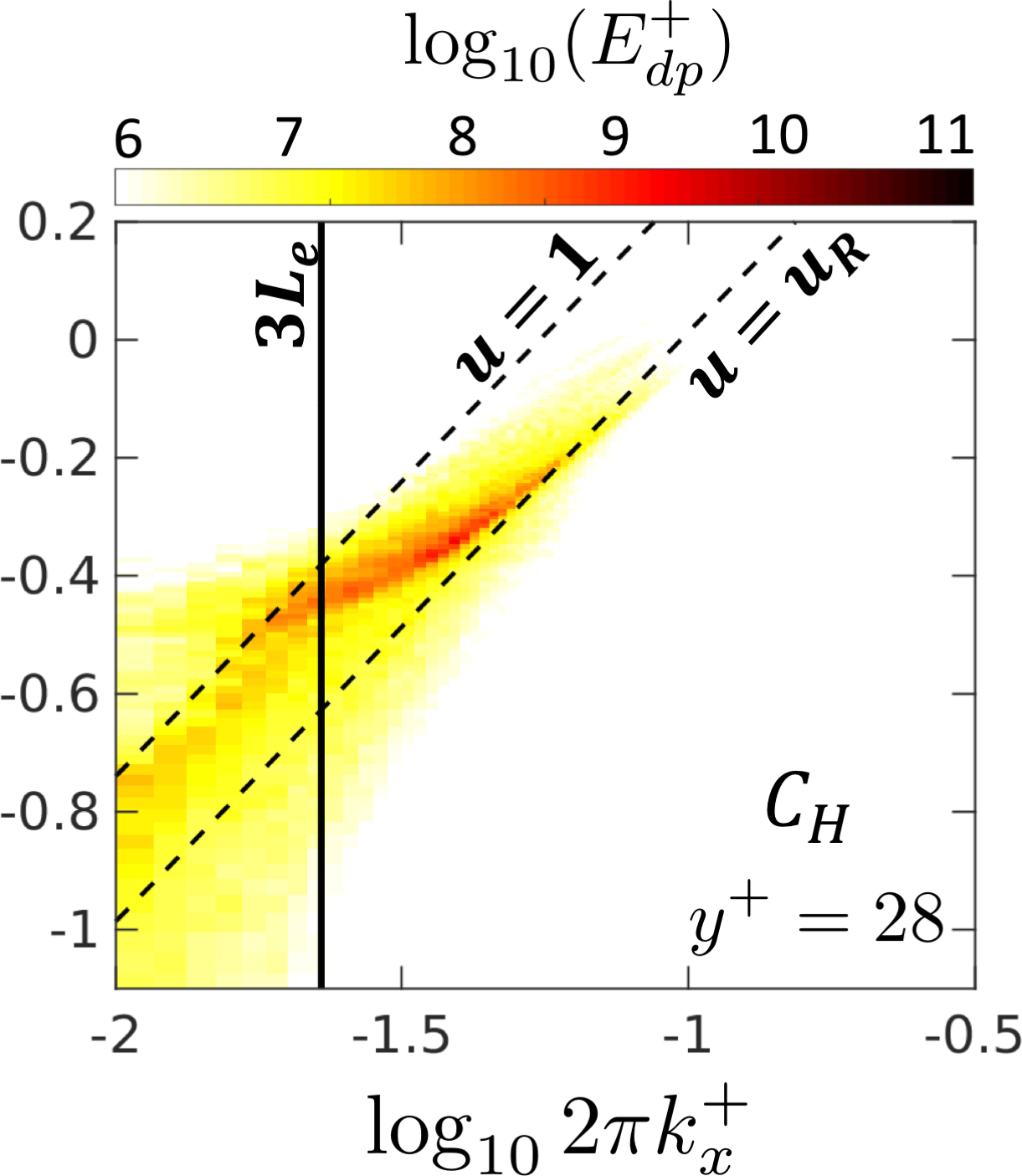}
        \end{minipage}  } \\ 
            \subfigure[]{\label{fig:spectra_pd_C0_Log}
    \begin{minipage}[b]{0.255\textwidth}
        \includegraphics[width =\textwidth,scale=1,trim={0cm 0cm 0cm 0}]{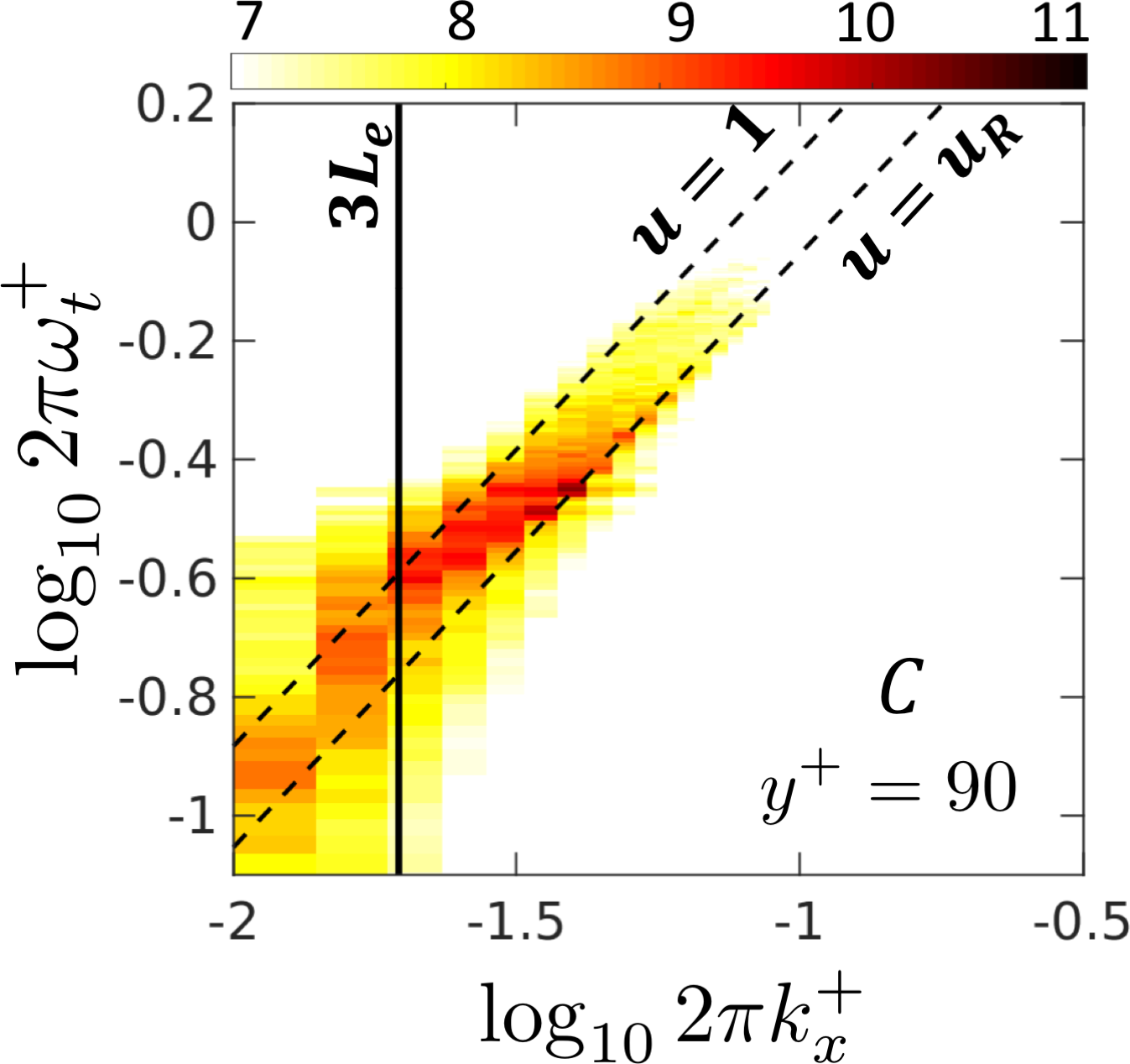}
     \end{minipage}          }
         \hspace{-10pt}
    \subfigure[]{\label{fig:spectra_pd_CL_Log}
    \begin{minipage}[b]{0.23\textwidth}
        \includegraphics[width =\textwidth,scale=1,trim={0cm 0cm 0cm 0}]{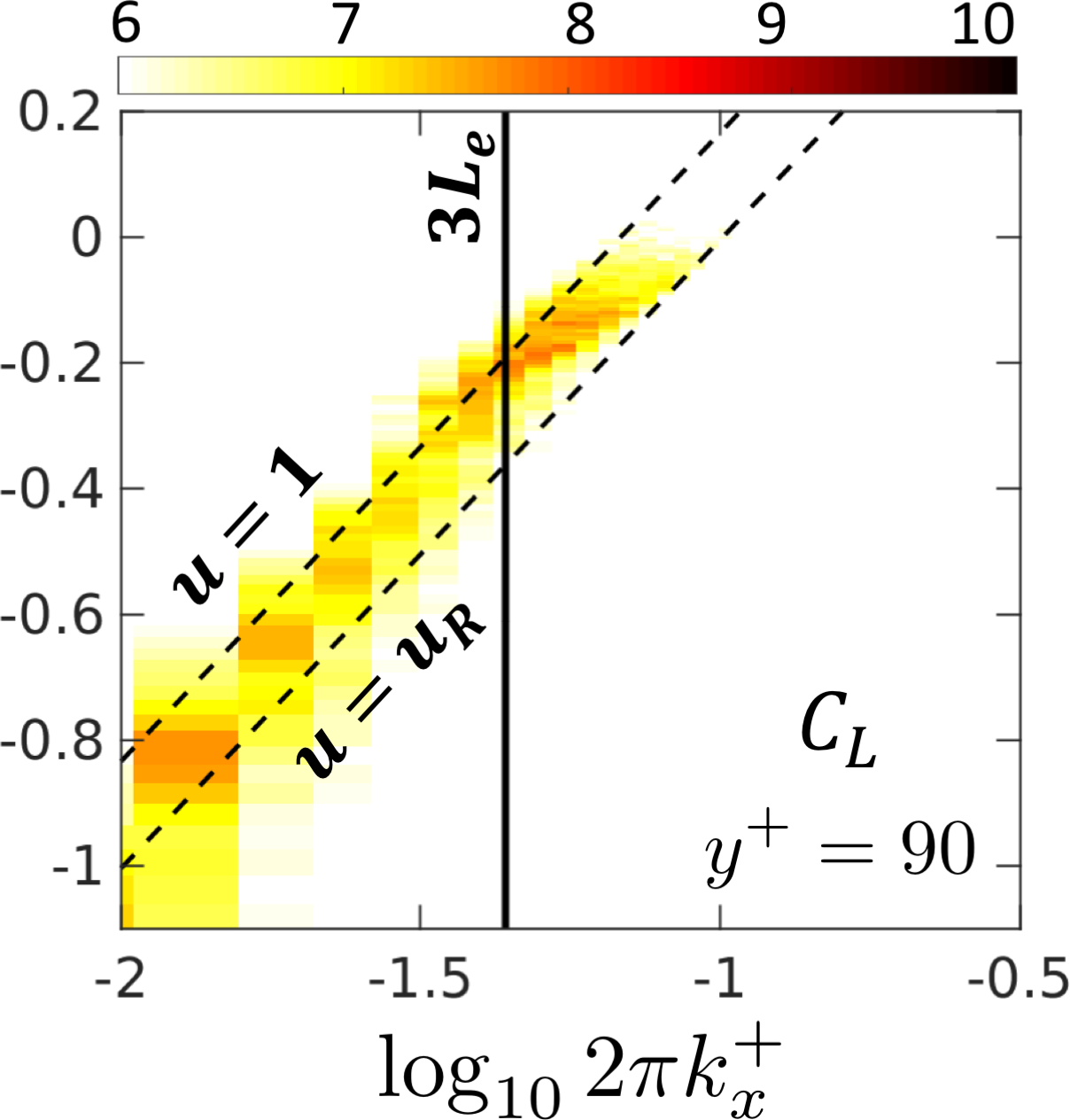} 
        \end{minipage}  }
        \hspace{-10pt}
    \subfigure[]{\label{fig:spectra_pd_CG_Log}
    \begin{minipage}[b]{0.23\textwidth}
        \includegraphics[width =\textwidth,scale=1,trim={0cm 0cm 0cm 0}]{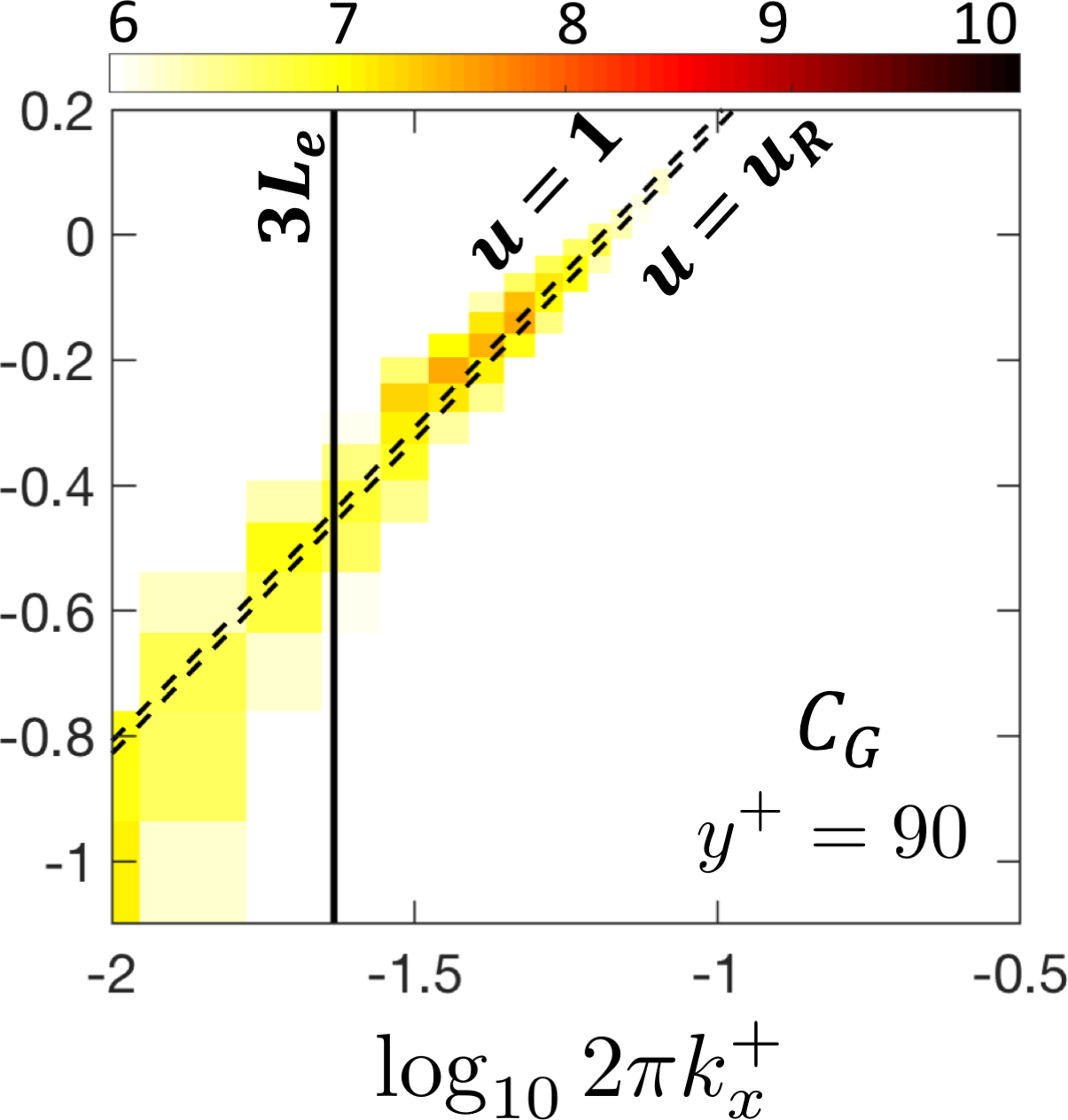}
        \end{minipage}  }
        \hspace{-10pt}
    \subfigure[]{\label{fig:spectra_pd_CH_Log}
    \begin{minipage}[b]{0.23\textwidth}
        \includegraphics[width =\textwidth,scale=1,trim={0cm 0cm 0cm 0}]{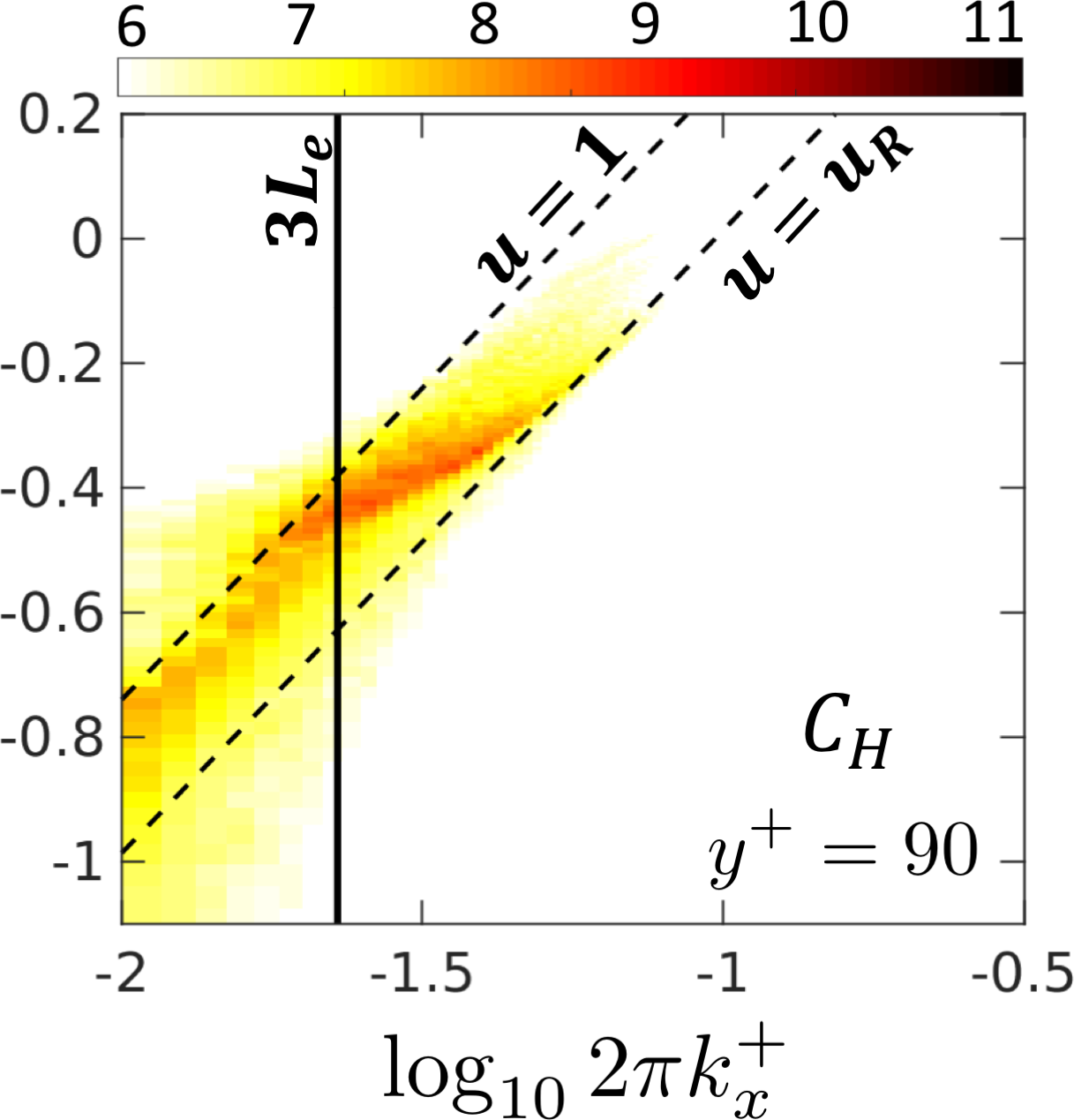}
        \end{minipage}  }
    \caption{Pressure-deformation wavenumber-frequency cross spectra for cases (a,e) $C$, (b,f) $C_L$, (c,g) $C_G$ and (d,h) $C_H$. In the top row (a,b,c,d) the closest $y$ location to the compliant surface without cutting the surface is selected, while the bottom row (e,f,g,h) is estimated at a fixed $y$ location. Vertical solid lines indicate the wavenumber corresponding to $3L_e$ and inclined dashed lines indicate the advection velocity of Rayleigh wave in elastic material, $0.954\sqrt{G^\star/\rho_s^\star}$.  \label{fig:cross_spectra}	}   
\end{figure}

In order to probe the impact of the travelling surface waves on the flow, streamwise wavenumber-frequency spectra of pressure are shown in figure \ref{fig:spectra_pp}. The $y^+$ location for each case is selected to be near the interface and beyond the wave crest, in order to avoid sampling the pressure inside the material. For comparison, the lower panels show the spectra from the rigid-wall cases at the same Reynolds numbers and $y^+$ locations. The spectra shows elevated energy in all compliant cases compared to the rigid ones, and this effect is accentuated in cases with larger surface deformations in wall units. The amplification of pressure spectra is particularly noticeable at advection velocities equal to the Rayleigh wave speed, which is approximately $0.68$ in cases $C$, $C_L$ and $C_H$, and unity in case $C_G$. Hence, the modulus of elasticity in wall units can affect both the magnitude and the advection speed of pressure fluctuations near the wall. The amplification of the spectra does not, however, appear to be confined by the wavenumber corresponding to $3L_e$, i.e.~the energy is elevated across a wider range of wavenumbers travelling with the same advection speed. Furthermore, despite the minimal differences between the mean flow velocities of cases $C_G$ and $R_{180}$ (figure \ref{fig:umean_180}), the pressure spectra are still altered due to wall compliance, particularly at high frequencies. Similar trends are also observed in the spectra of the wall-normal velocity (not shown). 
Therefore, the interactions between the flow and the surface is not merely one way; they are two-way coupled. Experiments by \citep{wang2020interaction} also showed that compliant surface deformations on the order of the viscous lengthscale can impact the near-wall flow features, when rigid roughness on the same scale would be considered hydrodynamically smooth.  The present results, as well as those experiments, underscore the important influence of the surface motion in the interaction between the flow and a compliant wall. 

To identify the pressure disturbances that correlate with the surface deformation, cross-spectra were evaluated in the $k_x$-$\omega_t$ plane and are plotted in figure \ref{fig:cross_spectra}. The cross-spectra are shown at two different $y^+$ locations for each case. In the near wall region (top panels), i.e.~$y^+ \le 28$, a clear advection band is observed in all cases with maximum magnitude coinciding with the peak mode in the surface spectra (compare \ref{fig:cross_spectra}(a-d) with \ref{fig:spectra_surf_x}(a-d)). Below the wavenumber corresponding to $3L_e$, the amplitudes of the cross-spectra reduce and shift to phase-speeds closer to the bulk velocity, which correspond to wall-signature of larger structures that travel at higher phase-speeds. In the log-layer (bottom panels), the overall amplitudes of the cross-spectra diminish, and the peak values are shifted to lower wavenumbers (\crefrange{fig:spectra_pd_C0_Log}{fig:spectra_pd_CH_Log}). This trend is expected since relatively larger eddies in the log-layer have a wall signature that can cause deformation of the compliant material. 

\subsection{Wave-correlated motions \& flow instabilities \label{sec:instability}}
\begin{figure}		
    \centering
    \includegraphics[width =0.6\textwidth,scale=1]{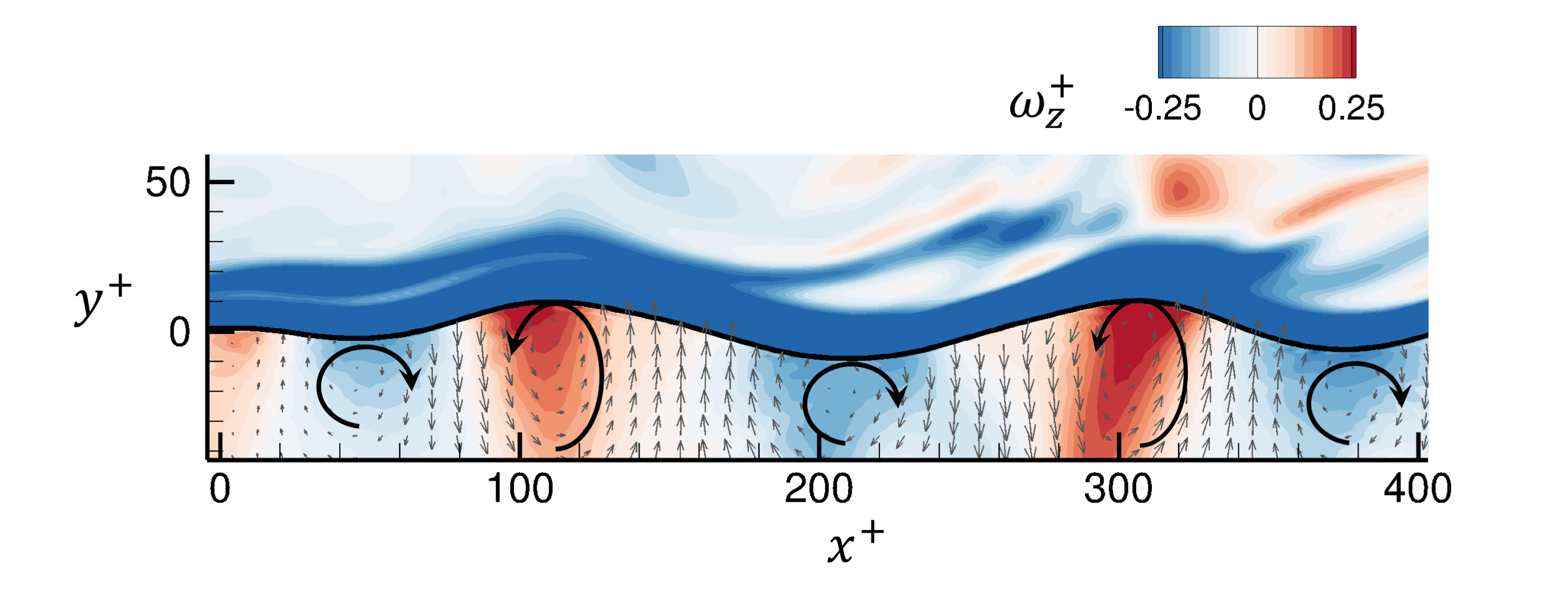}			
    \caption{Instantaneous contour plot of spanwise vorticity near the compliant surface and the in-plane velocity vectors inside the compliant wall. Counter-rotating spanwise rolls inside the compliant wall are induced by the Rayleigh wave propagating in the streamwise direction.  \label{fig:vis_omz_inst}	} 
\end{figure}

So far we described the wave propagation at the surface of the compliant wall, and its impact on integral flow properties, e.g.~drag, mean-flow profiles and pressure spectra. In this section, we closely examine the wave-correlated motions and important features of the pressure and velocity fields in the vicinity of the interface. In doing so, it is helpful to know the velocity field associated with the wave inside the compliant material. Figure \ref{fig:vis_omz_inst} shows an instantaneous $x$-$y$ plane with contours of spanwise vorticity near the interface, and the in-plane velocity vectors inside the compliant material. Similar to the waves propagating in free-surface elastic solid \citep{rayleigh1885waves}, the wave motion consists of counter-rotating spanwise rolls, with positive and negative spanwise vorticity below the crest and the trough, respectively. This wave motion is similar to wind-generated gravity waves, except for the tangential motion which is in the opposite direction, i.e.~the surface experiences negative and positive streamwise velocity below the crest and trough. In the fluid phase, a shear layer with negative vorticity is mostly attached to the interface, with occasional instabilities which will be herein described in detail. 

\begin{figure}		
    \centering
    \makebox[\linewidth][c]{%
    \subfigure[]{\label{fig:p_d_correl_C}
    \includegraphics[height=120pt,scale=1,trim={0 0 1cm 0},clip]{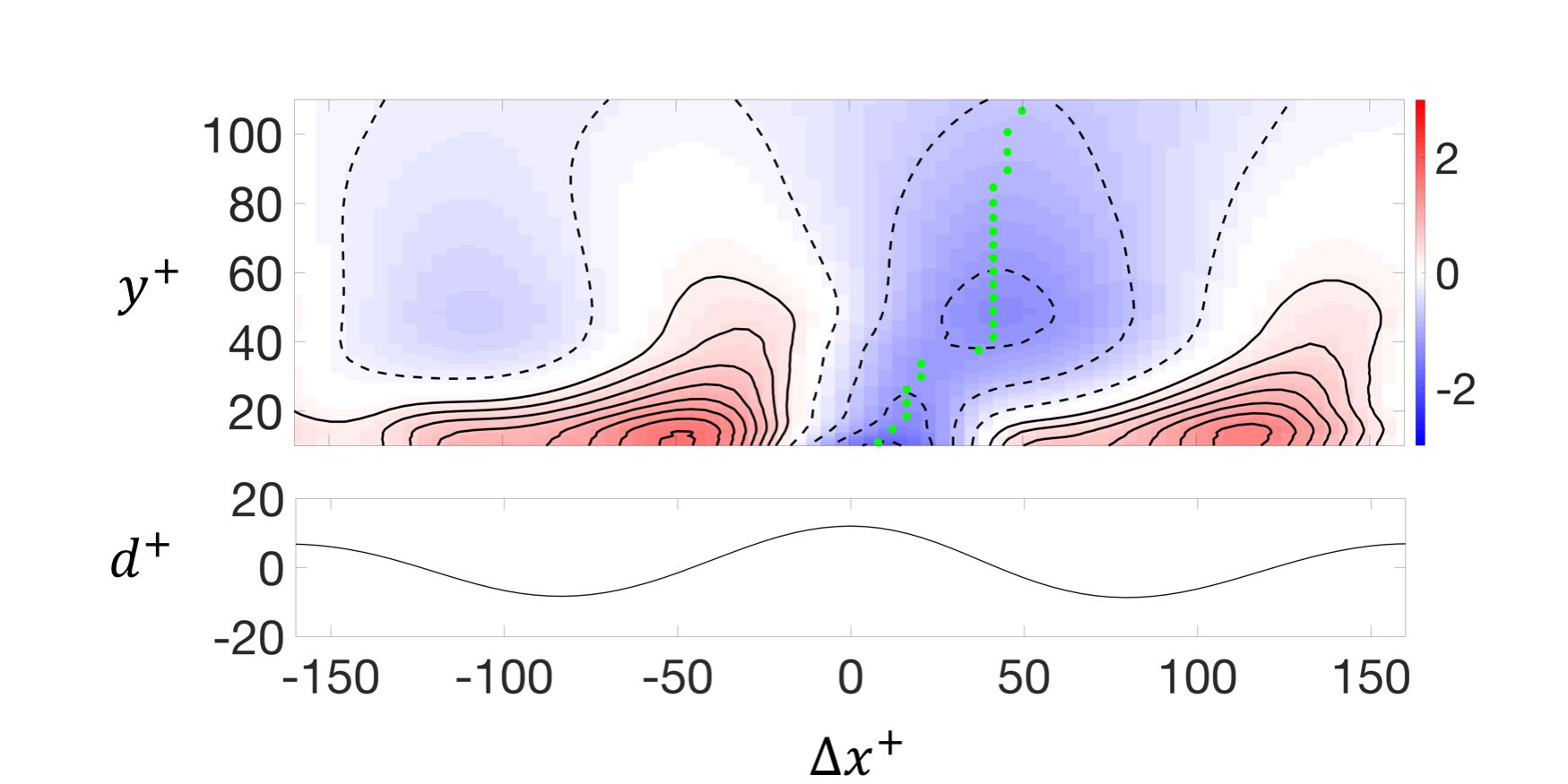} } 
    \subfigure[]{\label{fig:p_d_correl_CG}
    \includegraphics[height=114pt,scale=1,trim={2.3cm 0cm 0cm 0},clip]{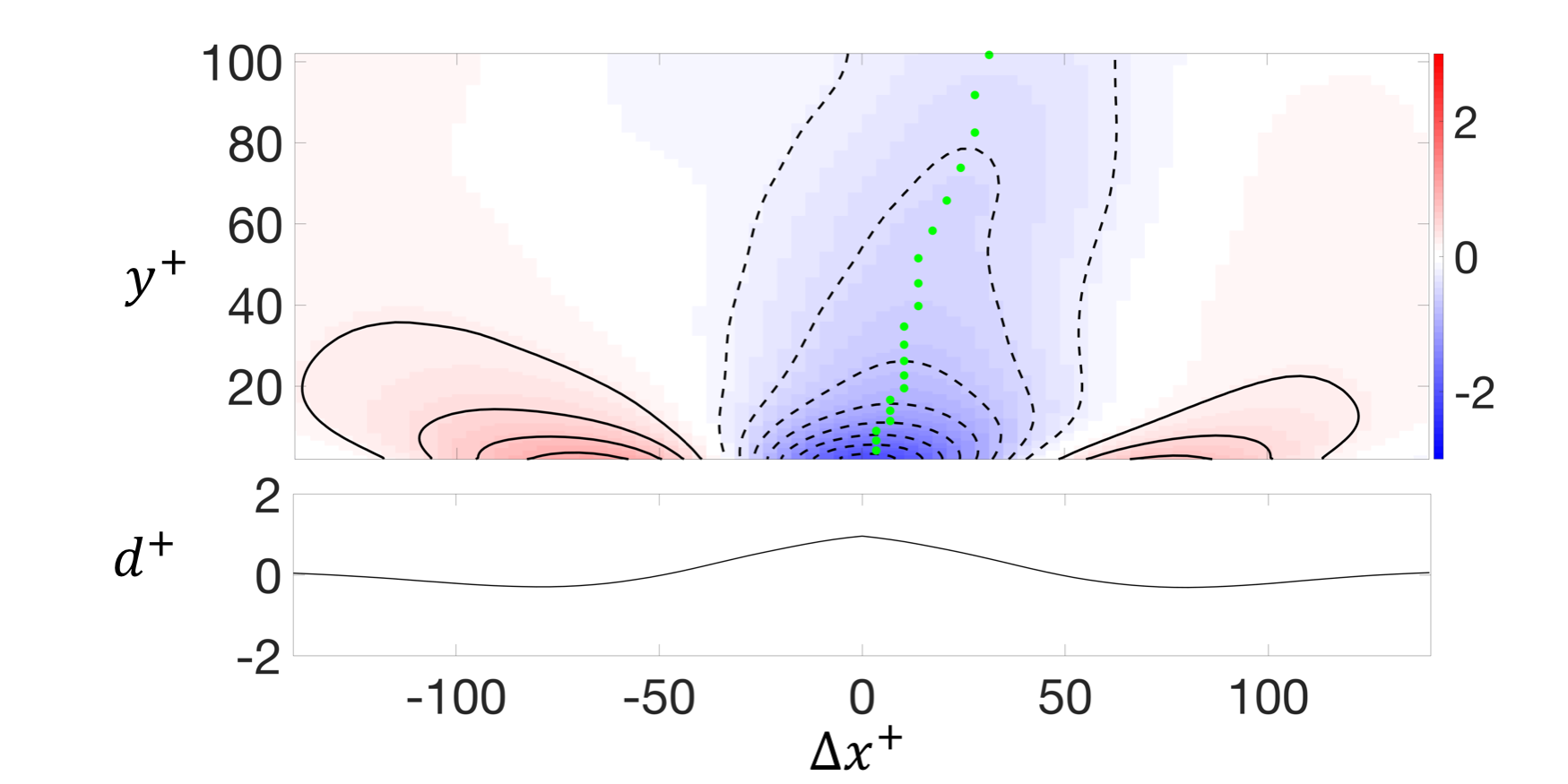} } } 
    \caption{(Top) Spatial correlation between surface displacement and pressure, $R(\Delta x, y, 0)$, defined in \eqref{eq:spatial_correl}. Dashed line contours mark negative values, and the positions of minimum value at each $y$ location are indicated by green dots. (Bottom) Phase-averaged surface displacement. All plots are conditioned on strong positive displacement, $d>d_\text{rms}$. Cases (a) $C$ and (b) $C_G$.     \label{fig:p_d_correl}}
\end{figure}
\begin{figure}
    \begin{center}
    \includegraphics[width =0.75\textwidth]{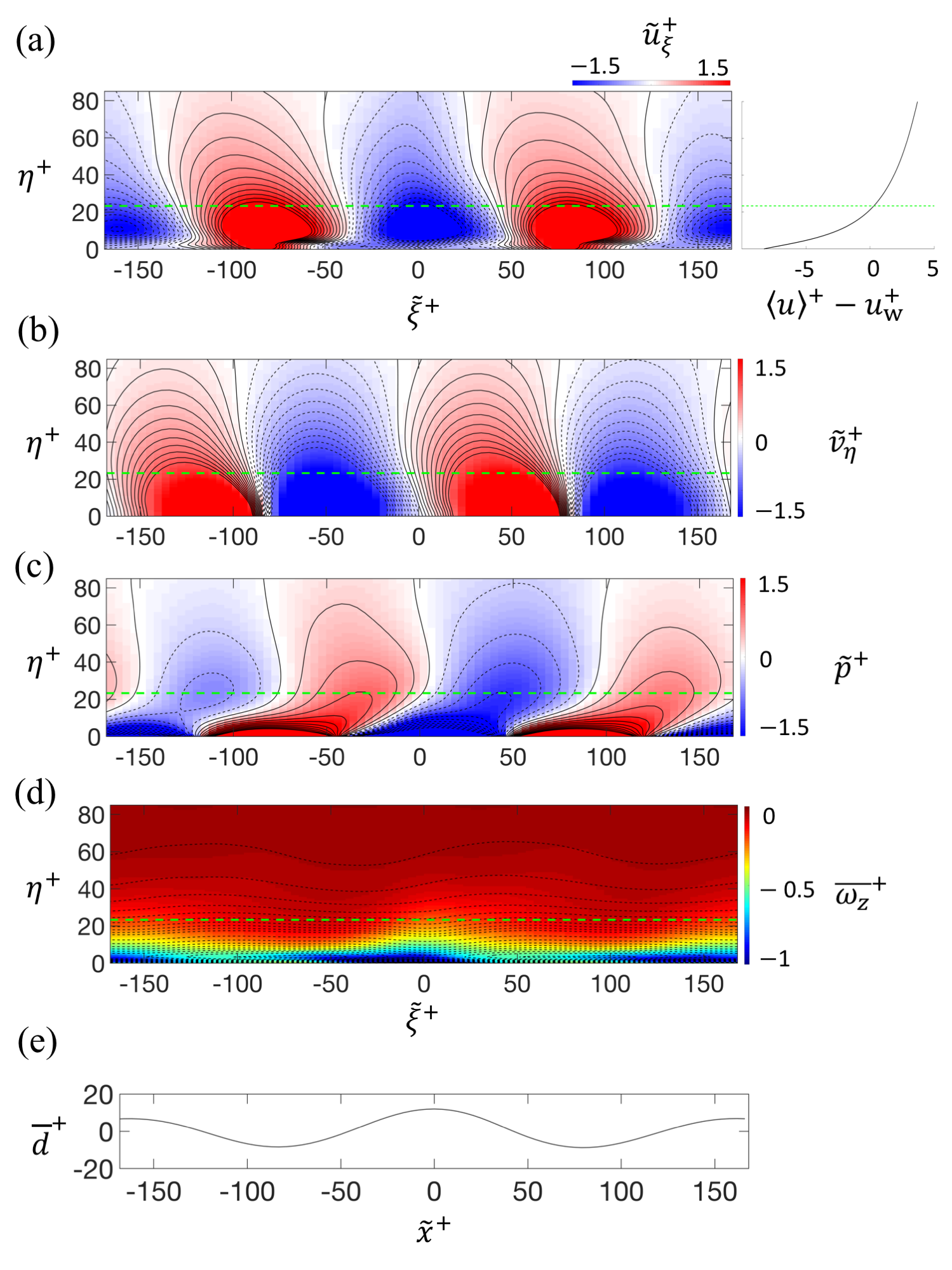}
    \end{center}
    \caption{Line and color contours in surface-fitted coordinates of wave-correlated (a) streamwise and (b) wall-normal contravariant velocities, (c) pressure, (d) phase-averaged spanwise vorticity and (e) surface displacement.  In (a-d), negative contour values are plotted with black dashed lines, and the green dashed line shows the mean critical-layer height. In (a), the line plot in the right panel shows the mean streamwise velocity in the frame of the wave and averaged over all phases. \label{fig:phase_avg_uvpomz}	}   
\end{figure}

We start by evaluating the conditional two-point correlation between the surface displacement and the pressure, 
\begin{equation}
    R_{dp}(\Delta x, y, \Delta z) = \dfrac{\langle d(x_0,z_0,t) p(x_0 + \Delta x, y, z_0 + \Delta z, t) \rangle}{(\langle d(x_0,z_0,t)^2 \rangle\langle p(x_0 + \Delta x, y, z_0 + \Delta z, t)^2 \rangle)^{1/2}}. 
\label{eq:spatial_correl}
\end{equation}
The adopted condition is strong positive displacement, specifically $d(x_0,z_0,t)>d_\text{rms}$, and only points in the fluid are sampled. Figure \ref{fig:p_d_correl} shows $R_{dp}(\Delta x,y,0)$ for cases $C$ and $C_G$, which feature the largest and smallest surface deformation in wall units. A positive pressure at the surface induces a depression in the compliant material, while a pressure deficit gives rise to a protrusion. Therefore, the displacement is expectedly anti-correlated with the pressure at the surface in both cases $C$ and $C_G$. A phase shift between the displacement and pressure at higher locations is observed, which increases with $y$ and saturates in the log-layer. \citet{zhang2017deformation} also reported a phase shift between the near-wall pressure and surface displacement for much stiffer compliant material with surface displacements smaller than one wall unit. 
In our simulations, the increase in this phase shift with $y$ is more gradual and smooth in case $C_G$ where $d^+_\text{rms}= 0.55$, while a sharp increase is observed near $y^+\approx 35$ in case $C$ where $d^+_\text{rms}= 5.6$. Also, while the pressure is minimum at the reference position $(y^+, \Delta x)=(d_\text{rms},0)$ in both cases, a local minimum is observed only in case $C$ at $(y^+,\Delta x^+)=(48,41)$. Therefore, it is possible that in case $C$ there are additional unsteady effects due to large surface displacements and strong two-way coupling that lead to a pressure minimum away from the surface. We will therefore focus the analysis on this case.  

Phase-averaged flow quantities were evaluated for case $C$ in the surface-fitted coordinate, and are plotted in figure \ref{fig:phase_avg_uvpomz}. 
The contravariant wave-correlated velocities, $\tilde{u}_\xi$ and $\tilde{v}_\eta$, are significant near the surface, and penetrate up to $\eta^+$ locations in the log-layer. Near the surface, the velocity contours are tilted upstream when viewed in the lab frame, and are slightly adjusted further away from the wall. In a frame travelling with the wave speed, $\langle u\rangle - u_\text{w}$ is negative, i.e.~the mean flow is in the opposite direction to the wave propagation near the surface; far from the surface the relative flow is positive. The $\eta$ location where the mean velocity matches the wave speed, $\langle u \rangle = u_\text{w}$, is the critical layer which is marked by a green dashed line in figure \ref{fig:phase_avg_uvpomz}.
Below the critical layer, surface-induced velocity perturbations are advected upstream relative to the wave due to the negative sign of $\langle u\rangle - u_\text{w}$, which gives rise to velocity contours that are tilted upstream.
In contrast, the tilt in the pressure contours is in the forward direction at all heights because it is due to a different mechanism. In congruence with the pressure-deformation correlations (figure \ref{fig:p_d_correl_C}), pressure contours away from the surface exhibit a phase lead compared to the pressure at the surface. It is interesting to note that the $\eta^+$ location of the pressure minimum coincides with the height of the critical layer. One explanation is that unsteady effects in the lee side of the wave, which will be further discussed below, give rise to additional turbulent motions and a pressure drop. The positive pressure gradient on the lee side, $0 \le \tilde{x}^+ \le 100$, increases the chance of flow destabilization and shear-layer detachment. Figure \ref{fig:phase_avg_uvpomz} also shows the contours of spanwise vorticity. Although the phase-averaged vorticity is negative everywhere, its magnitude is smaller on the lee side of the wave relative to the wind-ward side. Again, flow destabilization on the lee side can lead to occasional shear-layer detachment and instantaneous positive vorticity, and as a result a lower magnitude yet negative $\overline{\omega_z}$.  

\begin{figure}		
    \centering
    \subfigure[]{\label{fig:vis_omz_inst_detachment_pressure}
        \includegraphics[width =0.48\textwidth]{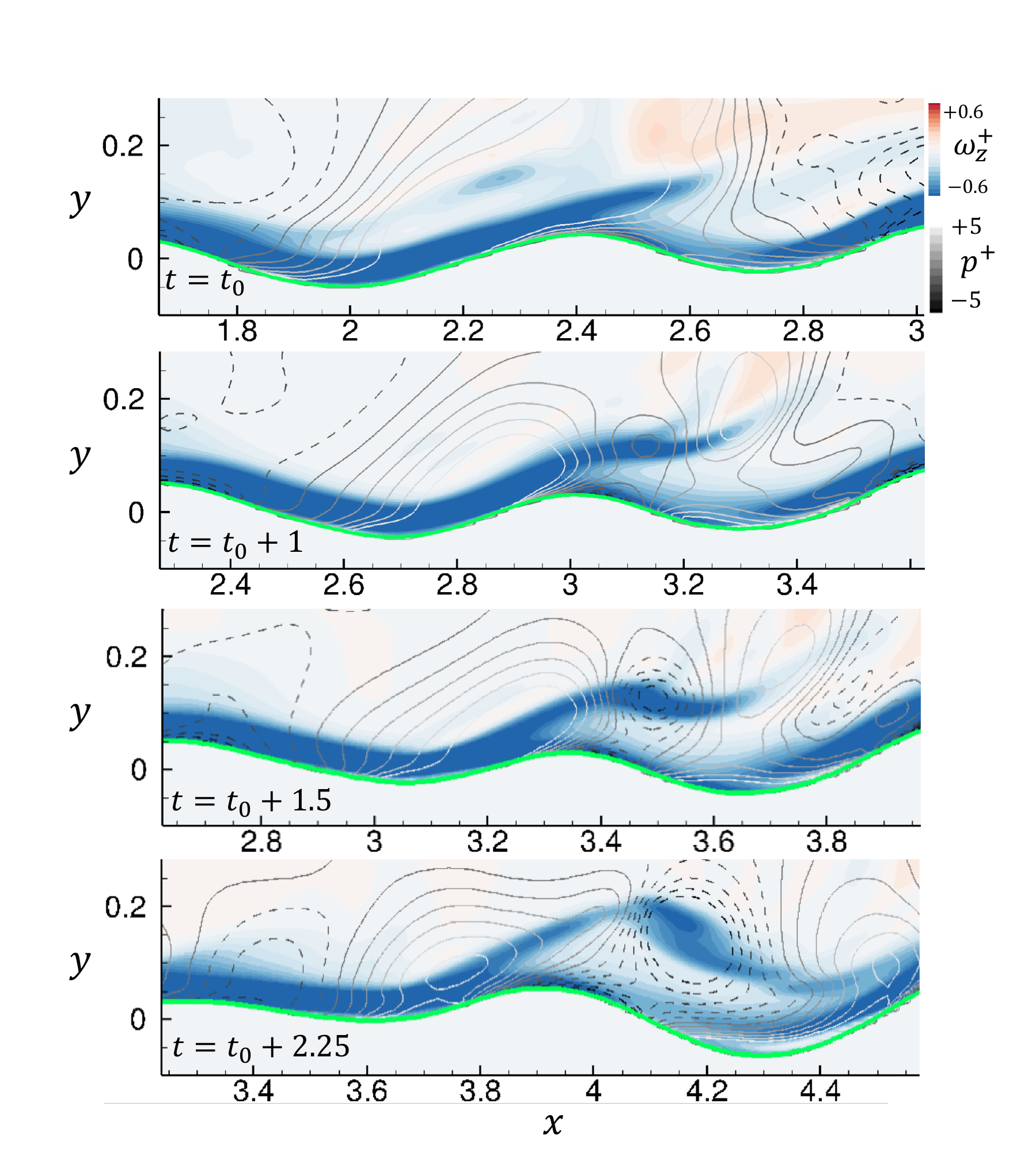} } 
    \subfigure[]{\label{fig:vis_omz_inst_detachment_velocity_vector}
        \includegraphics[width =0.48\textwidth]{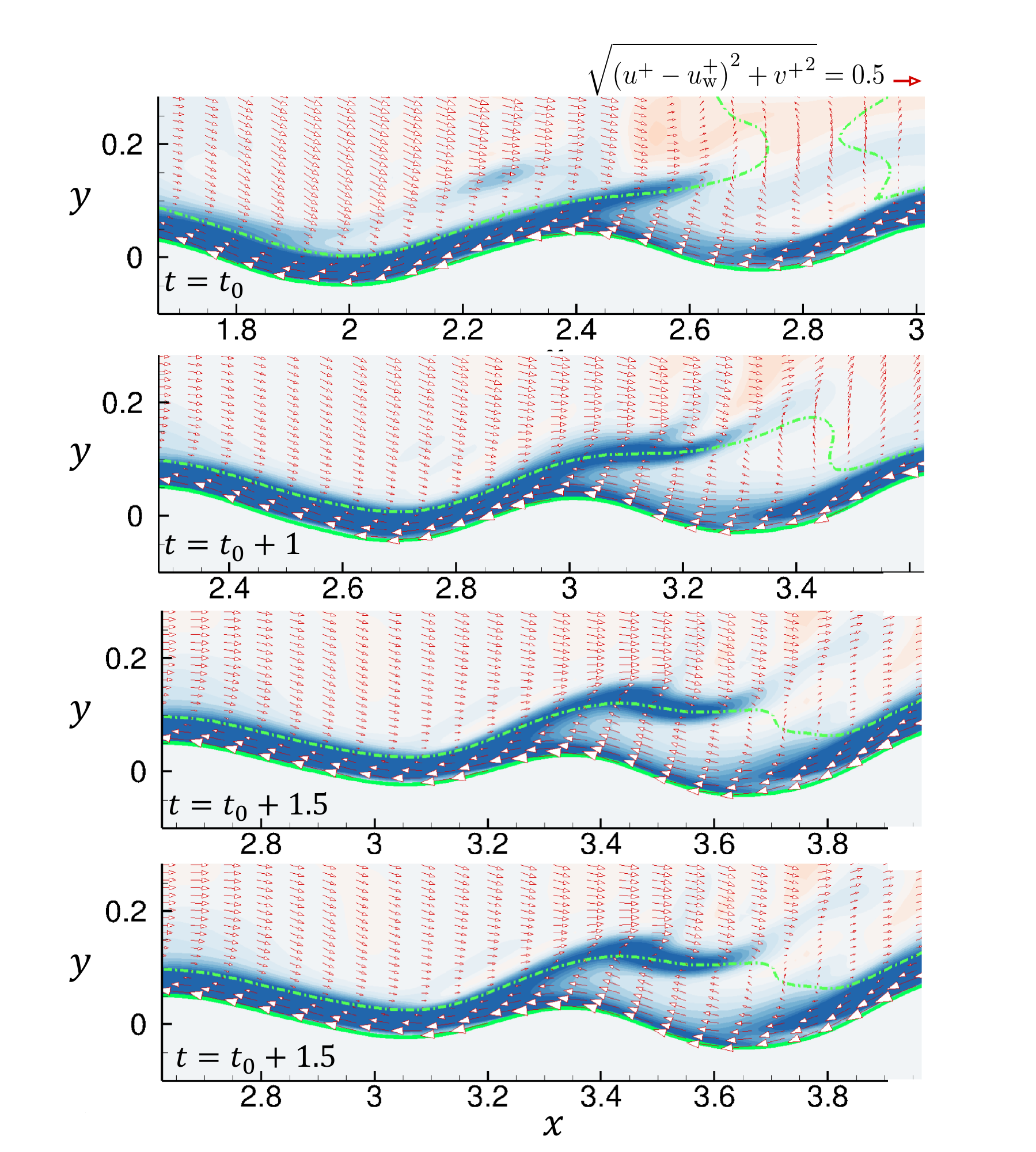} } 
    \caption{Instantaneous visualizations of spanwise vorticity contours in $xy$ plane over a sample wave crest overlaid with (a) line contours of pressure with dashed lines for negative values and (b) velocity vectors in the frame of the wave $(u-u_\text{w},v)$. Green dashed and solid lines are the instantaneous critical layer height and the fluid/solid interface, respectively.    \label{fig:vis_omz_inst_detachment}	} 
\end{figure}

Instantaneous flow visualizations in the frame of the wave clarify the unsteady flow features that are not visible in phase-averaged plots. Figure \ref{fig:vis_omz_inst_detachment} shows a series of snapshots of the vorticity field over a wave crest, which capture the shear-layer detachment near the surface. The vorticity contours are overlaid with line contours of pressure (figure \ref{fig:vis_omz_inst_detachment_pressure}) and velocity vectors in the wave frame (figure \ref{fig:vis_omz_inst_detachment_velocity_vector}). Below the critical layer, where the flow is reversed in this wave frame, the negative vorticity layer on the lee side is lifted up resulting in a significant amount of low-speed fluid being ejected from the wall region. The vertical component of the surface velocity is positive on the lee side (c.f.\,figure \ref{fig:phase_avg_uvpomz}), which contributes to this lifting process. Once the negative-vorticity crosses the critical layer, it is exposed to velocities that are faster than the wave speed, and is detached and transported downstream. The detachment process is reminiscent of a two-dimensional Kelvin-Helmholtz instability and roll-up. The pressure line contours show a drop at the core of the detached vortex, which explains the existence of a pressure minimum in phase-averaged fields (figure \ref{fig:phase_avg_uvpomz}) at the critical layer downstream of the crest.

\begin{figure}
    \centering
    \includegraphics[width =0.6\textwidth]{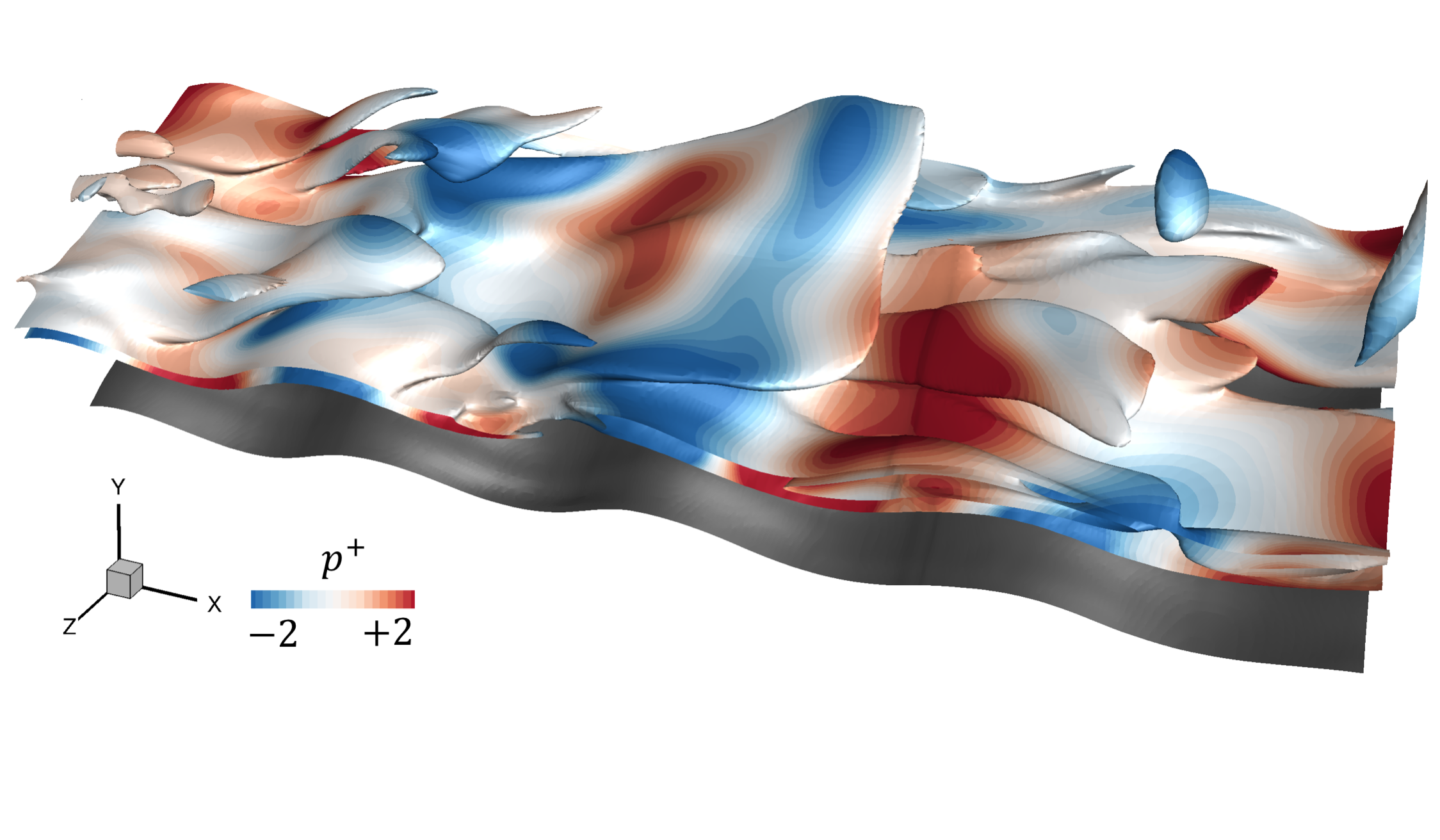}
    \caption{Instantaneous isosurface of spanwise vorticity, $\omega_z^+ = 0.275$, colored by pressure; visualization is a subset of the domain. The fluid/solid interface is also shown and is displaced vertically for clarity.}
    \label{fig:vis_omz_inst_3d}
\end{figure}

Figure \ref{fig:vis_omz_inst_3d} shows a three-dimensional view of another instance of this unsteady phenomenon. The iso-surface of vorticity colored by the pressure shows a lifted shear-layer which is about to detach. The shape of the layer is locally two-dimensional, which is consistent with the spanwise-aligned rolls formed at the compliant surface. Once the layer is completely detached from the surface, more complex and three-dimensional structures form. This unsteady phenomenon is frequently repeated on the lee side of the wave crest in cases with strong two-way coupling, i.e.~$C$, $C_L$, and $C_H$. Although the instantaneous visualizations show that the lift-up takes place in the lee side, the exact location at which the instability is triggered is not evident from these snapshots. We will investigate the origin of these events, describe the role of surface accelerations, and estimate the locations where the velocity profile is most prone to instability.

The near-interface velocity profile is directly related to the flux of vorticity at the surface. For a solid surface, the ‘vorticity flux density’ is directly related to the tangential pressure gradients \citep{lighthill1963introduction}. \citet{morton1984generation} extended Lighthill's theory to include the effect of wall acceleration. Since then, active flow control strategies exploited the relation between vorticity flux, pressure gradient and surface acceleration with the aim of reducing drag \citep{koumoutsakos1999vorticity,zhao2004turbulent}. Since the spanwise vorticity is primarily due to the wall-normal gradient of the streamwise velocity, the vorticity flux at a moving boundary can be approximated by, 
\begin{align}
    \left. \dfrac{1}{\Rey}\dfrac{\partial \omega_z}{\partial \eta}\right\vert_{\eta = 0} \approx \underbrace{-\dfrac{\partial p }{\partial \xi}|_{\eta = 0}}_{S_p} \quad \underbrace{-\DCP{u_{s,\xi}}{t}}_{S_u}. \label{eq:vorsource}
\end{align}
The two source terms on the right-hand side are the contributions from the pressure gradient and surface acceleration, where $\text{d} u_{s,\xi}/\text{d}t$ is the material derivative of the tangential surface velocity. 
When these terms lead to ${\partial \omega_z}/{\partial \eta}|_{\eta = 0}>0$, the surface is a source of negative vorticity and the velocity profile is stabilized. 
In contrast, when ${\partial \omega_z}/{\partial \eta}|_{\eta = 0}<0$, the surface is a sink of negative vorticity. As a result, a peak negative vorticity is established near the wall, and the sign of the vorticity gradient changes across it; the location of the peak is therefore an inflection point in the velocity profile.

\begin{figure}		
    \centering
    \includegraphics[width =\textwidth]{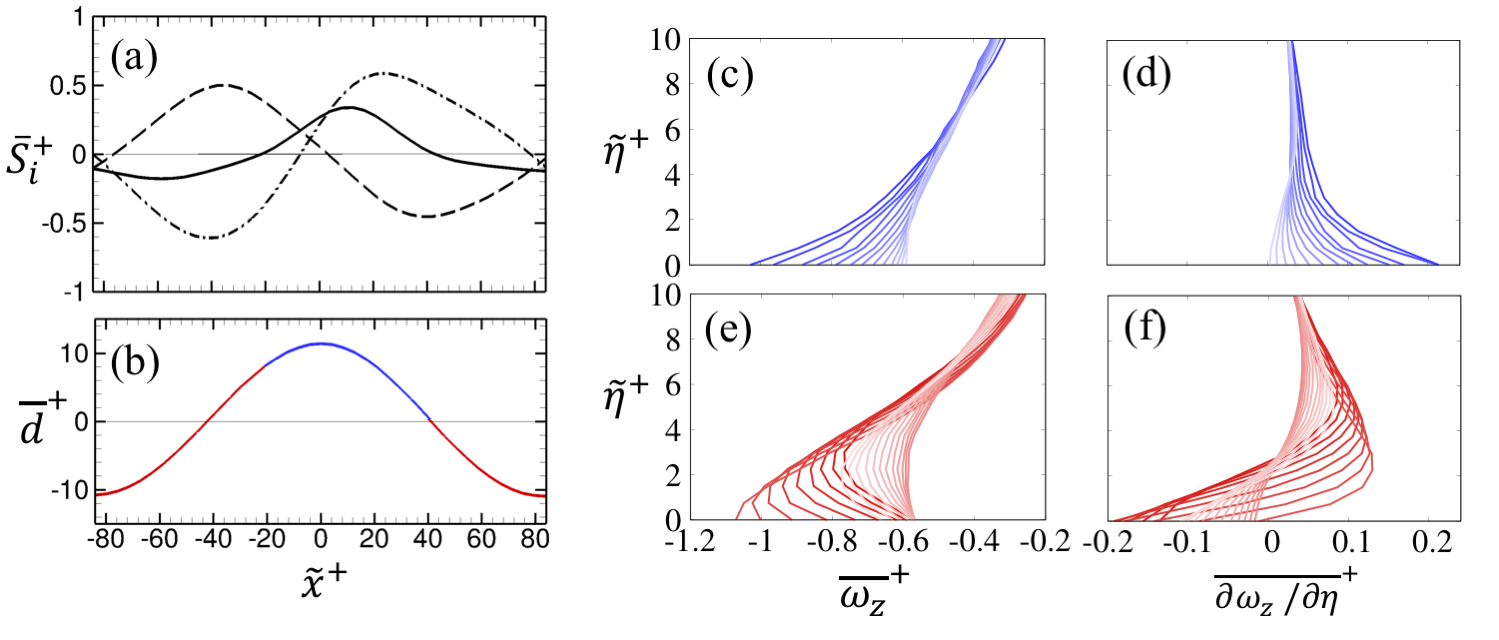}
    \caption{(a) Phase-averaged source terms of vorticity flux \eqref{eq:vorsource}: surface acceleration term $\overline{S}_u$ ({\protect\dashdotlt}, black); 
    pressure gradient term $\overline{S}_p$ ({\protect\dashlt}, black);
    $\overline{S}_u+\overline{S}_p$ ({\protect\solidlt}, black). (b) Phase-averaged surface deformation colored blue and red for $\overline{S}_u+\overline{S}_p$ positive and negative, respectively. (c-f) Profiles of $\overline{\omega_z}^+$ and $\overline{\partial{\omega_z}/\partial{\eta}}^+$ are conditioned on (c,d) $\overline{S}_u+\overline{S}_p < 0$ and (e,f) $\overline{S}_u+\overline{S}_p > 0$; Dark-to-light correspond to (c-d) $-20<\tilde{x}^+<40$ and (e-f) $-84<\tilde{x}^+<-20$ and $40<\tilde{x}^+<84$.   \label{fig:vorsource}} 
\end{figure}

\begin{figure}		
    \centering
    \includegraphics[width =0.90\textwidth]{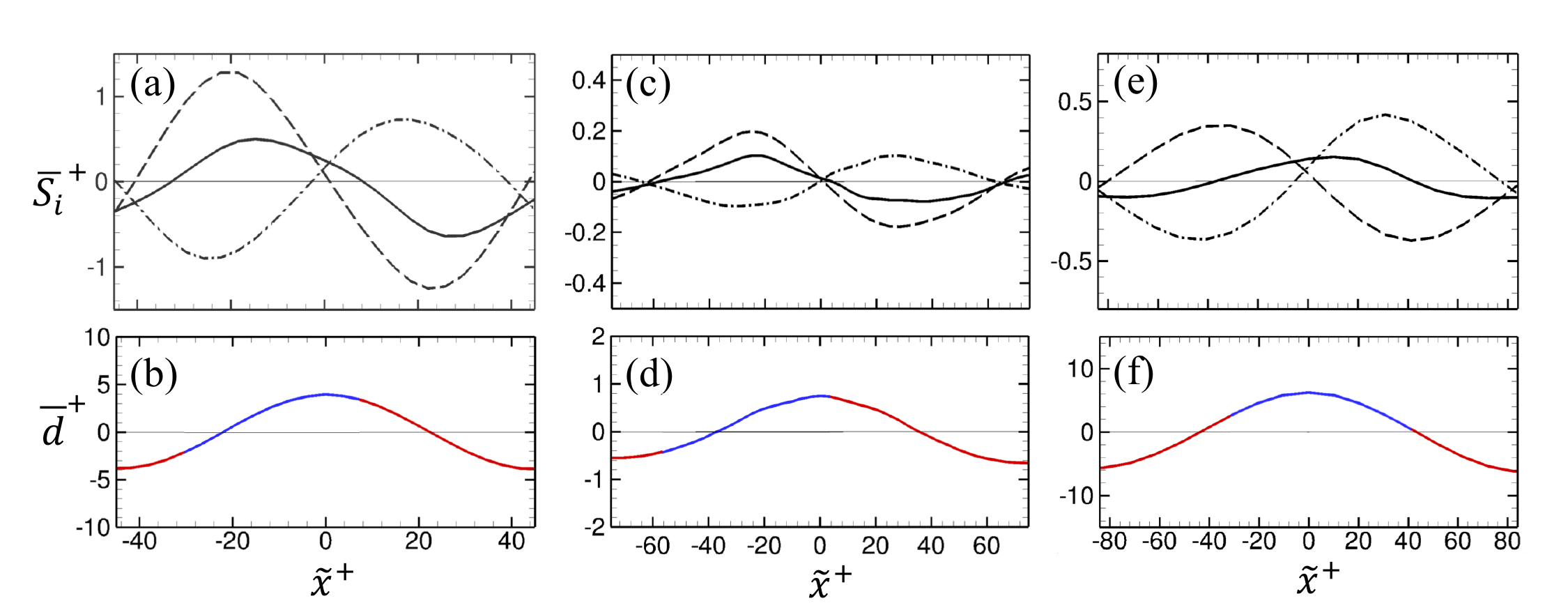} 
    \caption{
    Similar to figure \ref{fig:vorsource}(a,b), except for cases (a,b) $C_L$, (c,d) $C_G$, and (e,f) $C_H$.  \label{fig:vorsource_param}}
\end{figure}

The sources of vorticity flux were phase-averaged and are plotted in figure \ref{fig:vorsource}(a) for the case $C$. The pressure gradient contribution $\overline{S}_p$ is positive (stabilizing) on the wind-ward side and negative (destabilizing) on the lee side of the wave. This picture is consistent with intuition. Interestingly, the surface acceleration source term $\overline{S}_u$ has nearly the same amplitude as $\overline{S}_p$, and is almost out of phase. Due to asymmetries, however, the two contributions do not completely negate each other. The net source term is positive (stabilizing) over the crest and negative (destabilizing) near the troughs.
In figures \ref{fig:vorsource}(c-f), the profiles of $\overline{\omega_z}^+$ and $\overline{\partial{\omega_z}/\partial{\eta}}^+$ are shown for the phase locations that experience positive (blue) and negative (red) source terms. The blue curves show the enhanced vorticity in the stabilized phases.  The red curves show that $\overline{\partial{\omega_z}/\partial{\eta}}^+$ changes sign for a number of phases that are associated with net negative source term, $\overline{S}_p+\overline{S}_u < 0$. 
While the instabilities are initiated near the troughs where $\overline{S}_p+\overline{S}_u$ is minimum, the lifted shear layer is typically transported backward and towards the lee side of the wave because the local near-interface velocities are negative in the wave frame (c.f.\,figure \ref{fig:vis_omz_inst_detachment}). Also, despite the mean negative values of $\overline{\omega_z}^+$ at the interface, brief instances of $\overline{\omega_z}^+>0$ are common when an instability is triggered, although a complete separation event with flow reversal is extremely rare. 

We now direct our focus to the impact of wall properties and Reynolds number on the the competing effects of pressure gradient and surface acceleration (figure \ref{fig:vorsource_param}). In both cases $C_L$ and $C_G$, the two sources of vorticity flux remain out of phase. However, the pressure source is larger, and the net term $\overline{S}_p+\overline{S}_u$ is mostly negative (stabilizing) in the wind-ward side of the wave and \emph{vice versa}. Note that in these two compliant cases the surface displacements, and in turn the surface accelerations, are small in wall units relative to the main case $C$. Therefore, the compliant walls for cases $C_L$ and $C_G$ approach the behaviour of rigid roughness where the pressure gradient contribution is the only source of vorticity flux. In the case with higher Reynolds number $C_H$, the net source term and individual contributions are similar to those of case $C$. The wave amplitude in the wall unit, which is almost equal in cases $C$ and $C_H$, is therefore a controlling parameter in the balance between $\overline{S}_p$ and $\overline{S}_u$.

\subsection{Form drag, pressure work and near-wall stresses \label{sec:formdrag}}

In this section we examine the impact of the compliant wall on the flux of streamwise momentum in the surface-normal direction and the energy exchange with the flow. Due to the surface deformation, form drag becomes relevant.  While the impact of form drag is unambiguous in rough walls \citep{jimenez2004turbulent}, studies of air flow over surface gravity waves suggest a more nuanced role, e.g.~form drag can reverse sign and drive the wind for fast waves \citep{gent1977numerical,sullivan2000simulation}. 

\begin{figure}		
    \centering
    \includegraphics[width =\textwidth]{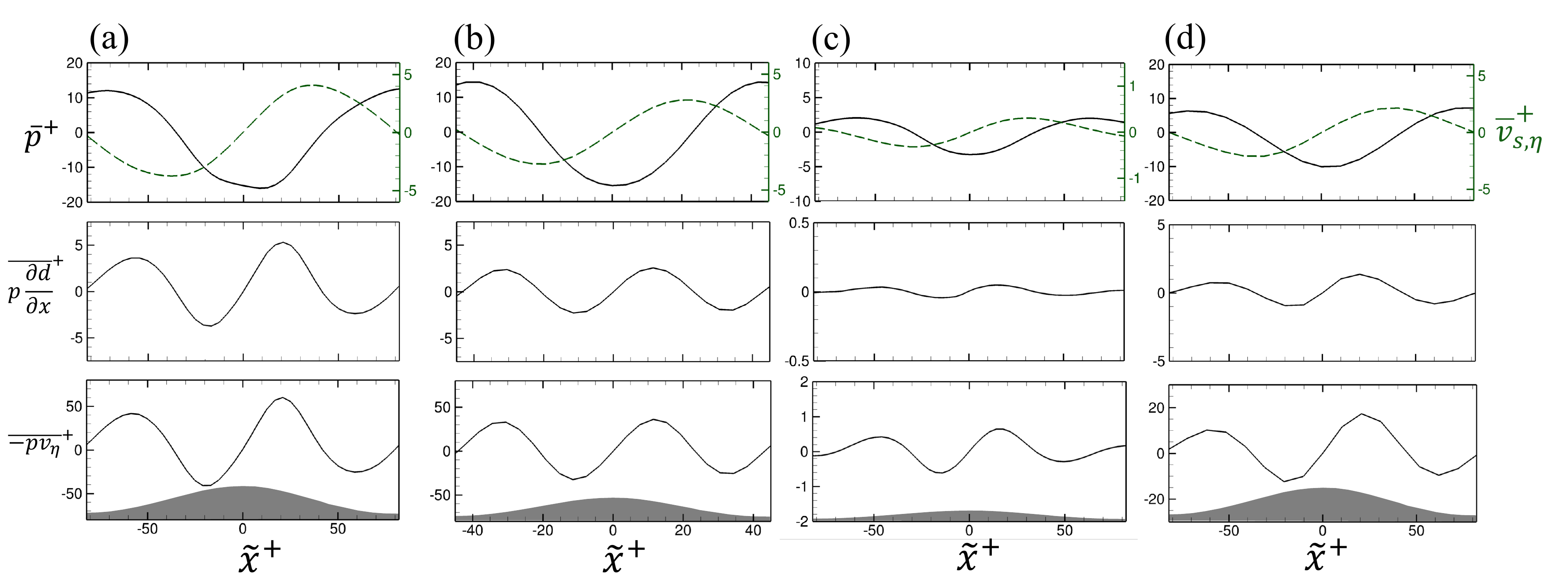} 

    \caption{Top panels: phase-averaged pressure $\overline{p}^+$ ({\protect\solidlt}, black) and surface-normal velocity $\overline{v_{s,\eta}}^+$ ({\protect\greendashlt}, green). 
    Middle panels: form drag $\overline{p\partial{d}/\partial{x}}^+$.
    Bottom panels: pressure-work exerted by the surface onto the fluid $\overline{-p v_{s,\eta}}^+$.
    Cases (a) $C$, (b) $C_L$, (c) $C_G$, and (d) $C_H$. 
    Surface displacement is schematically marked in the bottom panels.  \label{fig:pwork}}
\end{figure}

The phase-averaged pressure $\overline{p}^+$ and form drag $\overline{ p \partial d /\partial x}^+$ for different cases are shown in figure \ref{fig:pwork}. As discussed \S\ref{sec:instability}, the pressure is out-of-phase with the surface displacements. There is an asymmetry in the pressure relative to the wave crest, and the minimum pressure is slightly shifted towards the lee side of the wave.  The pressure drag onto the surface $\overline{ p \partial d /\partial x}^+$ (middle panels) is also asymmetric relative to the crest.  The net effect is a positive drag on the interface, which is most pronounced in case $C$ (see also table \ref{tab:totavg}). In cases with smaller surface displacements, such as $C_G$, not only the wave-correlated pressure is smaller, but also the pressure is more symmetric with respect to the crest. Both effects result in a reduced form drag. 

\begin{figure}		
    \centering
    \makebox[\linewidth][c]{%
    \subfigure[]{\label{fig:wavestress_C0}
    \includegraphics[height=120pt,scale=1,trim={0 0 1.14cm 0},clip]{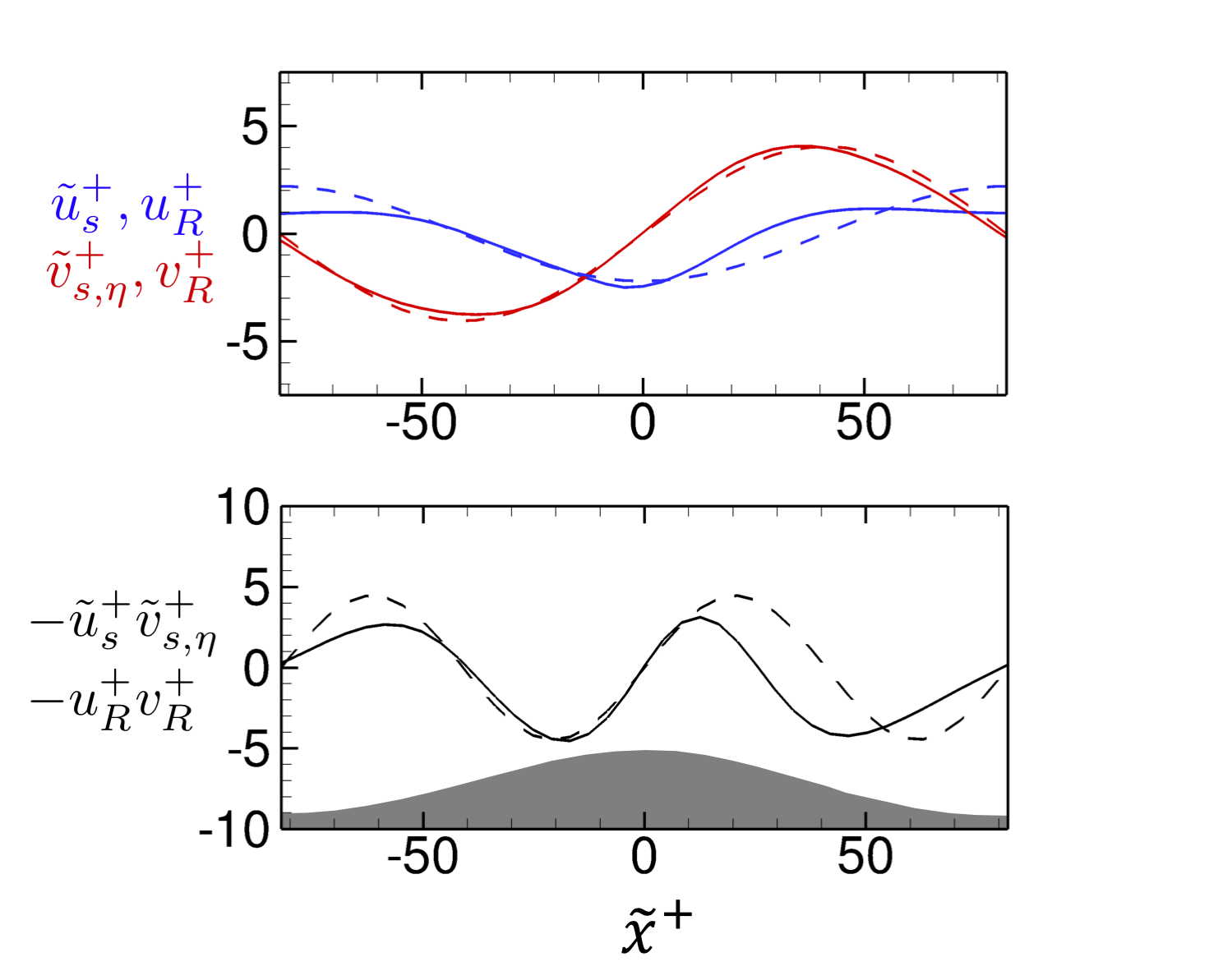} } 
    \subfigure[]{\label{fig:wavestress_CL}
    \includegraphics[height=120pt,scale=1,trim={1.8cm 0 1.14cm 0},clip]{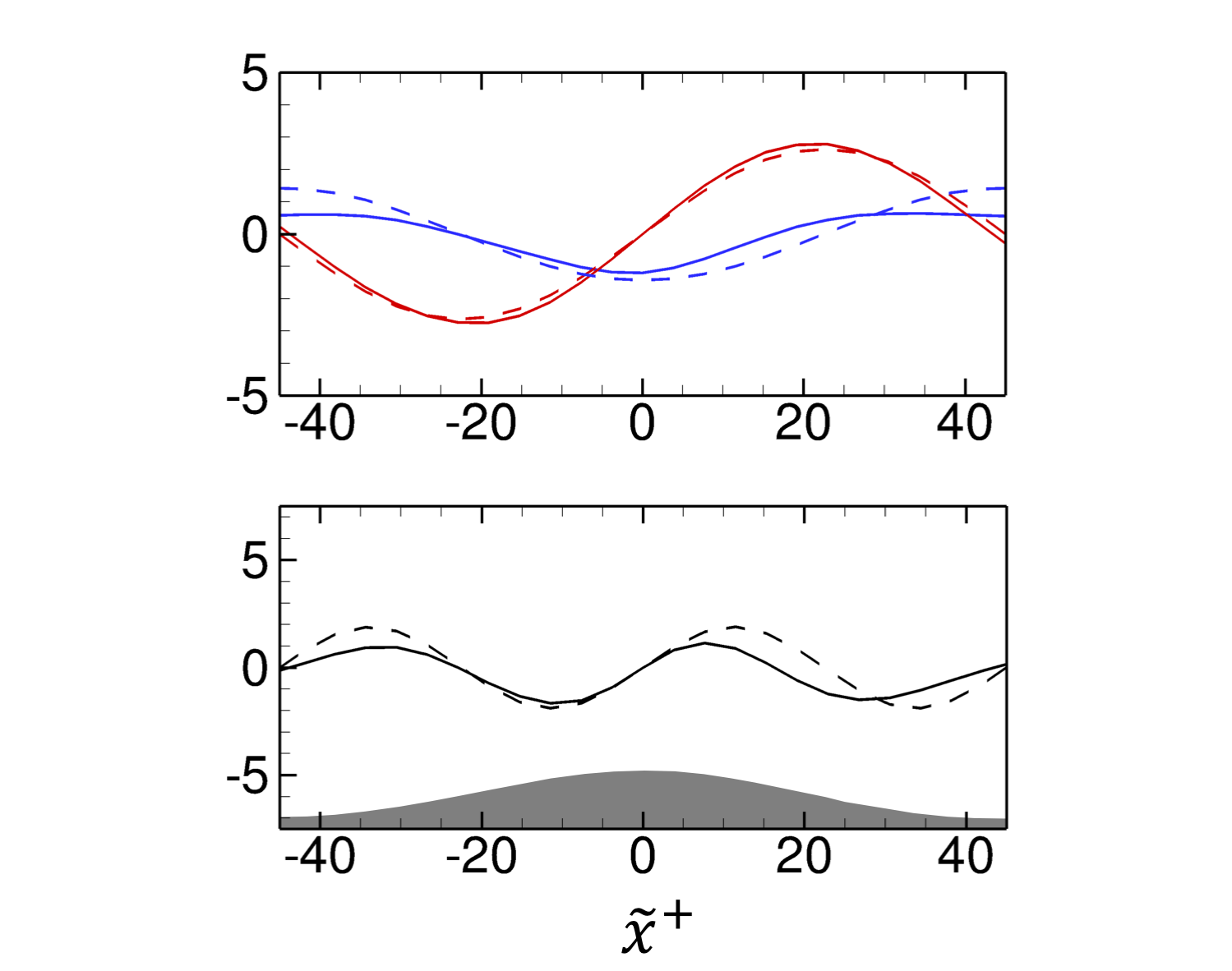} } 
    \subfigure[]{\label{fig:wavestress_CH}
    \includegraphics[height=120pt,scale=1,trim={1.8cm 0 0 0},clip]{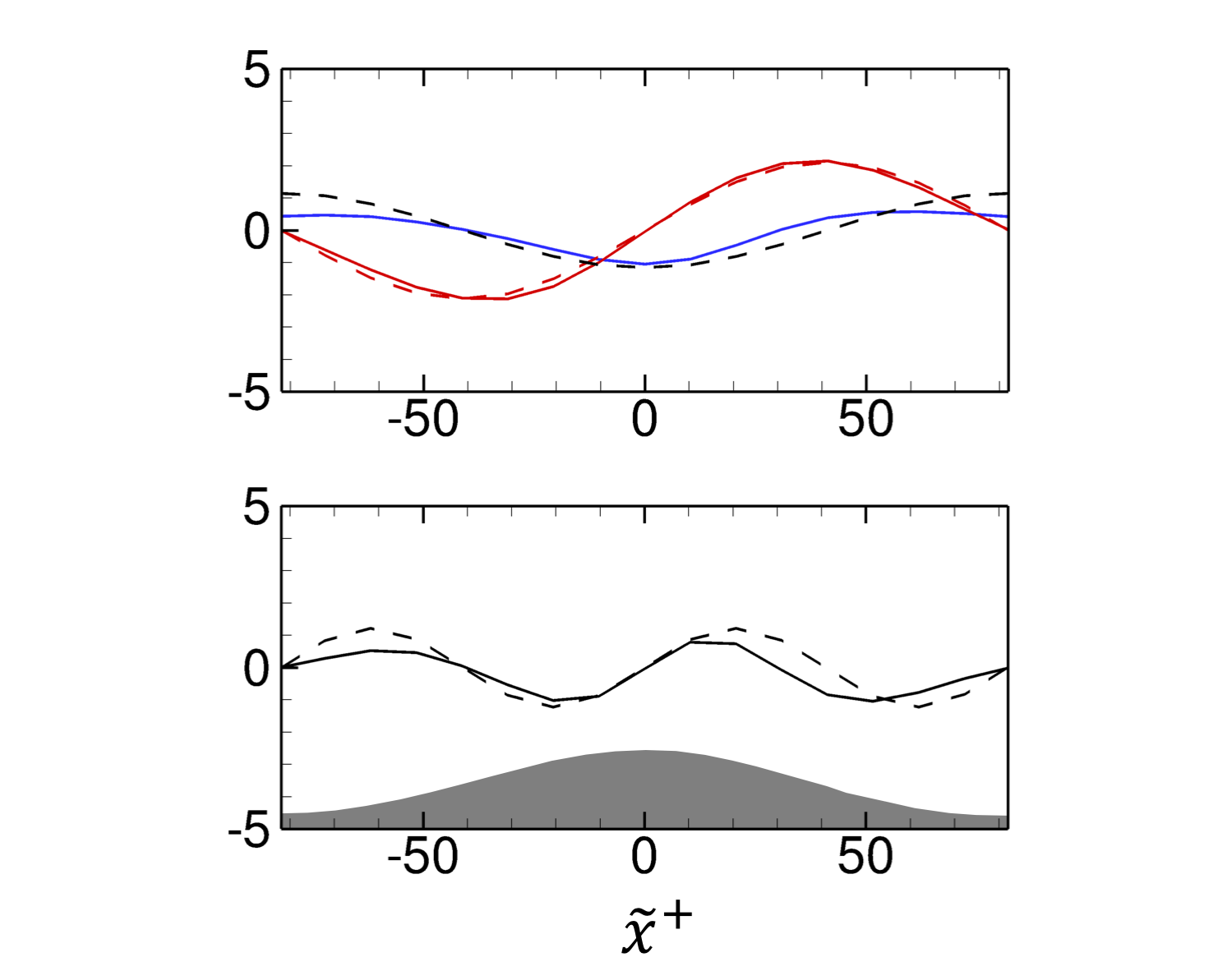} }  } 
        \caption{Top panels: phase-averaged surface velocities in streamwise $\tilde{u}^+_s$ ({\protect\bluelt}, blue) and surface-normal $\tilde{v}^+_{s,\eta}$ ({\protect\redlt}, red) directions. 
        Bottom panels: phase-averaged wave-induced shear stress $-\tilde{u}^+_s\tilde{v}^+_{s,\eta}$ ({\protect\solidlt}, black). 
        In all panels, velocities and stresses are also plotted from solution of the Rayleigh wave equations: horizontal velocity $u_R^+$ ({\protect\bluedashlt}, blue), vertical velocity ({\protect\reddashedlt}, red), and shear stress $-u_R^+v_R^+$ ({\protect\dashlt}, black). Cases (a) $C$, (b) $C_L$, (c) $C_H$. Surface displacement is schematically marked in the bottom panels.     \label{fig:wavestress}}
\end{figure}

Unlike for rigid wall, the compliant material can exert pressure work onto the fluid, $\langle -p v_{s,\eta} \rangle^+$.  The phase-averaged interface-normal velocity is included in the top panel of figure \ref{fig:pwork} along with the pressure. The surface motion is downward on the wind-ward side and upward on the lee side of the wave, and lags the pressure by nearly $\pi/2$ phase shift. 
The correlation is plotted in the bottom panel, and inherits the asymmetry of the pressure ($\overline{v}_{s,\eta}^+$ will be shown to closely match the symmetry of Rayleigh waves, c.f.~figure\ref{fig:wavestress}). 
Again, the net effect integrated over the entire surface is positive in all cases (table \ref{tab:totavg}). The fact that the compliant surface exerts work onto the fluid may seem in contradiction with the increase in drag (c.f.\,figure \ref{fig:stress_barchart}). It should be noted that this energy exchange is primarily due to the vertical motion of the wave, which is not desirable from a drag reduction perspective. Instead, this pressure work onto the fluid enhances the turbulence kinetic energy, which is eventually dissipated without any favorable impact on drag.  

\begin{table}
\begin{center}
\def~{\hphantom{0}}
\setlength\extrarowheight{2	pt}
\begin{tabular}{l|c|c|c}	
Case & $\langle -p(\bm{u}_s \cdot \bn) \rangle^+$ & $\langle p \partial {d}/\partial{x}\rangle^+$  & $-\langle u'_s v'_{s,\eta} \rangle^+$ \\ [3pt] \hline  
$C$ & 2.811 & 0.296 & -0.364 \\ 
$C_L$ & 0.771 & 0.069 & -0.089 \\
$C_G$ & 0.0259 & 0.0011 & -0.0007 \\
$C_H$ & 0.6116 & 0.0496 & -0.0981 \\
\end{tabular}
\end{center}
\caption{Space- and time-averaged pressure work exerted by the surface onto the fluid $\langle -p v_{s,\eta} \rangle^+$, form drag $\langle p \partial {d}/\partial{x}\rangle^+$ and surface shear stress $-\langle u'_s v'_{s,\eta} \rangle^+$.   \label{tab:totavg} }
\end{table}

In \S\ref{sec:global_changes} we showed that the turbulent shear stress, evaluated in Cartesian coordinates, changes sign near the surface and is negative at $y=0$. We recall that Rayleigh waves at the free surface of an elastic material have zero shear stress because its sinusoidal horizontal and vertical velocities are $\pi/2$ out of phase. Therefore, in our configuration the turbulent flow alters the surface wave motion in a manner that gives rise to finite net shear stress. 
We compare the phase-averaged surface velocities to Rayleigh waves in figure \ref{fig:wavestress}. Since our interest is in the surface-normal flux of streamwise momentum, we plot the streamwise and surface-normal velocities. It is clear in all cases that $\tilde{v}^+_{s,\eta}$ is essentially sinusoidal, while $\tilde{u}^+_s$ deviates from the Rayleigh wave motion in particular on the lee side. This deviation is due to the reaction force of the fluid onto the surface, in response to the total drag. As will be discussed below, turbulent and pressure drag are particularly large in the lee side of the wave near $0<\tilde{x}^+<50$. The reaction force on the material is, therefore, also large and in the positive $x$ direction, which decreases the magnitude of the negative tangential velocity of the surface in that $\tilde{x}^+$ range. 
Ultimately, $-\tilde{u}^+_s\tilde{v}^+_{s,\eta}$ also deviates from the Rayleigh sinusoidal pattern, with the largest difference on the lee side. Consistent with our earlier observations in Cartesian coordinates (\S\ref{sec:global_changes}), $-\langle u'_s v'_{s,\eta} \rangle^+$ averaged over the entire surface is negative in all cases (table \ref{tab:totavg}). 

\begin{figure}		
\begin{center}
        \subfigure[]{\label{fig:PA_uv_wave}
            \includegraphics[width =0.75\textwidth]{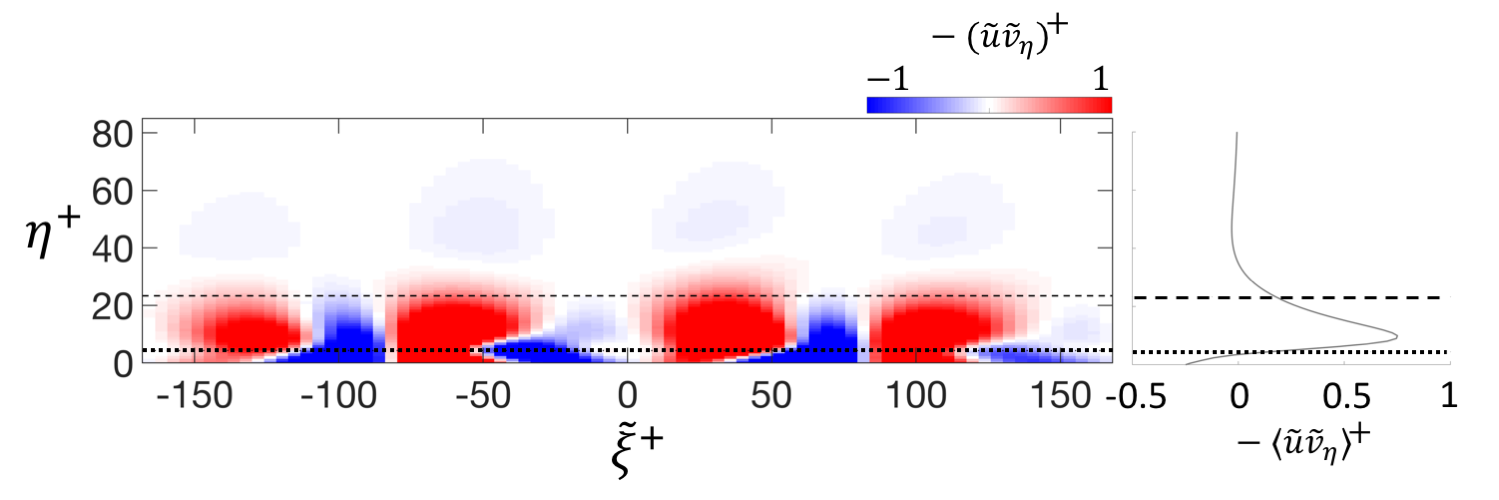}
             }\\ 
             \hspace{-20pt}
        \subfigure[]{\label{fig:PA_uv_stochastic}
            \includegraphics[width =0.75\textwidth]{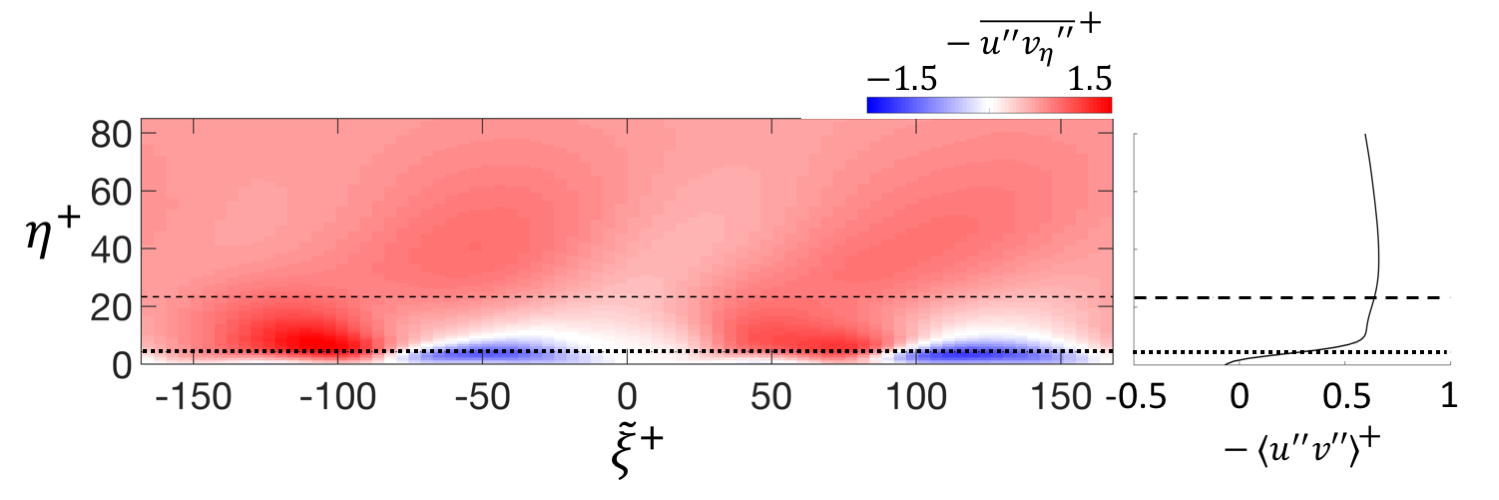}
            } \\ 
        \subfigure[]{\label{fig:PA_form}
            \includegraphics[width =0.75\textwidth]{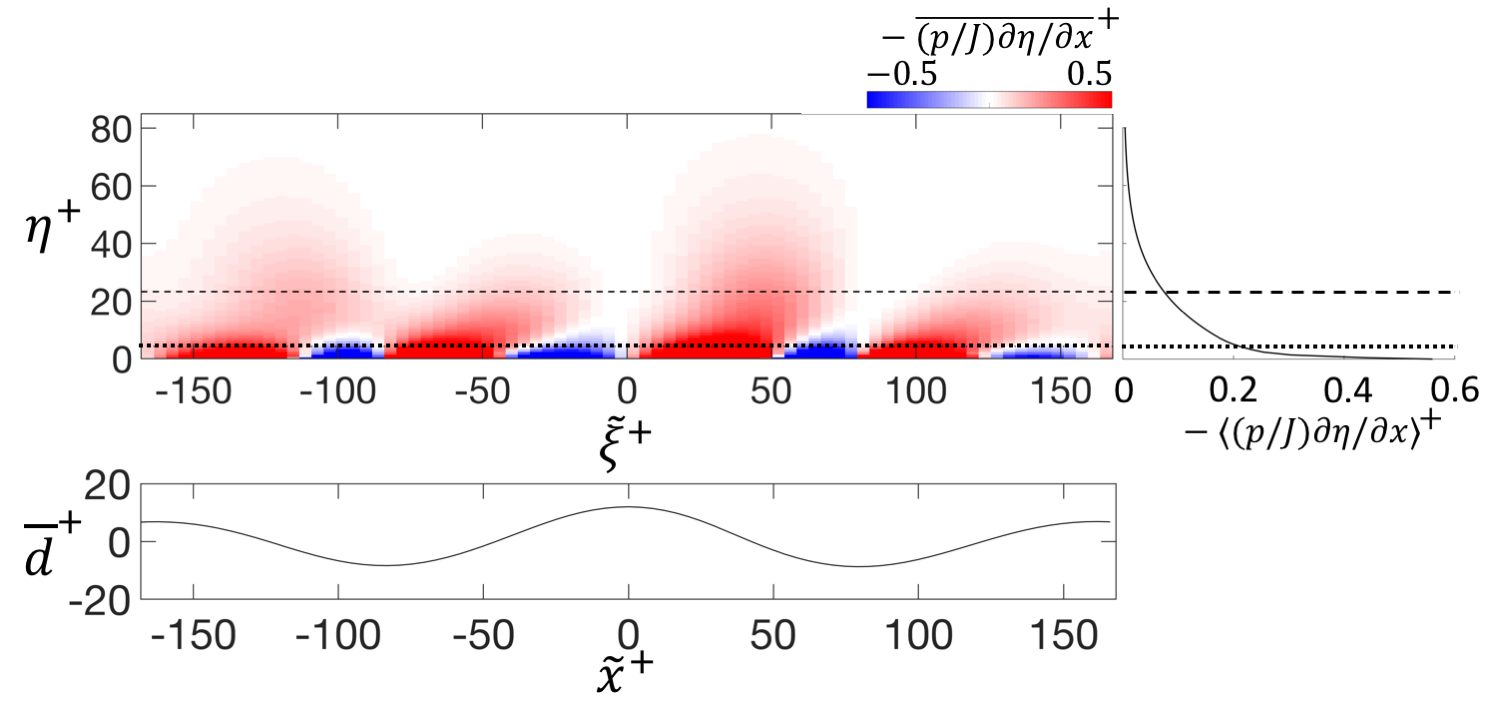}
             }\\ 
\end{center}        
\caption{Phase-averaged stresses in surface-fitted coordinates for case $C$. 
(a) Wave-correlated $- (\tilde{u}\tilde{v}_\eta)^+$ stress, (b) stochastic Reynolds shear stress $- \overline{u'' v''_\eta}^+$, and (c) pressure stress $- \overline{(p/J) \partial \eta/\partial x}^+$ and phase-averaged surface displacement $\overline{d}^+$. The line plots in the right panels show the quantities averaged over one wavelength. The horizontal lines show the height of the wave boundary layer ({\protect\dottedlt}, black), and the height of the critical layer ({\protect\dashlt}, black). \label{fig:phase_avg_stress}	}   
\end{figure}

The extent to which the negative shear stress at the surface persists into the flow is examined in figure \ref{fig:phase_avg_stress}.  We report the phase-averaged shear and pressure stresses in surface-fitted coordinates. We only present case $C$ because the results are qualitatively similar for cases $C_L$ and $C_H$, and the wave-correlated stresses are negligible for case $C_G$.  
The contours of the wave-correlated component of the Reynolds shear stress $- \tilde{u}\tilde{v}_\eta$ reflect the importance of the critical layer and the wave boundary-layer height (figure(\ref{fig:PA_uv_wave}). The latter is defined as $h_\text{w} \equiv \sqrt{\nu \lambda_x /u_\text{w}}$, where $\lambda_x$ is the dominant streamwise wavelength, and is approximately $h_\text{w}^+ \approx 4$. Below this height the motion of the fluid is appreciably influenced by the wave, and hence the patterns of $- \tilde{u}^+\tilde{v}_\eta^+$ are very similar to those reported at the surface (compare with \ref{fig:wavestress_C0}); the net contribution $\langle- \tilde{u}\tilde{v}_\eta\rangle^+$ is therefore negative. Between $h_\text{w}^+\approx 4$ and the critical layer height, the negative stress decays and the positive stress increases substantially, resulting in a change of sign of the net contribution, $\langle - \tilde{u}\tilde{v}_\eta\rangle^+$. The flow below the critical layer, which spans the viscous sub-layer and part of the buffer layer, is simultaneously influenced by the wave motion and the turbulence. Above the critical layer, the magnitude of $- \tilde{u}\tilde{v}_\eta$ decays quickly and is practically negligible. As such, the critical-layer height demarcates the region of impact of the wave-correlated stresses above the surface, similar to turbulent air flow above gravity waves \citep{sullivan2000simulation,yousefi2020momentum}.

The stochastic turbulent stresses $- \overline{u'' v''_\eta}^+$ are shown in figure \ref{fig:PA_uv_stochastic}. Except for the region below $h_\text{w}$ on the windward side of the wave, $- \overline{u'' v''_\eta}^+$ is positive everywhere. The stresses peak between the trough and the critical layer, where the unsteady shear-layer detachment events described in \S\ref{sec:instability} are most probable. Above the critical layer, the patters of strong positive $- \overline{u'' v''_\eta}^+$ are tilted forward due the higher local velocities.

The pressure stress $- \overline{(p/J) \partial \eta/\partial x}^+$, where $J$ is the Jacobian of the coordinate transformation, is shown in figure \ref{fig:PA_form}. Below the wave boundary layer, the pressure drag is equal to the form drag at the surface, and alternates between negative and positive stress on the windward and lee sides of the wave. Above the wave boundary layer, however, only positive drag is observed. This positive pressure stress has a phase lead above $h_\text{w}$, and is due to the unsteady shear-layer detachment events that take place on the trough. Note that the pressure stress by definition depends on the curvature of the iso-levels of $\eta$, and therefore it gradually decays to zero as the impact of surface undulations diminishes away from the surface. The pressure itself, however, remains correlated with the wave at much higher $\eta$ locations, as previously seen in figure \ref{fig:phase_avg_uvpomz}(c).

\begin{figure}		
    \centering
    \includegraphics[width =0.85\textwidth,scale=1.25,trim={0cm 0cm 0cm 0},clip]{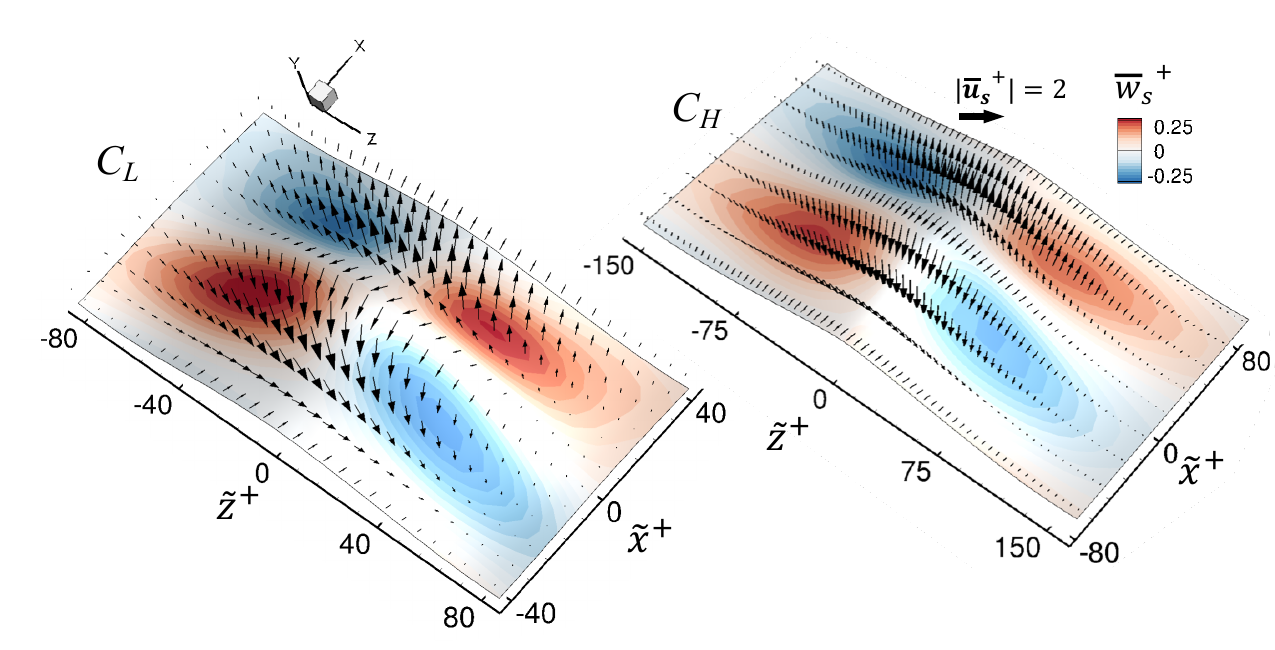}
    \caption{Phase-averaged surface displacement, velocity vectors and contours of the spanwise component of surface velocity near the crest.  (left) $C_L$ and (right) $C_H$. The surface displacement is multiplied by a factor of two for clarity.   \label{fig:surface_vel_3d}}
\end{figure}

\subsection{Three-dimensional effects \label{sec:3deffects}}
Since the propagating waves in the compliant layer are primarily spanwise-oriented rolls, we have so far focused on a two-dimensional analysis of the wave-correlated velocities. However, as shown previously in spanwise wavenumber-frequency spectra (figure \ref{fig:spectra_surf_z}), finite $k_z$ deformations are present although at lower wavenumbers compared to the streamwise ones. In this section we investigate the impact of three dimensionality of the surface on the wave-correlated velocities and Reynolds shear stress. 

We start by analysing the surface velocities associated with the wave propagation (figure \ref{fig:surface_vel_3d}). 
Despite the interface undulation in the span, the surface-velocity vectors are mostly two-dimensional and dominated by their $x$ and $y$ components. These two components are associated with the streamwise-propagating rolls, which dominate the surface motion. The spanwise interface velocity (color contours) is smaller in magnitude, and has a wave-correlated pattern:
the sign of $\overline{w}_s$ matches that of $\overline{v}_{s,\eta}\overline{n}_z$, where $\overline{n}_z$ is the spanwise component of the surface-normal vector. 
The observed pattern is better understood considering the combined effects of streamwise wave propagation and spanwise surface deformation. At $\tilde{z}=0$, the wave propagation is sustained by upward velocities on the lee side giving way to downward velocity on the windward side of the crest (vector plots in figure \ref{fig:surface_vel_3d}). In the region ($0<|\tilde{z}|<|\lambda_z/2|$), where $\lambda_z$ is the dominant spanwise wavelength, the interface-normal velocity has a component in the spanwise direction.  The product of $\overline{v}_{s,\eta}$ and $\overline{n}_z$ determines the sign and magnitude of $\overline{w}_s$. Due to the phase relation between $\overline{v}_{s,\eta}$ and $\overline{n}_z$, the maximum amplitude of $\overline{w}_s$ occurs at $\tilde{z} = \pm \lambda_z/4$. 
Physically, the interface on the lee side is stretched outward in the span as it is pulled away from the wall, and on the windward side it is squeezed in the span towards the middle as it retracts towards the wall.  
While these three-dimensional features of the surface are more clearly observed in cases $C_L$ and $C_H$, similar patterns are also observed in cases $C$ and $C_G$ (not shown). In case $C$, the spanwise-oriented roles are dominant and hence the structures are less three-dimensional. In case $C_G$, the wave-correlated velocities are in general smaller due to the weaker coupling between surface and the flow.

\begin{figure}		
    \centering
    \makebox[\linewidth][c]{%
    \subfigure[]{\label{fig:PA_surface_3d}
    \includegraphics[width =0.45\textwidth,scale=1,trim={0 0 0cm 0},clip]{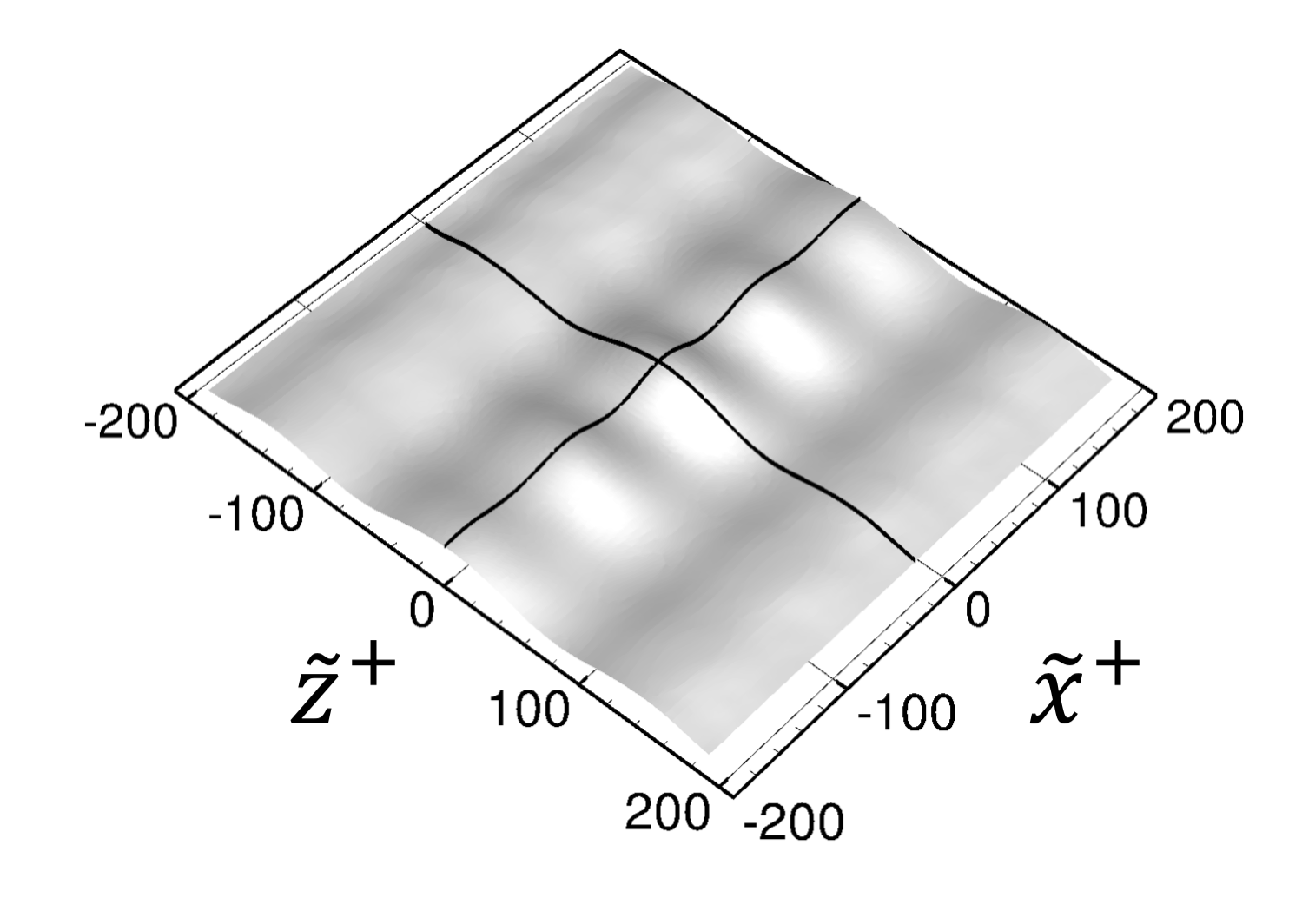} } 
     \subfigure[]{\label{fig:PA_vfluid_3d}
    \includegraphics[width =0.45\textwidth,scale=1,trim={0cm 0 0cm 0},clip]{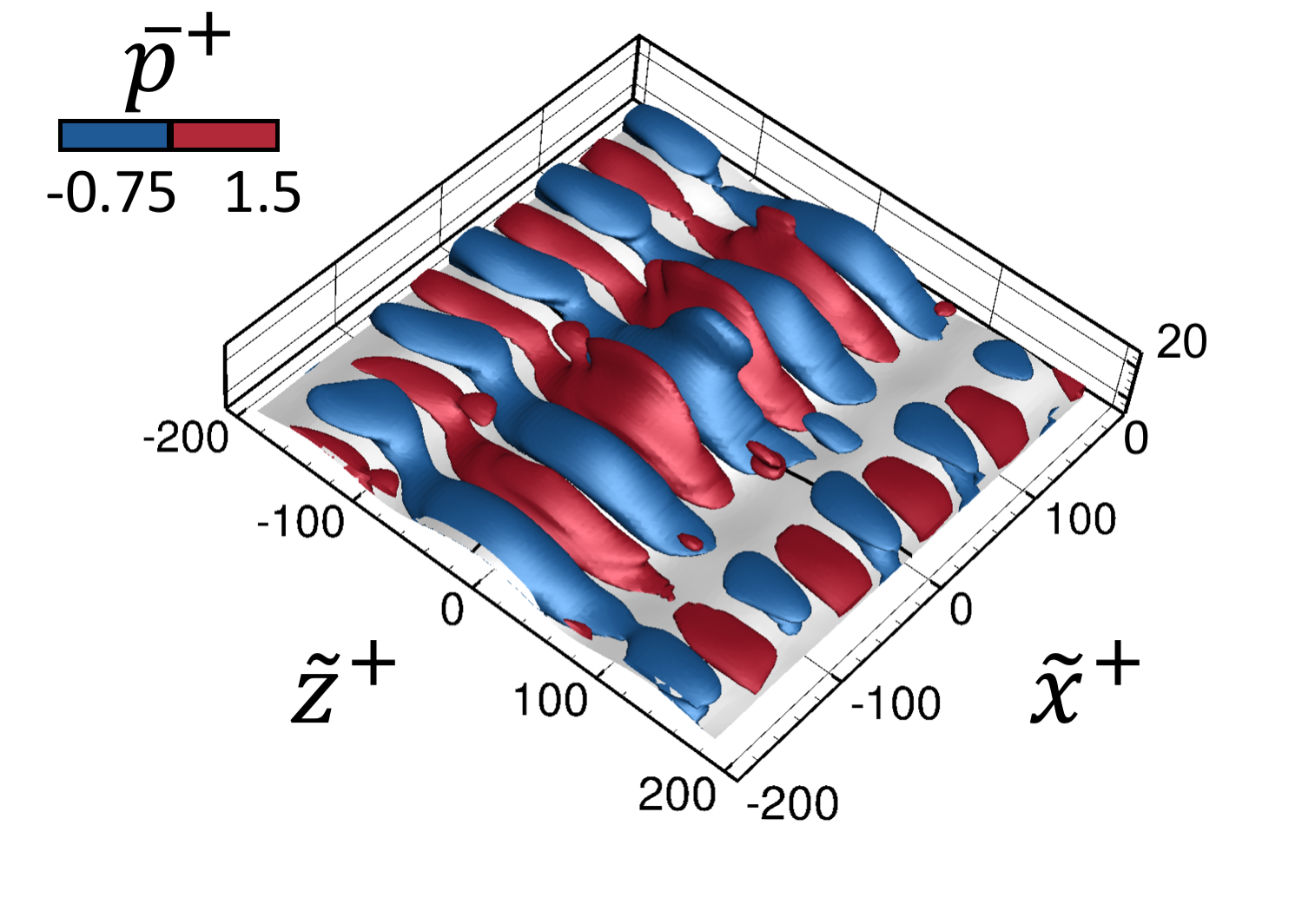} }  } \\
    \makebox[\linewidth][c]{%
    \subfigure[]{\label{fig:PA_p_3d}
    \includegraphics[width =0.45\textwidth,scale=1,trim={0cm 0 0cm 0},clip]{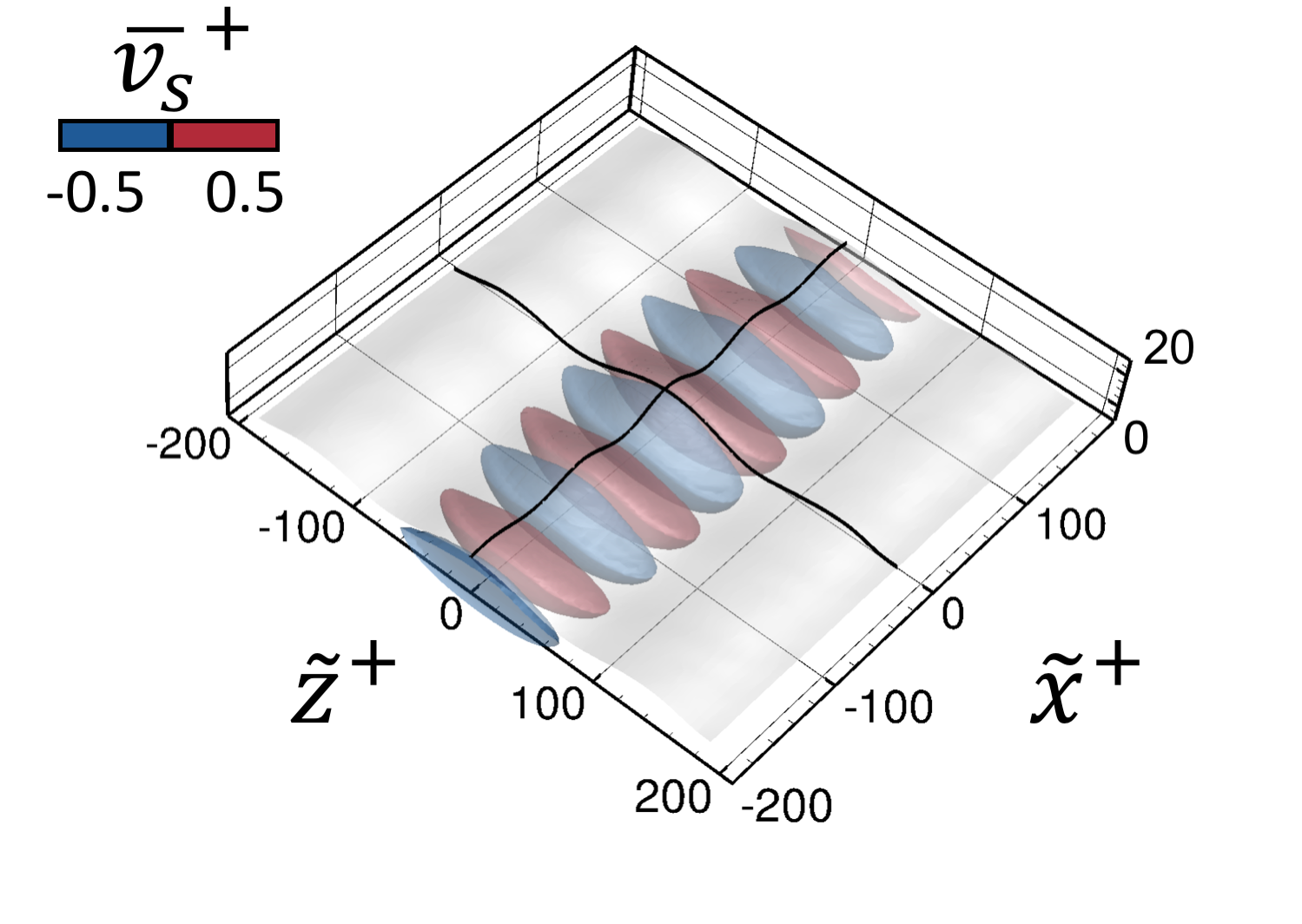}} 
    \subfigure[]{\label{fig:PA_vsolid_3d}
    \includegraphics[width =0.45\textwidth,scale=1,trim={0 0 0cm 0},clip]{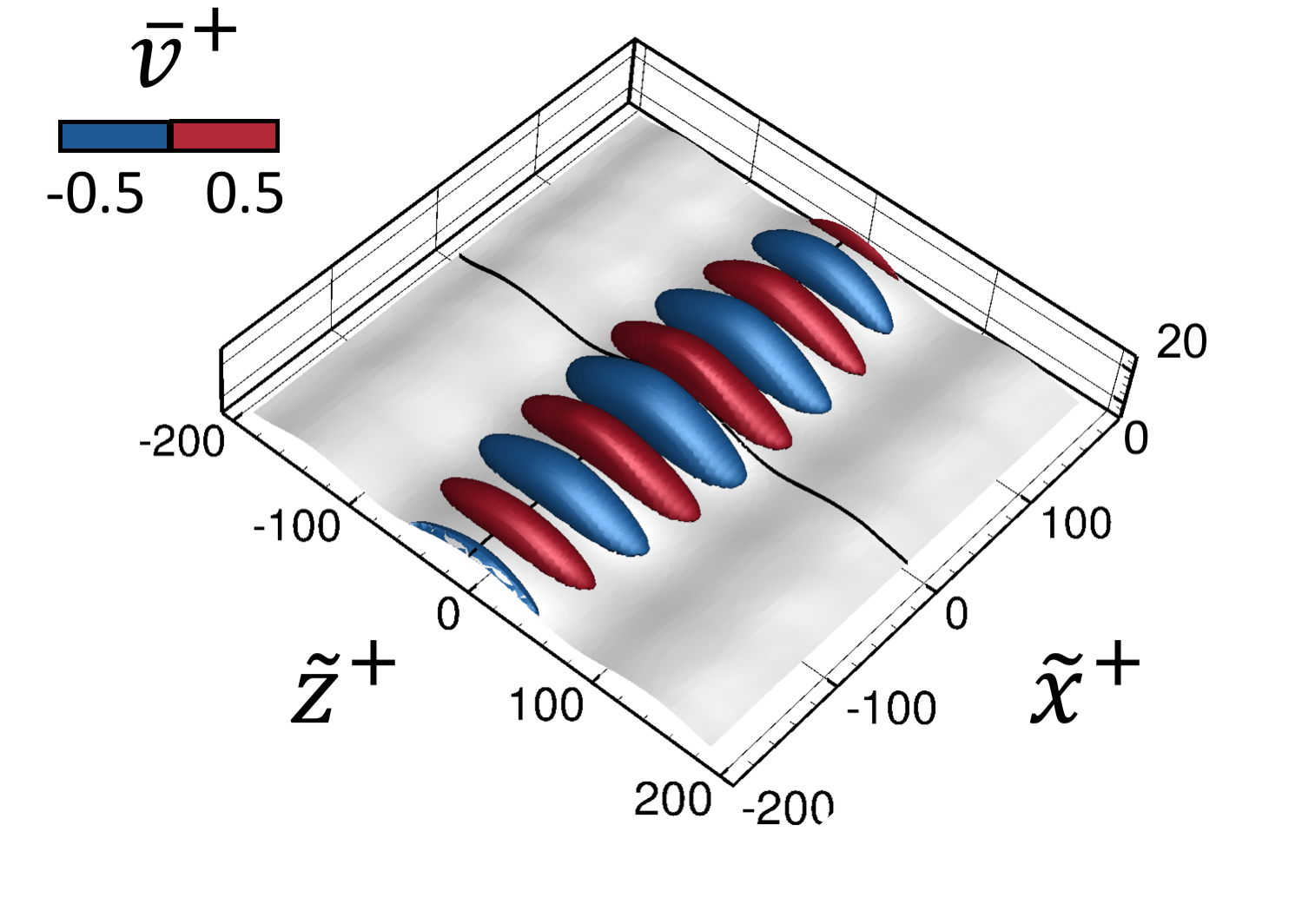} } } \\
    \makebox[\linewidth][c]{%
    \subfigure[]{\label{fig:PA_wfluid_3d}
    \includegraphics[width =0.45\textwidth,scale=1,trim={0cm 0 0cm 0},clip]{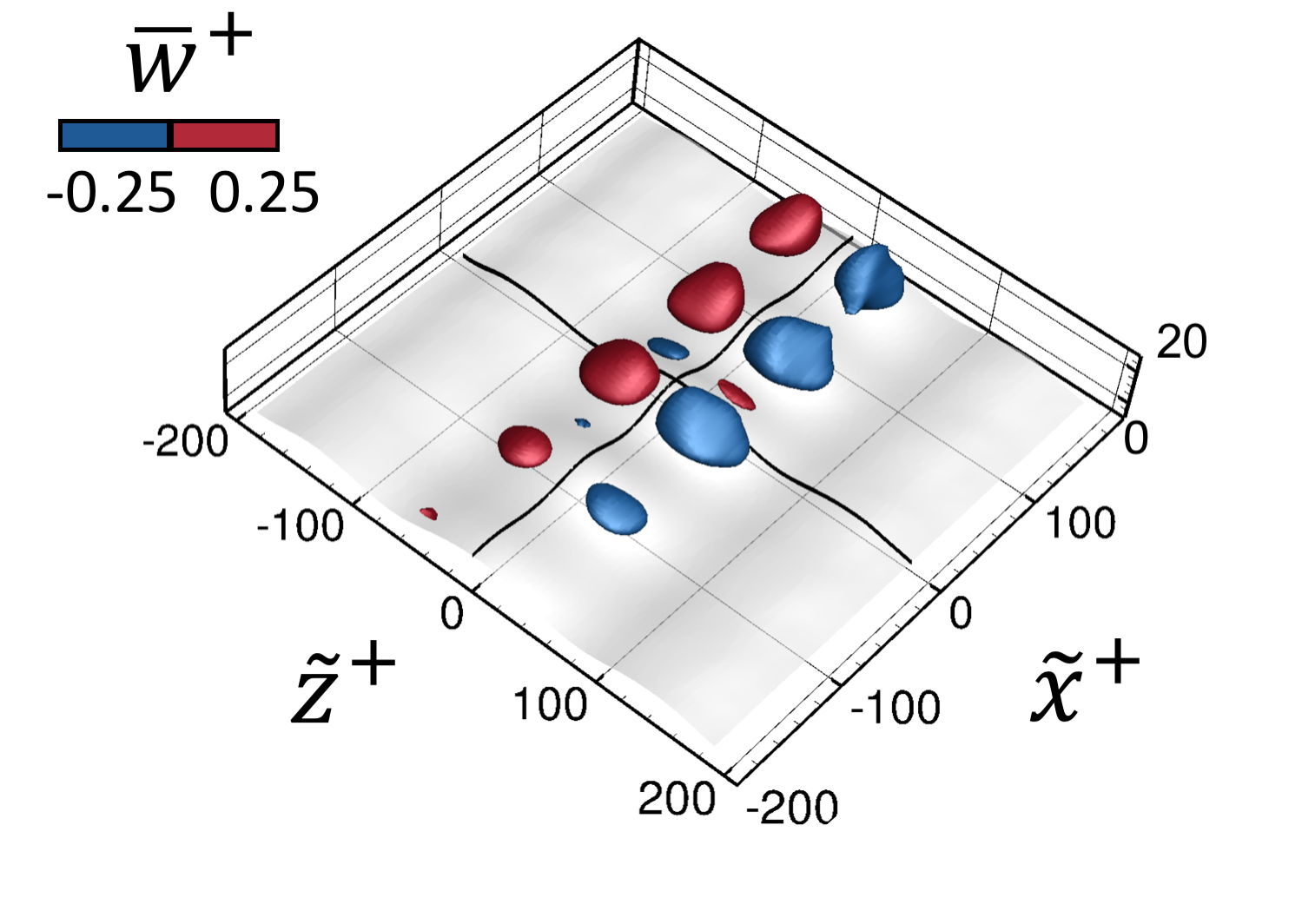} }
    \subfigure[]{\label{fig:PA_ufluid_3d}
    \includegraphics[width =0.45\textwidth,scale=1,trim={0cm 0 0cm 0},clip]{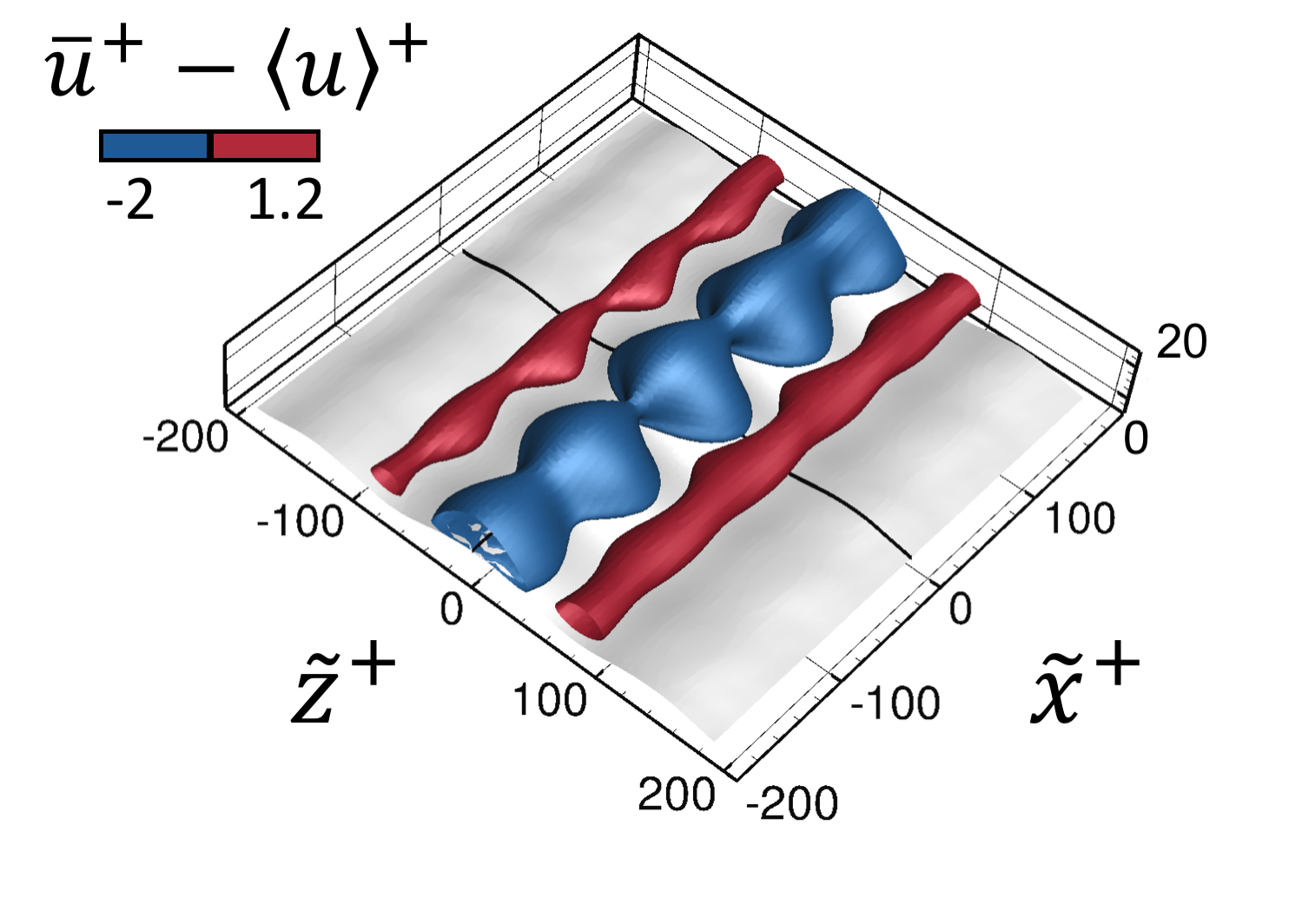} } }   
    \caption{Phase-averaged isosurfaces from case $C_L$.  (a) Fluid/solid interface, (b) pressure $\overline{p}$, (c) wall-normal velocity inside the compliant wall $\overline{v_s}^+$, (d) wall-normal velocity in the fluid $\overline{v}^+$, (e) spanwise velocity $\overline{w}$ and (f) streamwise velocity fluctuations $\overline{u}^+-\langle u\rangle^+$.  \label{fig:PA_3d}}
\end{figure}

Figure \ref{fig:PA_3d} shows phase-averaged quantities below and above the surface for case $C_L$. The phase-averaged pressure iso-surfaces are shown in figure \ref{fig:PA_p_3d}. They are predominantly two-dimensional, echo the wave motion in the streamwise direction, and their magnitude is largest in the span above the crest.  The vertical velocity component inside the compliant layer and in the fluid is clearly dominated by the wave motion (figures \ref{fig:PA_vfluid_3d} and \ref{fig:PA_vsolid_3d}). 
The spanwise fluid velocity $\overline{w}$ in figure \ref{fig:PA_wfluid_3d} matches the above description of figure \ref{fig:surface_vel_3d};  the velocity structures are, however, asymmetric in the direction of wave propagation, specifically stronger on the windward side of the wave. This asymmetry is due to the presence of a secondary flow which we will discuss below. The streamwise velocity fluctuations with respect to the Cartesian average $\overline{u}^+-\langle u\rangle^+$ are shown in figure \ref{fig:PA_ufluid_3d}. While influenced by the interface wave motion in the streamwise direction, there is also a clear spanwise dependence due to the three-dimensionality of the interface. 
Specifically, the streamwise velocity is slower above the spanwise crests and faster above the spanwise troughs. Momentum transport across this spanwise gradient, $-\overline{w}{\partial \overline{u}}/{\partial z}$, can contribute to drag \citep{jelly2014turbulence}. In particular, on the windward side of the traveling wave where the magnitude of $\overline{w}$ is larger, the term $-\overline{w}{\partial\overline{u}}/{\partial z}$ is positive and, therefore, a drag penalty.

\begin{figure}		
    \centering
    \makebox[\linewidth][c]{%
    \subfigure[]{\label{fig:omx_yz_vector_scaled_windward}
    \includegraphics[height=100pt,scale=1,trim={0 0 0cm 0},clip]{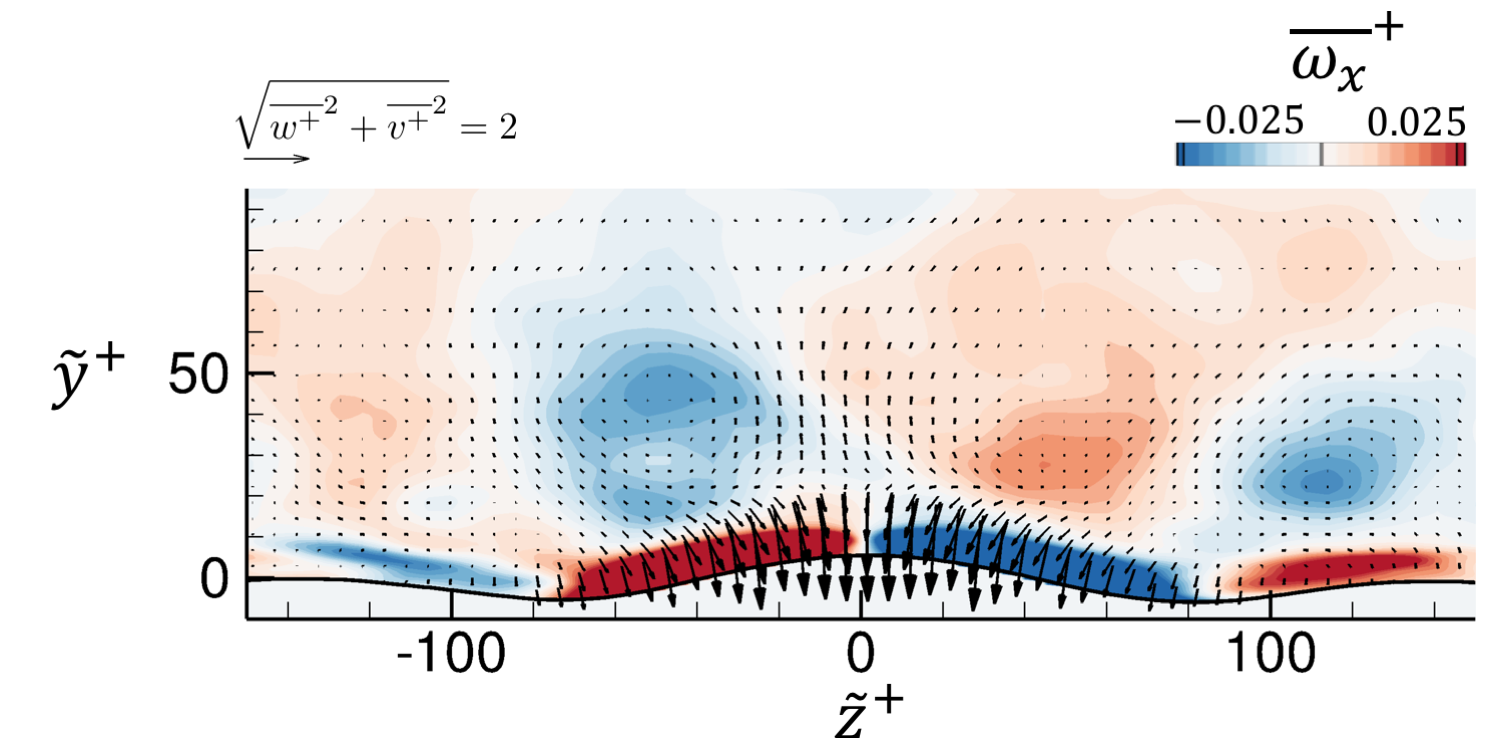} } 
    \vspace{-10pt}
    \subfigure[]{\label{fig:omx_yz_vector_uniform_windward}
    \includegraphics[height=100pt,scale=1,trim={2cm 0cm 0cm 0},clip]{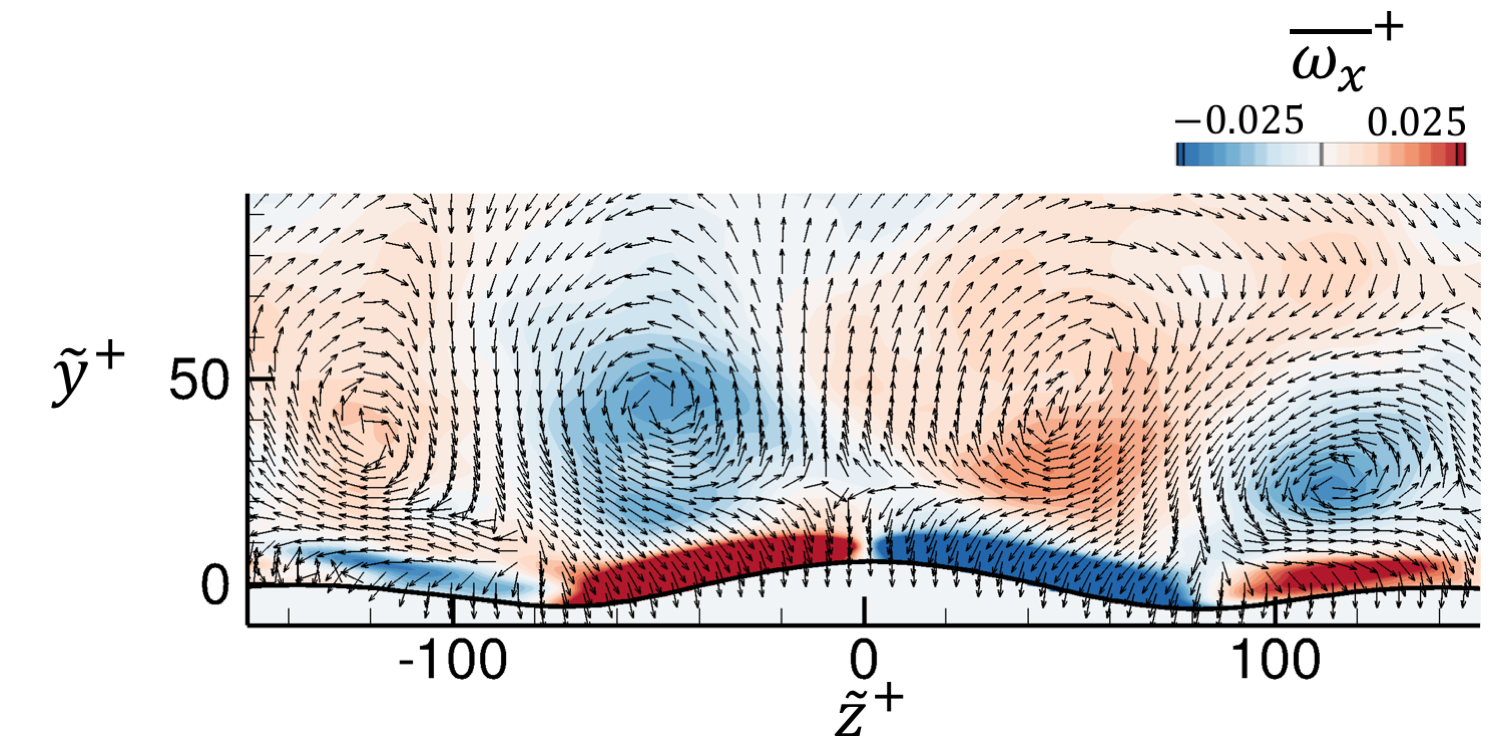} } }
    \makebox[\linewidth][c]{%
    \subfigure[]{\label{fig:omx_yz_vector_scaled_leeside}
    \includegraphics[height=100pt,scale=1,trim={0 0 0cm 0},clip]{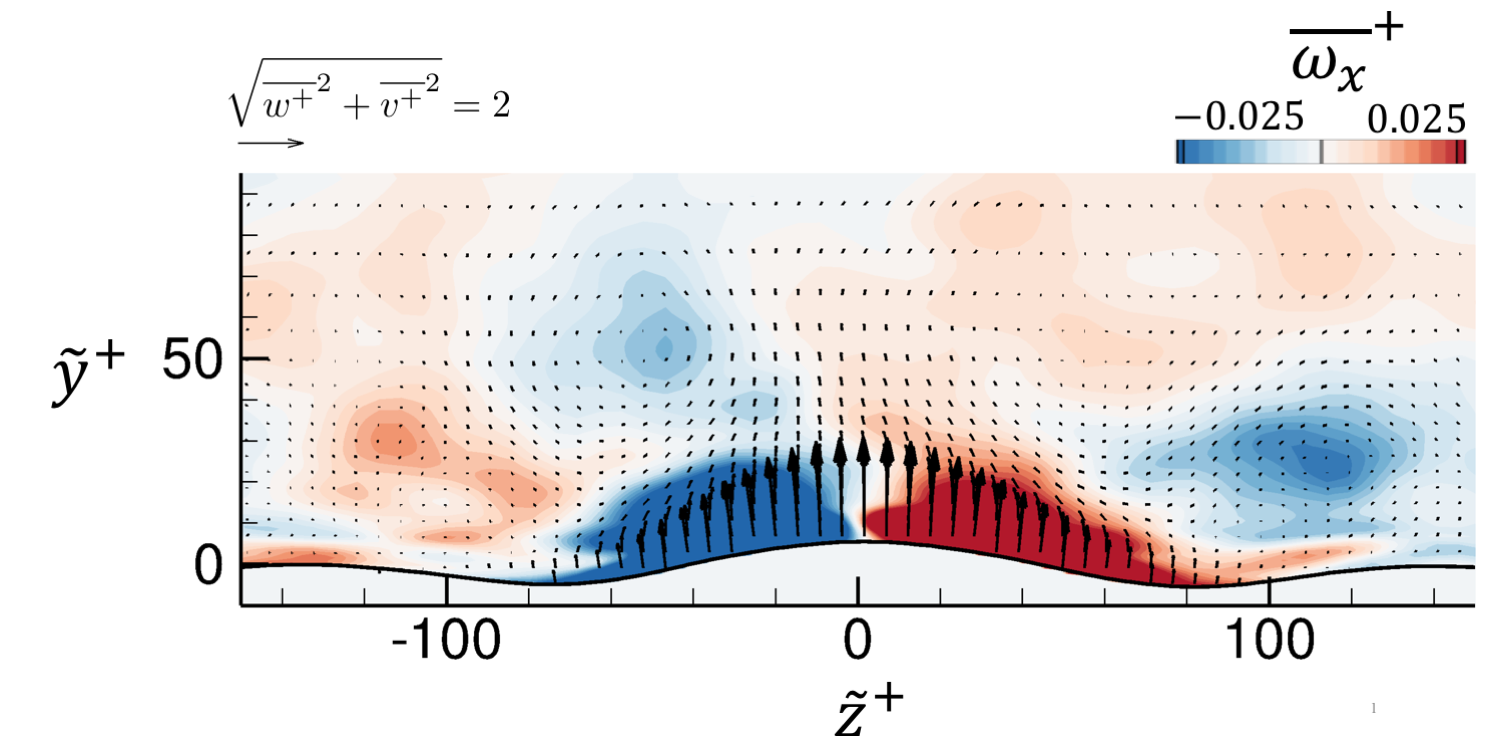} } 
    \subfigure[]{\label{fig:omx_yz_vector_uniform_leeside}
    \includegraphics[height=100pt,scale=1,trim={2cm 0cm 0cm 0},clip]{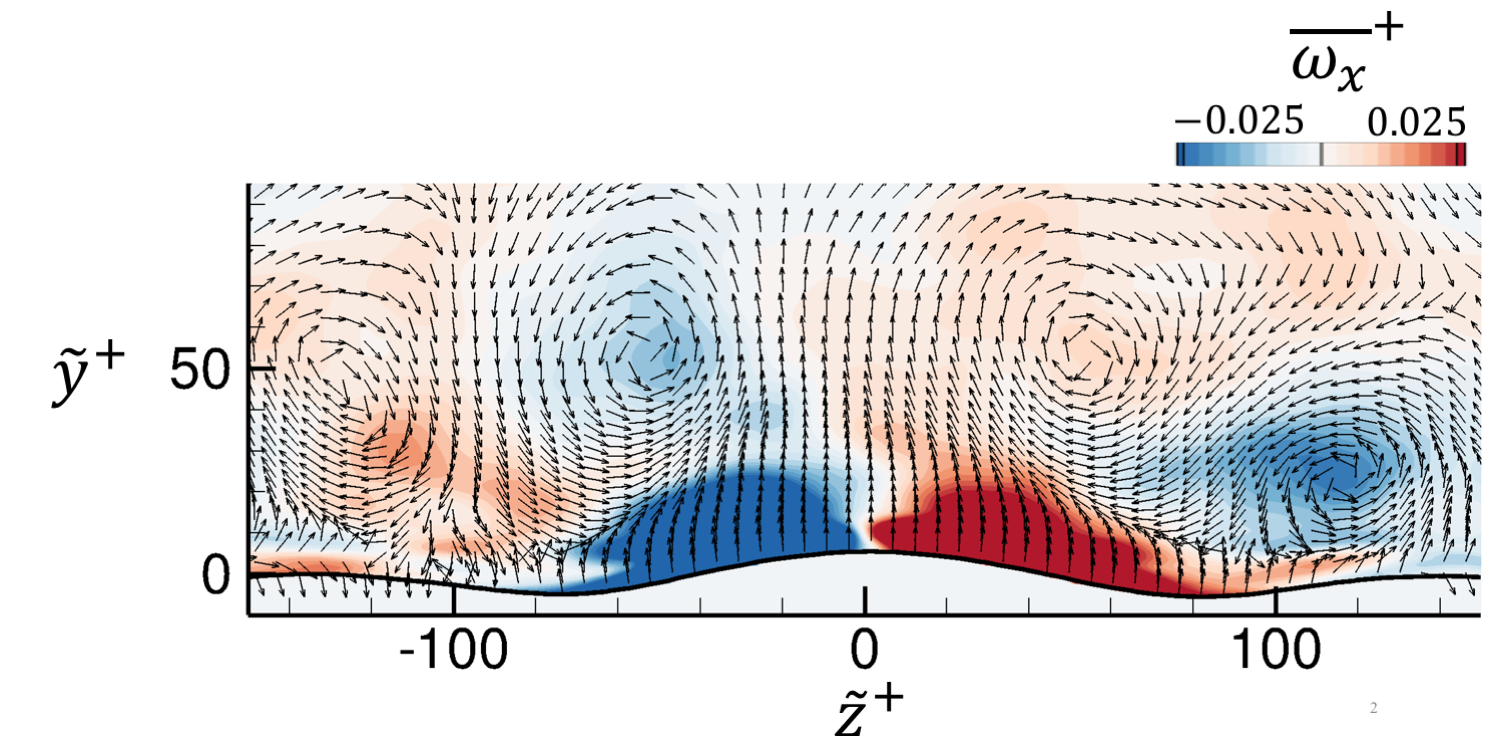} } } 
    \makebox[\linewidth][c]{%
    \subfigure[]{\label{fig:ux_yz_leeside}
    \includegraphics[height=100pt,scale=1,trim={0 0 0cm 0},clip]{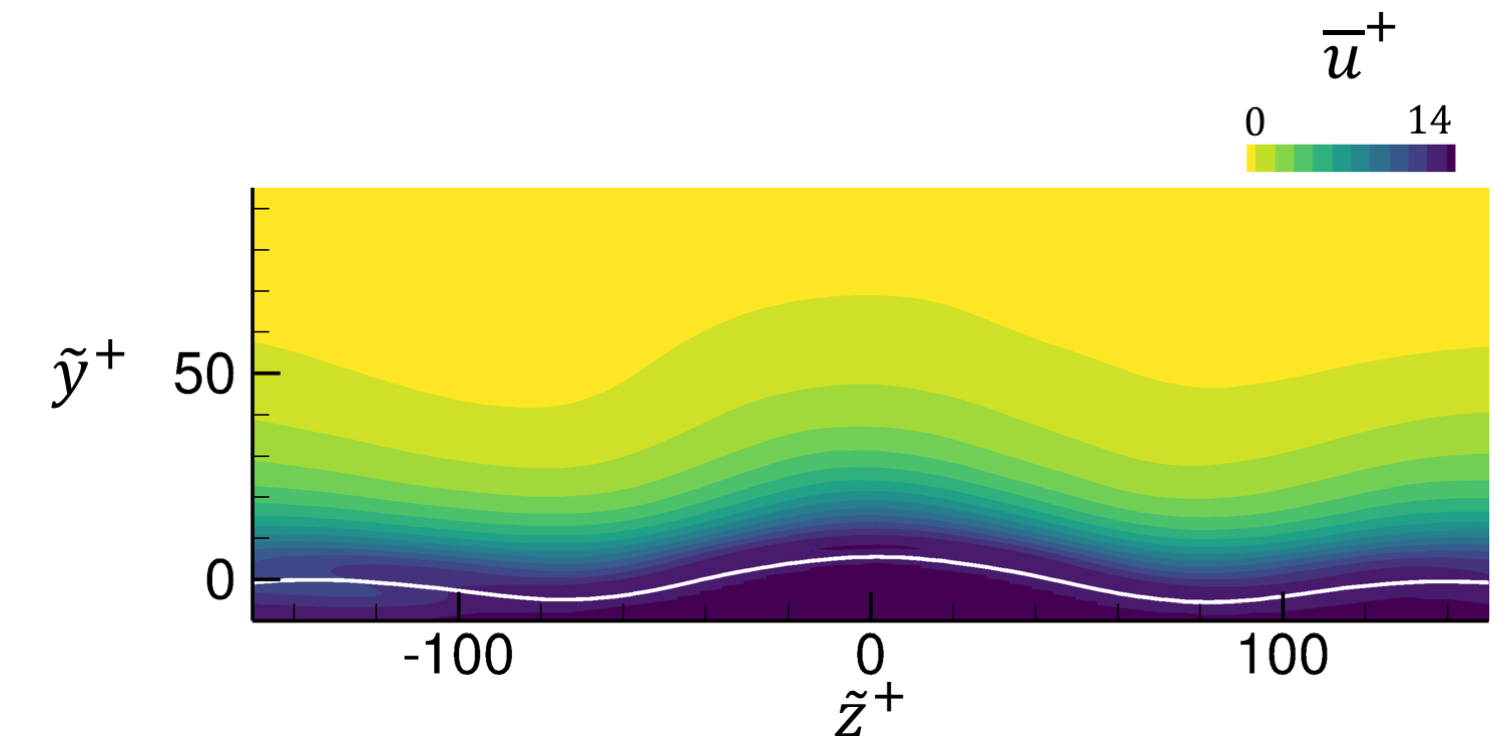} } 
    \subfigure[]{\label{fig:uprimewprime_yz_leeside}
    \includegraphics[height=100pt,scale=1,trim={2cm 0cm 0cm 0},clip]{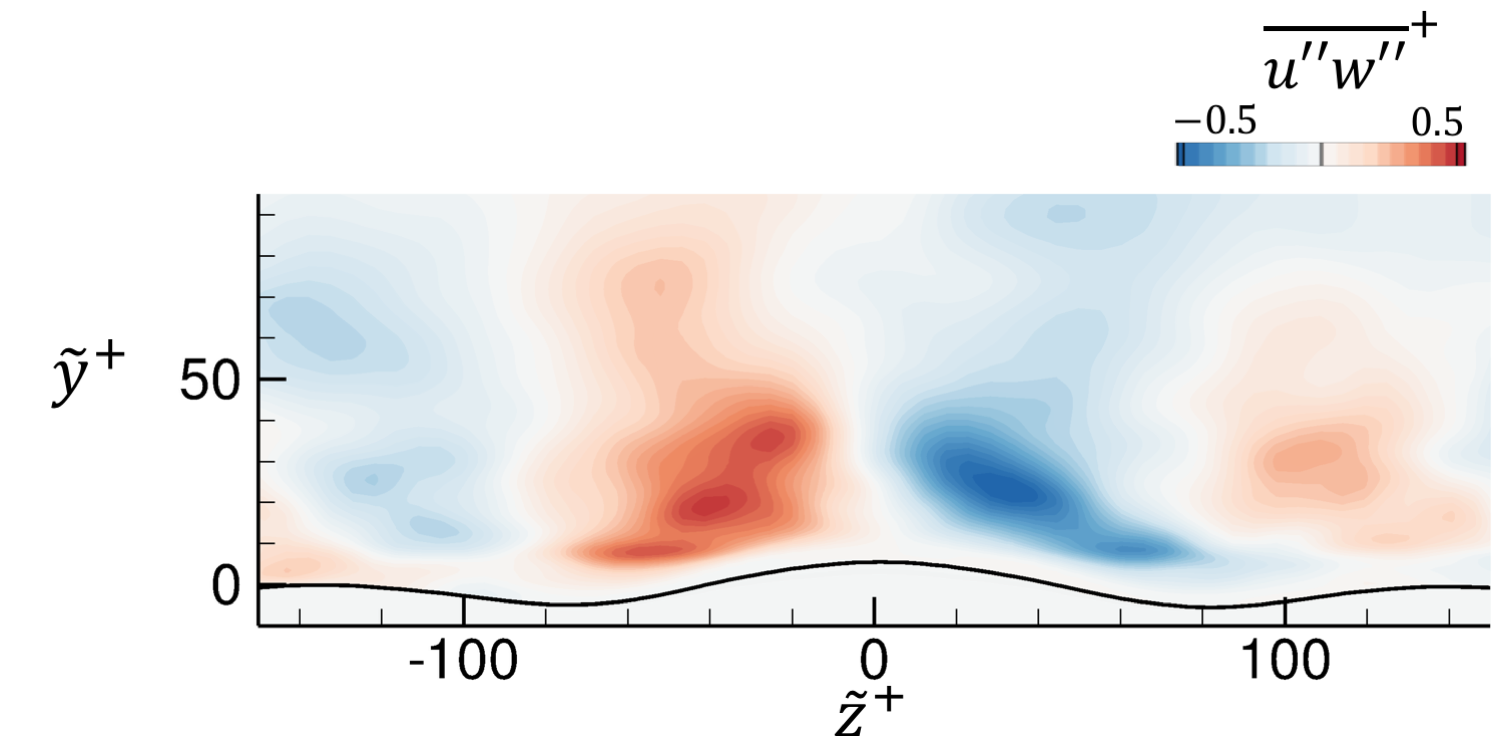} } }
    \caption{Phase-averaged contours of (a-d) streamwise vorticity $\overline{\omega_x}^+$, (e) streamwise velocity $\overline{u}^+$ and (f) $\overline{u''w''}^+$. Vectors are in-plane phase-averaged velocity $(\overline{w}^+,\overline{v}^+)$. Vector lengths in (a,c) are scaled by the magnitude of the in-plane velocity and in (b,d) are uniform. The streamwise locations are (a-b) in the windward side, $\tilde{x}^+ = -17$, and (c-f) in the lee side, $\tilde{x}^+=17$, of the streamwise wave.   \label{fig:omx_ux_uprimewprime}}
\end{figure}

The three-dimensionality of the flow is further examined by plotting phase-averaged fields in cross-flow planes on the windward (figures \crefrange{fig:omx_yz_vector_scaled_windward}{fig:omx_yz_vector_uniform_windward}) and lee  (\crefrange{fig:omx_yz_vector_scaled_leeside}{fig:uprimewprime_yz_leeside}) sides of the wave. 
Of particular interest is the phase-averaged streamwise vorticity which can capture secondary flows that may arise due to inhomogeneity in near-wall turbulence \citep{perkins1970formation}. For compliant walls, an additional source of streamwise vorticity is the spatial variations in surface velocity, specifically $\partial \overline{v}_{s,\eta} / \partial {\zeta}$, where $\zeta$ is the unit-tangent vector to the surface in the cross-flow plane. In the figure, vectors show the in-plane components of the phase-averaged flow velocity. Since the range of magnitudes differs appreciably near to and away from the interface, the vector fields are plotted twice, with vector lengths proportional to the velocity magnitudes in figures \ref{fig:omx_yz_vector_scaled_windward} and \ref{fig:omx_yz_vector_scaled_leeside}, and using uniform vector lengths in figures  \ref{fig:omx_yz_vector_uniform_windward} and \ref{fig:omx_yz_vector_uniform_leeside}.  
On the windward side, strong patterns of $\overline{\omega_x}$ with opposite signs are observed at the surface, which are primarily generated due to the gradient of wall-normal surface velocities, $\partial \overline{v}_{s,\eta} / \partial {\zeta}$ (figure \ref{fig:omx_yz_vector_scaled_windward}). The sign of this near-surface vorticity is reversed on the lee side, again due to the surface motion.  We will focus our attention on the weaker pattern of $\overline{\omega_x}$ that is observed away from the surface, which does not reverse direction from the windward to the lee side. The uniform vector fields show an associated pair of counter-rotating streamwise vortices whose cores coincide with the local extrema of $\overline{\omega_x}$, and whose core-to-core spacing is roughly half the dominant wavelength of the spanwise surface undulations.

The outer vortex motion is best characterized as Prandtl’s secondary flows of the second kind, and is comparable to those observed above streamwise-aligned riblets \citep{goldstein1998secondary} and superhydrophobic textures \citep{jelly2014turbulence}. It is generated due to the inhomogeneity in Reynolds stresses in the span. In particular, $\partial^2 \overline{w''w''} /\partial{y} \partial{z}$ is a source term in the streamwise vorticity equation, and changes sign across the spanwise crest. The pair of counter-rotating vorticies are therefore generated by the turbulence above the spanwise surface undulation, and are largely independent of the streamwise phase: the secondary-flow vorticies away from the surface are similar in figures \ref{fig:omx_yz_vector_uniform_windward} and \ref{fig:omx_yz_vector_uniform_leeside}.

With the above description in mind, it is helpful to recall the pattern of the spanwise surface velocities (figure \ref{fig:PA_wfluid_3d}). The stronger $\overline{w}$ on the windward side arises because the spanwise motion of the surface (left panel in figure \ref{fig:surface_vel_3d}) has the same sign as the spanwise velocity of the outer secondary vortex near $\tilde{y}^+ \approx 25$.  In contrast, on the lee side of the wave, there is a reversal in the spanwise wall velocity alone while the outer secondary flow remains unchanged. As a result, their superposition leads to a weaker $\overline{w}$.

Another implication of the spanwise surface undulations is the generation of a stochastic Reynolds stress $\overline{u''w''}$ (figure \ref{fig:uprimewprime_yz_leeside}). Patterns of positive and negative $\overline{u''w''}$ on both sides of the crest are visible, and arise due to turbulence production against the spanwise shear, $-\overline{w''w''} \partial \overline{u}/\partial{z}$ (figures \ref{fig:uprimewprime_yz_leeside} and \ref{fig:ux_yz_leeside}). A fluid parcel which is transported by a stochastic perturbation from the crest towards the positive $z$ direction ($w''>0$) carries low-momentum fluid towards the high-momentum zone ($u''<0$). This and the opposite motion results in $\overline{u''w''}<0$. Conversely, $\overline{u''w''}>0$ is generated on the other side of the crest. The spanwise gradient of $\overline{u''w''}$ is therefore negative near $\tilde{z} = 0$, implying that the lateral turbulent transport above the crest results in a drag penalty. This effect, however, is expected to be canceled by the negative drag contribution over the spanwise troughs where ${\partial \overline{u''w''}}/{\partial z}>0$.

These results highlight the central role of the propagation of quasi two-dimensional surface waves in studying the interaction of turbulence with the compliant surface. In addition to the roughness effect and the incurred form drag, the surface velocities modify the flow structures near the interface and up to the log-layer, contribute to the flux of vorticity at the interface, and gives rise to a net positive pressure work onto the flow. While the surface waves are spanwise-elongated, they also express low-wavenumber spanwise undulations. The three-dimensionality of the wave motion generates streamwise vorticity near the interface, and a secondary outer flow with an associated spanwise inhomogeneity.

\section{Summary and conclusions \label{sec:conclusion}}
The interaction of channel-flow turbulence with a compliant wall was examined using direct numerical simulations in an Eulerian-Eulerian framework. The compliant layer was an incompressible viscous hyper-elastic material.  We considered layers with different thicknesses and elastic shear moduli, selected based on the response of compliant coatings in one-dimensional linear models \citep{benschop2019deformation,chase1991generation}, and also considered two Reynolds numbers. Consistent with the recent experimental \citep{wang2020interaction} and numerical \citep{rosti2017numerical} efforts, we observed enhanced turbulence intensity, which resulted in reduced streamwise momentum and a drag increase. We showed that, in a surface-fitted coordinate, the wall compliance gives rise to a downward shift of the log-layer without a significant impact on the viscous sublayer. Spanwise-elongated deformations of the surface propagated as waves in the streamwise direction, and their impact on the flow was investigated through the lens of wave-turbulence interactions.

The surface deformation spectra showed a band of streamwise-advected waves with phase-speed equal to Rayleigh waves. 
The range of energetic wavenumbers was shifted to higher values in the case with a thinner layer, and the stiffer compliant material sustained higher wave speeds. The design of the higher Reynolds-number case aimed at constant values of $G^+$ and $L_e^+$, which led to similar spectra thus demonstrating the relevance of the viscous scaling for the material parameters. 
In the spanwise direction, most of the surface energy was concentrated at low wavenumbers without a clear indication of spanwise-travelling modes. The range of excited frequencies in $k_z-\omega_t$ spectra coincided with that in $k_x-\omega_t$ spectra, which supports the view that the streamwise travelling waves set the frequency response. 
It is of note that spanwise wave propagation was also occasionally observed in the time series of the flow field, but much less discernible than the downstream propagating counterpart. The evolution of the surface directly impacted the pressure field, which was captured in the pressure spectra and the deformation-pressure cross-spectra.

The travelling Rayleigh waves in the compliant material were comprised of out-of-phase wall-normal and streamwise velocities whose influence penetrates deep into the flow. Visualizations of the instantaneous vorticity field in the frame of the wave showed frequent shear-layer detachment that was initiated near the trough. The detached layer rolled up near the critical layer (where the flow speed is equal to the Rayleigh wave speed) accompanied by a local pressure drop. The origin of these detachment events was studied by evaluating two sources of spanwise-vorticity flux at the surface: the pressure gradient and a nearly out-of-phase surface acceleration due to the Rayleigh waves. For small amplitude waves, the pressure gradient term was dominant and, similar to static roughness, was favorable on the windward side and adverse on the less side. For large amplitude waves, the contribution by surface acceleration became significant, and the stabilized flow region shifted towards the lee side of the wave. 

The wall motion exerted a net positive pressure work onto the fluid. The energy flux was due to surface-normal motion, and did not lead to streamwise momentum gain. In fact, the asymmetry of pressure with respect to the crest of the waves resulted in form drag that opposed the near-wall streamwise momentum. The Rayleigh waves were also affected by the flow, in particular their streamwise velocity component.  As a result, the compliant material sustained a negative wave-correlated Reynolds shear stress. This wave-correlated shear stress rapidly changed sign above the wave boundary later and below the critical layer. The stochastic shear stress was also substantial in this region due to the unsteady shear-layer detachment events. 

While the surface waves were primarily two-dimensional, the surface also exhibited low wavenumber spanwise undulations. 
The associated phase-averaged flow were examined: 
The streamwise velocity was relatively slow above the spanwise peaks and fast above the spanwise troughs.  
Strong streamwise vorticity was generated by the surface motion, specifically the spanwise gradient of the surface-normal velocity. The pair of opposite-sign vorticity on the windward side reversed sign on the lee side of the wave. Away from the surface, counter-rotating vortices signaled the presence of a secondary flow.
In addition, the spanwise gradient of the mean streamwise velocity led to the generation of stochastic $u''w''$ stresses. All these flow features in turn modified the wall-normal and lateral transport of momentum and, in turn, drag.

On the other hand, we also showed that the surface acceleration contributed a spanwise vorticity flux which is out-of-phase with respect to the pressure gradient along the surface topography.  
This flux can potentially be harnessed to stabilize the flow and mitigate the unsteady detachment of near-wall vorticity.  
Finally, the herein discussed results can guide future studies and designs of compliant coatings for turbulent-flow applications, and may motivate new innovative designs, for example that exploit anisotropic material properties to suppress or promote particular effects.

\par\bigskip
\noindent
\textbf{Funding.} This work was supported in part by the Office of Naval Research (grant N00014-20-1-2778). 
Computational resources were provided by the Maryland Advanced Research Computing Center (MARCC).

\par\bigskip
\noindent
\textbf{Declaration of interests.} 
The authors report no conflict of interest.
 
\par\bigskip
\noindent
\textbf{Author ORCIDs.} \\
Amir Esteghamatian, \url{https://orcid.org/0000-0002-5072-2605} \\
Joseph Katz, \url{https://orcid.org/0000-0001-9067-2473} \\
Tamer A. Zaki, \url{https://orcid.org/0000-0002-1979-7748}

\newpage 

\appendix
\section{Validation of the numerical algorithm \label{app:validation}}
The level-set algorithm for capturing the material-fluid interface was extensively validated \citep{jung2015effect}. An additional validation case is presented here, where the deformation of a neo-Hookean elastic particle subjected to a simple shear flow is simulated and compared with data by \citet{villone2014simulations}. 

An initially spherical neo-Hookean elastic particle with undeformed radius $r_0^\star$ is placed at the center of a cubical domain. The ratio between the undeformed radius of the particle and the wall-normal height of the domain $H^\star$ is $r_0^\star/H^\star=0.1$. The flow is induced by two parallel plates located at $y^\star = \{-H^\star/2, H^\star/2\}$, moving opposite to one another in the $x$ direction with the same speed, generating a constant shear rate $\dot{\gamma^\star}$. Periodicity is imposed in $x$ and $z$ directions, and the isotropic homogeneous grid resolution is set to $\Delta x^\star = r_0^\star/24$. 

\begin{figure}		
\centering
\subfigure[]{\label{fig:valid_Ca0p05}
\begin{minipage}[b]{0.3\textwidth}
\begin{center}
\includegraphics[width =\textwidth,scale=1]{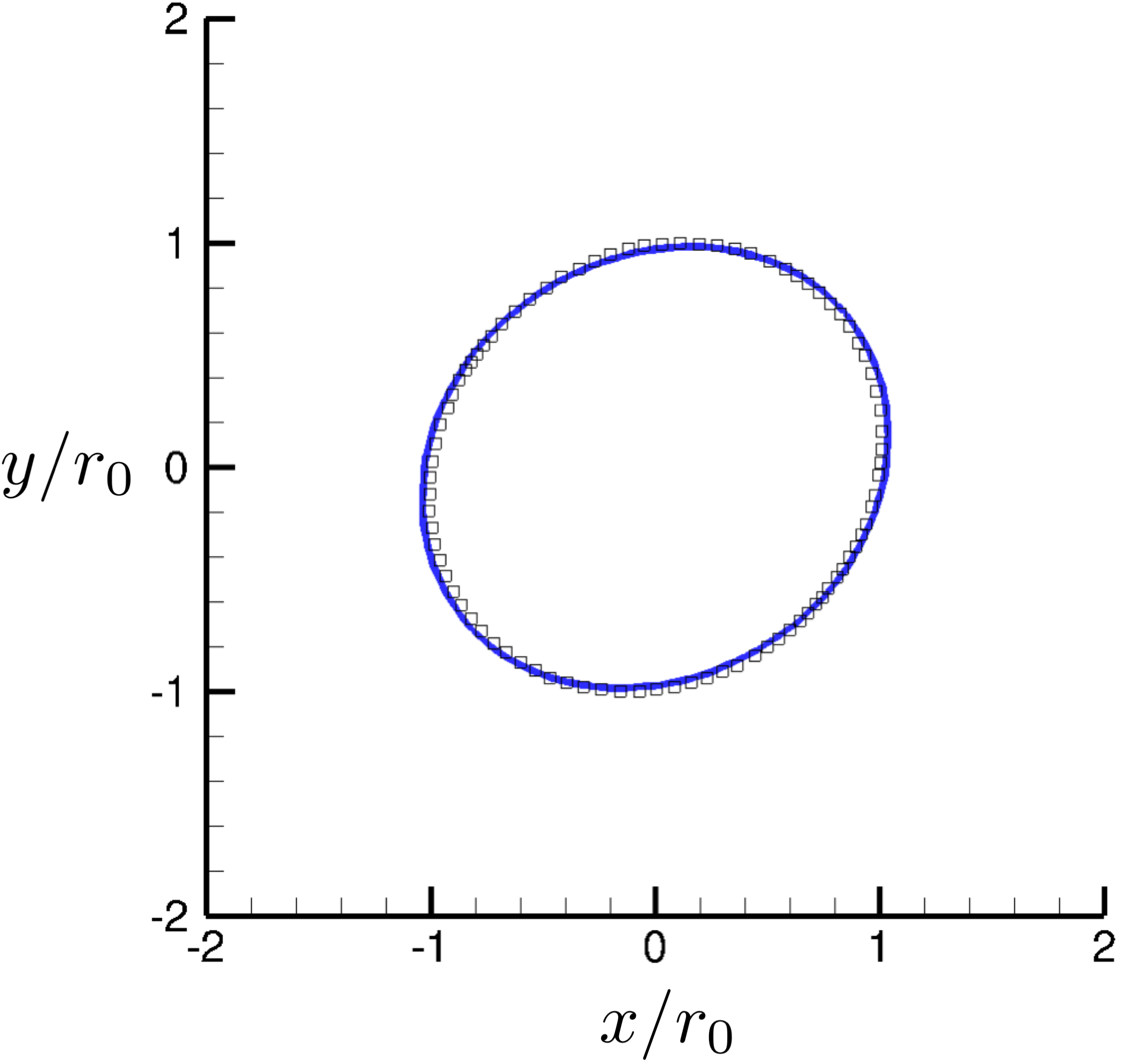}
\end{center}
\end{minipage}}	
\subfigure[]{\label{fig:valid_Ca0p2}
\begin{minipage}[b]{0.3\textwidth}
\begin{center}
\includegraphics[width =\textwidth,scale=1]{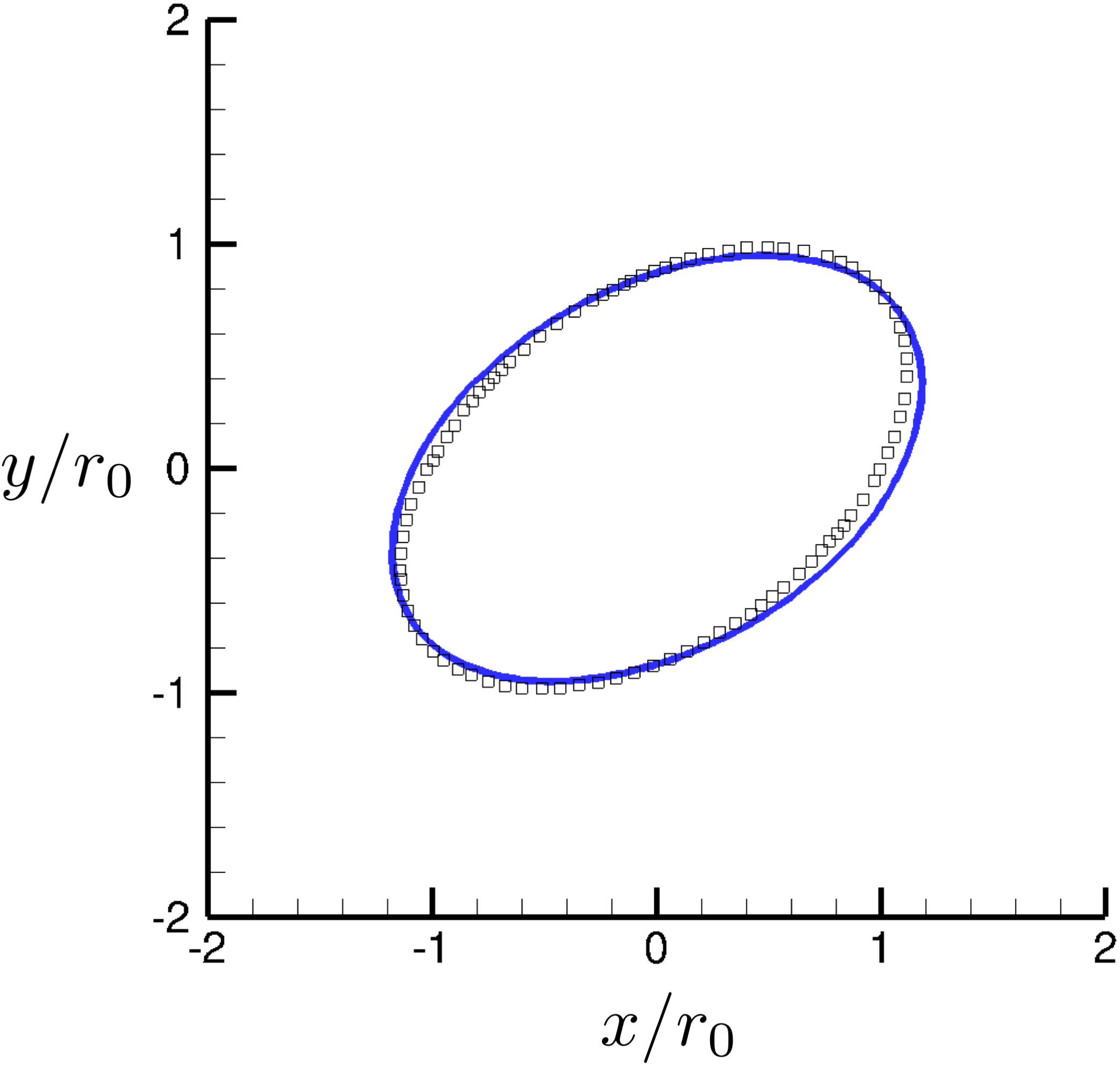}
\end{center}
\end{minipage}}		
\subfigure[]{\label{fig:valid_Ca0p35}
\begin{minipage}[b]{0.3\textwidth}
\begin{center}
\includegraphics[width =\textwidth,scale=1]{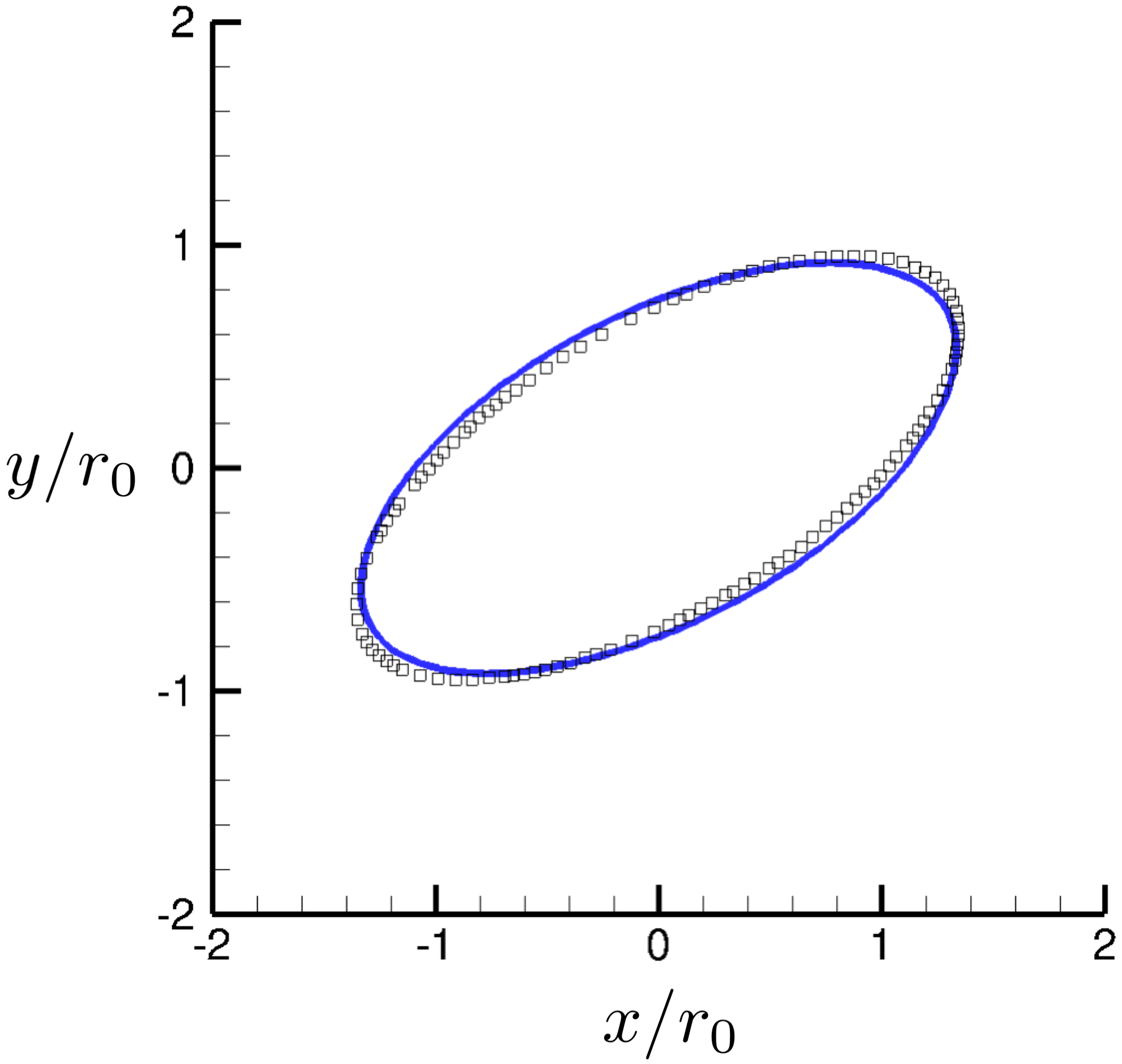}
\end{center}
\end{minipage}}	
\caption{Shape of a deformed neo-Hookean particle in a Newtonian fluid under confined shear flow. (a) $Ca=0.05$, (b) $Ca=0.2$, and (c) $Ca=0.35$. The fluid/solid interface is projected onto the plane of shear located at $z=0$, and the axes are normalized by the undeformed radius of the particle. Present simulations (solid blue line) are compared with the data by \citet{villone2014simulations} (square symbols).  }   \label{fig:validation}	
\end{figure}

The particle deforms due to the applied shear flow, until it reaches a steady-state ellipsoid-like shape. The Reynolds number is sufficiently small to avoid inertial effects ($\Rey_{r} \equiv {r_0^\star}^2\dot{\gamma^\star}/\nu^\star = 0.025$), and three different elastic capillary numbers $Ca \equiv \mu^\star \dot{\gamma^\star}/G^\star = \{0.05, 0.2, 0.35\}$ are simulated for comparison with data by \citet{villone2014simulations}. As shown in figure \ref{fig:validation}, agreement between the experimental results and our numerical predictions in particle deformation is satisfactory. 

\section{Wall-following coordinate \label{app:wallcoordinate}}
A surface-fitted coordinate is introduced in order to probe the wave-induced motions near the interface. We define a coordinate system that follows the interface near the compliant surface and smoothly transitions to a laboratory Cartesian coordinate away from the surface. Such coordinate system is particularly of interest for horizontal averaging, were the $y$ location of data points in a Cartesian coordinate is not an accurate measure of the distance to the surface. The adopted coordinate transformation is widely used in analysis of experimental measurements above ocean waves \citep{hara2015wave,yousefi2020boundary}. 

\begin{figure}		
    \centering
    \includegraphics[width =0.7\textwidth,scale=1]{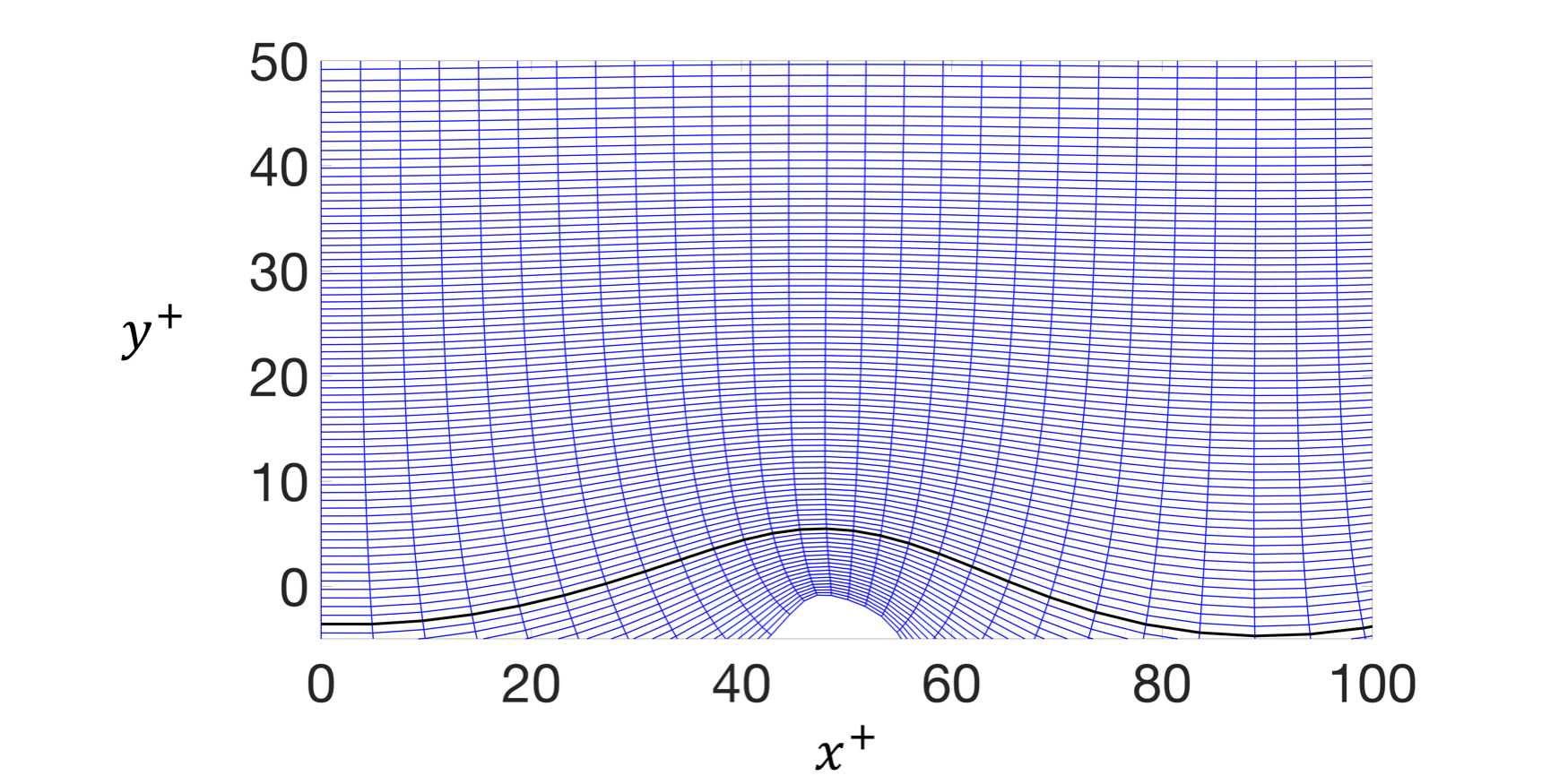}			
    \caption{Iso-surfaces of $(\xi,\eta)$ defined in (\ref{eq:xi}-\ref{eq:eta}), shown above a wave crest. Fluid/solid interface is marked by the black line.   \label{fig:vis_wall_coordinate}	} 
\end{figure}

Following \citet{benjamin1959shearing}, the surface displacement at each spanwise location is first decomposed into corresponding spatial Fourier components; i.e.~$d(x) =  \Sigma_n a_n\exp (i(k_n)x + \phi_n)$, where $a_n$, $k_n$, and $\phi_n$ are amplitude, wavenumber and phase of the $n^\text{th}$ mode. The orthogonal coordinate $(\xi,\eta)$ is defined by
\begin{align}
   & \xi = x - i\Sigma_n a_n\exp (i k_n x + \phi_n) \exp(-k_n \eta), \label{eq:xi} \\
   & \eta = y -\Sigma_n a_n\exp (i k_n x + \phi_n) \exp(-k_n \eta). \label{eq:eta}
\end{align}
To first order in $k_na_n$, the actual wave profile is given by $\eta = 0$. At higher distances from the surface, and proportional to the surface wavenumbers, iso-lines of $\eta$ smoothly tend to Cartesian horizontal lines of $y$. As an example, iso-surfaces of $(\xi,\eta)$ are shown above a crest in figure \ref{fig:vis_wall_coordinate}. 

\bibliographystyle{jfm}
\bibliography{CompliantRefs} 

\end{document}